\documentclass[fleqn,twocolumn,tighten,apj,twocolappendix]{aastex63} 
	
\usepackage[english]{babel}
\usepackage{blindtext}
\usepackage{CJKutf8}

\usepackage{amsmath,bm,nccmath,textcomp,gensymb}

\usepackage{scalerel}

\usepackage{hyperref,multirow,dsfont,balance,enumitem,tabularx,booktabs}

\usepackage[titles]{tocloft}
\usepackage[titletoc]{appendix}

\newcommand\HII{H\protect\scaleto{$II $}{1.2ex}}
\DeclareMathOperator\erfc{erfc}
\DeclareMathOperator{\arcsinh}{arcsinh}

\newcommand{\crowdsource}{\texttt{crowdsource }}
\newcommand{\cloudcoverr}{\textsc{CloudCovErr.jl }}
\newcommand{\flux}{\texttt{flux }}
\newcommand{\cflux}{\texttt{cflux }}

\newcommand{\crowdsourcens}{\texttt{crowdsource}}
\newcommand{\cloudcoverrns}{\textsc{CloudCovErr.jl}}
\newcommand{\fluxns}{\texttt{flux}}
\newcommand{\cfluxns}{\texttt{cflux}}

\shorttitle{DECaPS2}
\shortauthors{Saydjari et al.}

\graphicspath{{./}{figures/}}

\begin{document}

\title{The Dark Energy Camera Plane Survey 2 (DECaPS2): More Sky, Less Bias, and Better Uncertainties}

\correspondingauthor{Andrew Saydjari}
\email{andrew.saydjari@cfa.harvard.edu}

\author[0000-0002-6561-9002]{Andrew K. Saydjari}
\affiliation{Department of Physics, Harvard University, 17 Oxford St., Cambridge, MA 02138, USA}
\affiliation{Harvard-Smithsonian Center for Astrophysics, 60 Garden St., Cambridge, MA 02138, USA}

\author[0000-0002-3569-7421]{Edward F. Schlafly}
\affiliation{Lawrence Livermore National Laboratory, 7000 East Ave., Livermore, CA 94550-9234} 

\author[0000-0002-1172-0754]{Dustin Lang}
\affiliation{Perimeter Institute for Theoretical Physics, 31 Caroline Street N, Waterloo, ON N25 2YL, Canada}
\affiliation{Department of Physics and Astronomy, University of Waterloo, Waterloo, ON N2L 3G1, Canada}

\author[0000-0002-1125-7384]{Aaron M. Meisner}
\affiliation{NSF's National Optical-Infrared Astronomy Research Laboratory, 950 N. Cherry Ave., Tucson, AZ 85719, USA}

\author[0000-0001-5417-2260]{Gregory M. Green}
\affiliation{Max Planck Institute for Astronomy, K\"{o}nigstuhl 17, D-69117 Heidelberg, Germany}

\author[0000-0002-2250-730X]{Catherine Zucker}
\altaffiliation{Hubble Fellow}
\affiliation{Space Telescope Science Institute, 3700 San Martin Drive, Baltimore, MD 21218, USA}
\affiliation{Harvard-Smithsonian Center for Astrophysics, 60 Garden St., Cambridge, MA 02138, USA}

\author[0000-0002-7588-976X]{Ioana Zelko}
\affiliation{Harvard-Smithsonian Center for Astrophysics, 60 Garden St., Cambridge, MA 02138, USA}
\affiliation{Department of Physics and Astronomy, University of California-Los Angeles,
475 Portola Plaza, Los Angeles, CA 90095
}

\author[0000-0003-2573-9832]{Joshua S. Speagle (\begin{CJK*}{UTF8}{gbsn}沈佳士\ignorespacesafterend\end{CJK*})}
\altaffiliation{Banting \& Dunlap Fellow}
\affiliation{Department of Statistical Sciences, University of Toronto, 100 St George St, Toronto, ON M5S 3G3, Canada}
\affiliation{David A. Dunlap Department of Astronomy \& Astrophysics, University of Toronto, 50 St George Street, Toronto ON M5S 3H4, Canada}
\affiliation{Dunlap Institute for Astronomy \& Astrophysics, University of Toronto, 50 St George Street, Toronto, ON M5S 3H4, Canada}

\author[0000-0002-6939-9211]{Tansu Daylan}
\affiliation{Department of Astrophysical Sciences, Princeton University, 4 Ivy Lane, Princeton, NJ 08544}

\author[0000-0002-1441-770X]{Albert Lee}
\affiliation{Institute for Disease Modeling, Bill \& Melinda Gates Foundation, 500 Fifth Ave N, Seattle, WA 98109}

\author[0000-0001-5567-1301]{Francisco Valdes}
\affiliation{NSF's National Optical-Infrared Astronomy Research Laboratory, 950 N. Cherry Ave., Tucson, AZ 85719, USA}

\author[0000-0002-5042-5088]{David Schlegel}
\affiliation{Physics Division, Lawrence Berkeley National Laboratory, 1 Cyclotron Road, Berkeley, CA, 94720}

\author[0000-0003-2808-275X]{Douglas P. Finkbeiner}
\affiliation{Department of Physics, Harvard University, 17 Oxford St., Cambridge, MA 02138, USA}
\affiliation{Harvard-Smithsonian Center for Astrophysics, 60 Garden St., Cambridge, MA 02138, USA}

\begin{abstract}
Deep optical and near-infrared imaging of the entire Galactic plane is essential for understanding our Galaxy's stars, gas, and dust. The second data release of the DECam Plane Survey (DECaPS2) extends the five-band optical and near-infrared survey of the southern Galactic plane to cover $6.5\%$ of the sky, $|b|\leq 10\degree$ and $6\degree > \ell > -124°\degree$, complementary to coverage by Pan-STARRS1. Typical single-exposure effective depths, including crowding effects and other complications, are 23.5, 22.6, 22.1, 21.6, and 20.8 mag in $g$, $r$, $i$, $z$, and $Y$ bands, respectively, with around 1 arcsecond seeing. The survey comprises 3.32 billion objects built from 34 billion detections in 21.4 thousand exposures, totaling 260 hours open shutter time on the Dark Energy Camera (DECam) at Cerro Tololo. The data reduction pipeline features several improvements, including the addition of synthetic source injection tests to validate photometric solutions across the entire survey footprint. A convenient functional form for the detection bias in the faint limit was derived and leveraged to characterize the photometric pipeline performance. A new post-processing technique was applied to every detection to de-bias and improve uncertainty estimates of the flux in the presence of structured backgrounds, specifically targeting nebulosity. The images and source catalogs are publicly available at \href{http://decaps.skymaps.info/}{http://decaps.skymaps.info/}.
\end{abstract}


\tableofcontents

\section{Introduction} \label{sec:Intro}

\subsection{Background} \label{sec:back}

Most of the stars and dust in the Milky Way are located in the Galactic disk. Yet, the high density of stars makes the disk difficult to study, requiring analyses to simultaneously model many sources in order to optimally measure stellar positions and fluxes. Moreover, large column densities of dust (and gas) along the line of sight cause significant extinction, severely limiting the maximum distance of detectable stars in optical wavelengths \citep{Green:2019:ApJ:}. Measurements in near-infrared (NIR) wavelengths can partially mitigate this problem and reach greater distances, because dust extinction impacts the optical more than the NIR \citep{Draine_2003_ARA&A}. However, variations in dust extinction (reddening), thought to be related to varying chemical composition and/or grain size distributions \citep{Weingartner2001,Zelko2020}, are most prominent in optical wavelengths \citep{Cardelli:1989:ApJ:,Schlafly:2016:ApJ:}. Thus, deep photometric surveys spanning a broad wavelength range (optical to NIR) are essential to understanding the composition and three-dimensional structure of the Galaxy.  While Pan-STARRS1 (PS1) surveyed 75\% of the Galactic plane \citep{Chambers:2016:arXiv:}, 25\% ($\ell \sim -5\degree$ to $-95\degree$) remained unmeasured at a comparable photometric depth.

\begin{figure*}[t]
\centering
\includegraphics[width=\linewidth]{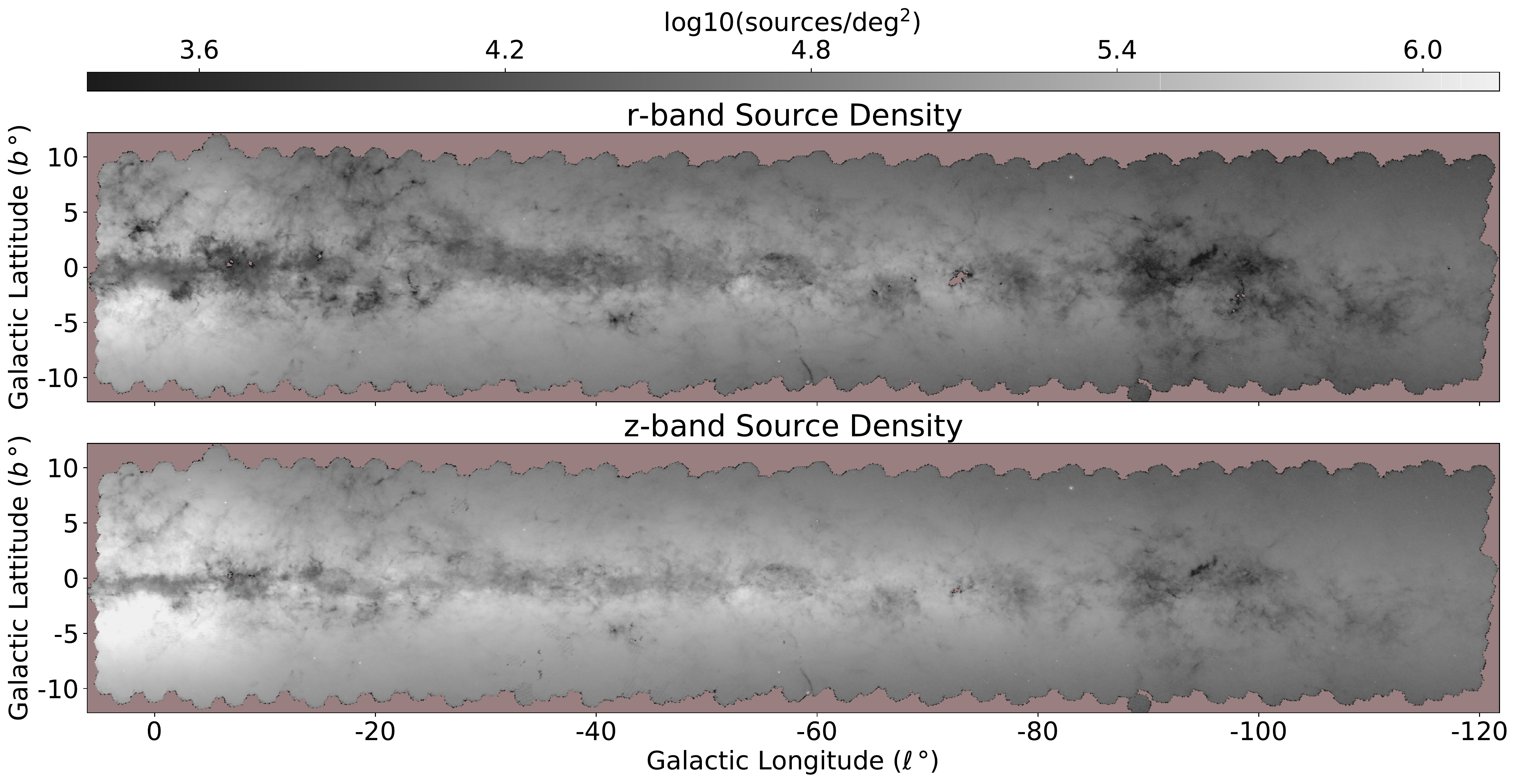}
\caption{DECaPS2 source density map (using a HEALPix grid at NSide = 512) in $r$-band (top) and $z$-band (bottom) using a common, logarithmic color scale. This figure employs a cut requiring sources be brighter than 19th magnitude in the reported band and be detected in at least one of the two immediately adjacent bands.}
\label{fig:density}
\end{figure*}

Measuring the fluxes of stars in images of the Galactic plane is complicated by nebulosity, crowding, and numerous faint sources. We broadly refer to the presence of structured emission and absorption regions, such as filaments or clouds of gas and dust (\HII regions, dark nebulae, reflection nebulae, etc.) as ``nebulosity.'' Most photometric pipelines model images as the sum of galaxies, point sources, and backgrounds that are smooth on scales much larger than the point spread function (PSF) \citep{Stetson:1987:PASP:,Bertin:1996:A&AS:,Lang:2016:ascl:,Schlafly:2018:ApJS:}. Thus, without explicit handling, nebulosity with fine spatial structure is often incorrectly modeled as the sum of many point sources or galaxies. Further, for real sources located near nebulosity, variations in flux associated with nebulosity can be incorrectly attributed to the source, biasing the estimated source flux \citep{Saydjari:2022:arXiv:}.

Images of the Galactic plane also suffer from ``crowding,'' where light from different stars substantially overlap one another in images.  At the single-exposure level, crowding complicates the identification of stars and the measurements of their fluxes. When sources have large angular separation, identifying stars via peaks in an image is easy (for large signal-to-noise ratios, or ``SNR'') and source modeling is likewise straightforward. When two sources have zero angular separation, they should be modeled as a brighter single source with sum of the flux from both sources.\footnote{In single-band, single-exposure imaging, the combined source solution at zero angular separation is the best one can do. In joint analysis of multi-band and/or multi-epoch imaging, one can do better, and the overlapping limit is less clear cut.} The objective of modeling crowded fields of stars is to accurately measure the fluxes and positions of stars in between these two limits.

Crowding also complicates catalog construction. In many surveys, like DECaPS2, modeling is performed on individual exposures, and the resulting catalogs must be combined to form a multi-band, multi-epoch catalog. Variation in the estimated source locations and the overall number of estimated sources in different exposures complicates the notion of a single object list. One solution is to create a super object list from stacks of multi-epoch images and then perform forced-photometry at those object locations \citep{Magnier:2020:ApJS:}. However, the creation of the super object list is hindered by the lack of a well defined PSF on the stacked images and by the astrometric precision of the stacking. Another solution is to identify and fit sources using all imaging touching a given location simultaneously \citep{Dey:2019:AJ:}. However, this prevents the massively parallel processing that can be used when individual exposures are processed independently. Progress has been made using transdimensional, probabilistic cataloging methods on single-exposures \citep{Brewer:2013:AJ:,Portillo:2017:AJ:} or multi-band imaging \citep{Feder:2020:AJ:,Liu:2021:arXiv:}, where the total number of sources in the image is not fixed. However, these methods are computationally expensive and have not yet been applied at scale (i.e., have only been applied to a tiny fraction of the sky).

As photometric surveys push deeper to observe fainter stars, a larger number of stars in the resulting catalog will be near the detection threshold. This is because the probability distribution of apparent stellar flux is approximately a power law, increasing in probability with decreasing stellar flux \citep{Gorbikov:2010:Ap&SS:}. The detection threshold is often an explicit cut on the SNR of a peak relative to the background used during source identification \citep[e.g.][]{Schlafly:2018:ApJS:}. Even probabilistic methods have an implicit detection threshold set by the evidence required to predict a source with reasonable probability. However, only identifying sources above a threshold introduces a selection bias in sources with flux near the threshold. For example, for a source with true flux below threshold, only observations of the source where the realization of the noise deviates high will be identified, biasing the flux estimate high. 

Another faint-limit bias arises from the common practice of using maximum-likelihood approaches to identify source locations \citep{Portillo:2020:AJ:}. The maximum-likelihood location prefers a source center capturing more of the flux in the image, even if that flux is noise, and thus biases flux estimates high. While both faint-limit biases described above can be partially mitigated by multi-epoch approaches, it is imperative to understand their form and realization in practice at the single-detection level to predict how these biases are modified by multi-epoch methods. Further, it is a pressing challenge to the community to correctly model these biases so that, when combined with the appropriate priors, better use can be made of the statistical power of the large number of faint sources in photometric surveys.

Synthetic injection tests are an important tool for evaluating the magnitude of bias introduced by nebulosity, crowding, and faint-limit selection effects. While the importance of synthetic injection tests has long been recognized, they have only recently been applied in large surveys. The Dark Energy Camera Legacy Survey (DECaLS, \citealt{Dey:2019:AJ:}) used \texttt{Obiwan} to inject synthetic galaxies into single-epoch images across multiple bands in a single patch \citep{Kong:2020:MNRAS:}. The Hyper-Suprime Cam Subaru Strategic Program (HSC-SSP) Survey \citep{Aihara:2018:PASJ:} used \texttt{SynPipe} to inject synthetic stars and galaxies into single-epoch images across multiple bands in two test tracts \citep{Bosch:2018:PASJ:,Huang:2018:PASJ:}. The Dark Energy Survey (DES, \citealt{TheDarkEnergySurveyCollaboration:2005:arXiv:}) used \texttt{Balrog} to inject synthetic galaxies into single-epoch images across multiple bands in a random 20\% of exposures in the Year 3 release \citep{Everett:2022:ApJS:}. In this work, we will describe how DECaPS2 uses \crowdsource to inject synthetic stars into single-epoch images for a single band in a random 2\% of exposures. While injecting into only a single band prevents analysis of single-object level color biases, this restriction allows \crowdsource to perform injection tests at runtime and achieve, what is to our knowledge, the first full survey characterization of injection tests of synthetic stellar sources.

\subsection{DECaPS2} \label{sec:DECaPS2intro}

We present the second data release of the Dark Energy Camera Plane Survey (DECaPS2), which provides optical and near-infrared photometry in the Galactic plane accessible in the Southern hemisphere with $\delta\leq-24\degree$. The combination of DECaPS2 with PS1 finalizes deep optical-NIR ($g \sim 24$th to $Y \sim 21$st mag) coverage of the entire Galactic plane. The source density in selected bands is shown in Figure \ref{fig:density}. The DECaPS2 catalog contains 3.32 billion sources built from 34 billion detections using \crowdsource \citep{Schlafly:2021:ascl:}, a photometric pipeline optimized to handle crowded-field photometry.\footnote{\crowdsource has been previously used to create two of the largest photometric catalogs, DECaPS1 \citep{Schlafly:2018:ApJS:} and the unWISE Catalog \citep{Schlafly:2019:ApJS:}.} A new synthetic injection module for \crowdsource was used uniformly throughout the survey footprint to rigorously benchmark photometric performance and constrain uncertainties so as to enable interpretable downstream statistical inference.

We further develop, validate, and implement a new method to handle structured backgrounds (nebulosity) ubiquitous in the Galactic plane \citep{Saydjari:2022:arXiv:}. To do this, we perform a statistical interpolation of nebulous structures to correct both the flux and flux uncertainties of stars. We use the injection tests to quantitatively characterize the photometry as a function of blending between two neighboring sources in order to better understand the intermediate separation regime. In the faint limit, we derive a convenient fitting form of the single-exposure threshold bias. Combined with the bias from using the maximum-likelihood position, we show that the faint-limit bias can be used as an empirical measure of the photometric depth. Applying corrections for these biases requires a more careful treatment of the combination of detections into objects than that performed in DECaPS2. However, we both introduce a model that captures the faint-limit behavior and demonstrate the utility of understanding these biases, which are important steps toward obtaining usable photometry near survey thresholds.

\begin{deluxetable}{lllll}[b]
\tablenum{1}
\tablecaption{Summary of DECaPS Imaging
\label{tab:exposures}}
\tablecolumns{5}
\tablehead{
\multirow{2}{2 em}{Filter} & \multirow{2}{5.2 em}{Wavelength Range (nm)} & \multirow{2}{4 em}{Exposure Time (s)} & \multirow{2}{5.5 em}{Number of Exposures\tablenotemark{a}} & \multirow{2}{4.6 em}{On Sky Time\tablenotemark{a} (h)} \\
}
\startdata
g & 398.0 - 548.5 & 96 & 4440 (3685) & 118 (98) \\ 
r & 565.5 - 717.0 & 30 [50]\tablenotemark{b} & 4386 (3644) & 39 \, (33) \\ 
i & 704.5 - 858.0 & 30 & 4345 (3550) & 36 \, (30) \\ 
z & 846.5 - 1000.0 & 30 & 4343 (3526) & 36 \, (29) \\ 
Y & 950.0 - 1034.0 & 30 & 3916 (3292) & 33 \, (27) \\ 
\enddata
\tablenotetext{a}{Numbers in parentheses are reflective of the reduced number of exposures actually included in the catalog due to cuts imposed during photometric calibration. See Section \ref{sec:cal}.}
\tablenotetext{b}{A secondary exposure time in brackets for $r$-band was used for observations when the Moon was up in observing run 7.\vspace{-11mm}}
\end{deluxetable}

This work presents the second data release, which we refer to as DECaPS2. When referring to imaging, we use DECaPS2 to refer to those images taken after the first data release of DECaPS \citep{Schlafly:2018:ApJS:}, which we refer to as DECaPS1 for both the imaging and catalog hereafter. When referring to the photometric catalog, DECaPS2 refers to a new reduction which processed both DECaPS1 and DECaPS2 imaging. The main differences compared to the first data release are increased sky coverage in Galactic latitude from $|b|<4\degree$ to $|b|\leq10\degree$ and improvements in the photometric reduction. 

The DECaPS2 catalog contains high-quality photometry with rich information about both the composition and structure of the dust and stellar populations in the Milky Way. Our work on nebulosity, crowding, and faint-limit biases may inform the next generation of photometric pipelines necessary to handle imaging of the Galactic plane planned for the Nancy Grace Roman Space Telescope \citep{Akeson:2019:arXiv:} and the Legacy Survey of Space and Time (LSST) at the Vera C. Rubin Observatory \citep{jones2020scientific}.


\section{Observations} \label{sec:Observe}

All observations associated with DECaPS were obtained using the Dark Energy Camera (DECam, \citealt{Flaugher_2015_AJ}) mounted on the 4m Victor M. Blanco telescope at the Cerro Tololo Inter-American Observatory (CTIO). The 2.2\degree\ diameter field of view, 0.26$"$/pixel plate scale, and arcsecond seeing (Section \ref{sec:inject}) make these observations well-suited to surveying and resolving even the extremely crowded inner galaxy. The survey imaged the Galactic plane $|b|\leq10\degree$, $6\degree > \ell > -124\degree$ (2700 deg$^2$, 6.5\% of the sky) in five broad photometric bands, $grizY$. The efficiency of DECam enabled observations to keep up with the survey footprint as it crossed the meridian and thus achieved very low airmass, with a mean of 1.17 and standard deviation of 0.15.

One of the goals of DECaPS is to capture a large fraction of stars with possibly high extinction in the Galactic midplane, probing distances out to the Galactic center. The exposure times were initially set to target the main sequence turn-off for a 10 Gyr, solar metallicity population of stars at eight kiloparsecs, through dust extinction of $E(B-V) = 1.5$~mag. These target depths are 24.5, 22.3, 21.2, 20.6, and 20.3 (AB) mag in $grizY$. We reduced the target of 24.5 mag in $g$ to 24.1 mag in order to better balance the exposure times among the dark- \& bright-time bands, given a minimum exposure time of 30s in each image in order to avoid being dominated by overheads. This led to exposure times of 96s in $g$-band and 30s in $rizY$-bands, and means that we only reach out to 1.4~mag $E(B-V)$ in $g$ but reach past the main-sequence turn-off in all other bands. We achieved these depths for all bands except for $g$-band (see Section \ref{sec:depth}).

\begin{figure*}[t]
\centering
\includegraphics[width=\linewidth]{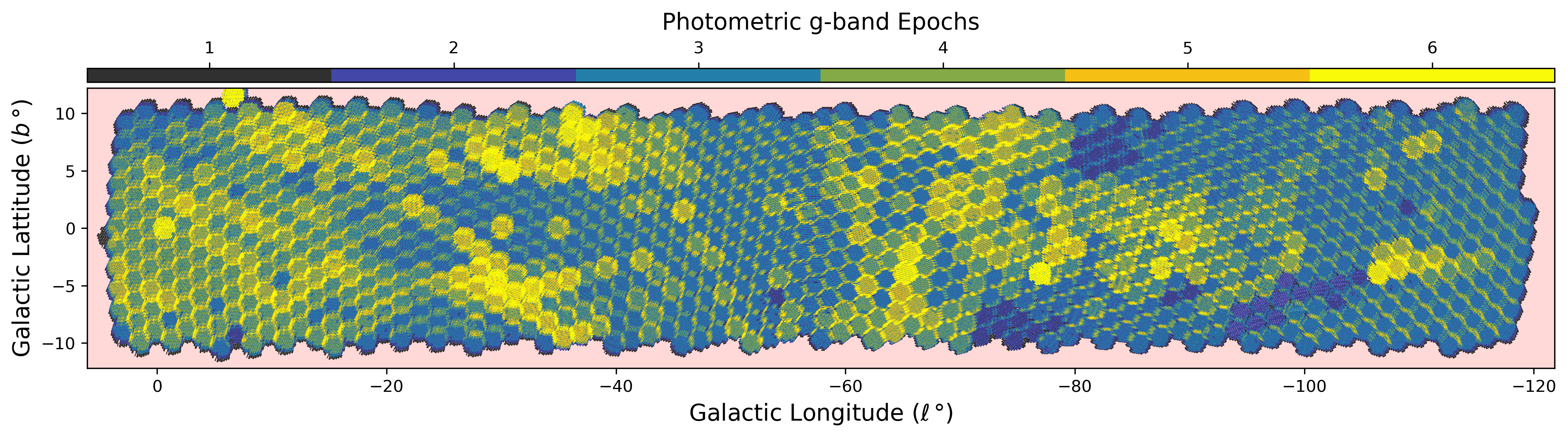}
\caption{High resolution coverage map of the DECaSP2 survey footprint showing the number of photometric $g$-band visits. Small scale variations in coverage resulting from the tiling strategy and CCD failures are visible.}
\label{fig:covg}
\end{figure*}

\begin{figure}[t]
\centering
\includegraphics[width=\linewidth]{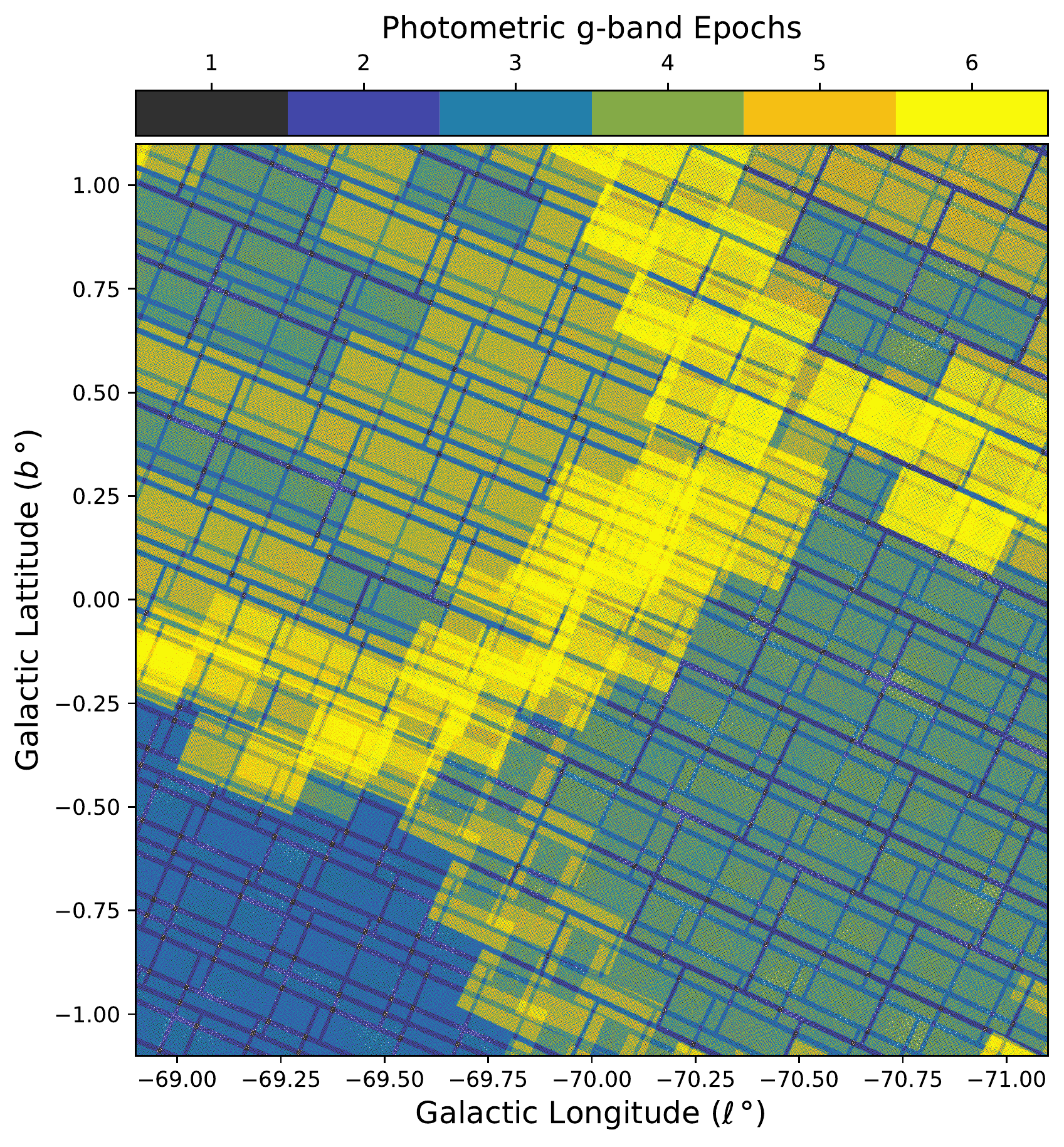}
\caption{Zoomed-in version of Figure \ref{fig:covg} at the scale of the focal plane of DECam.}
\label{fig:covg_small}
\end{figure}

The survey obtained a total of 21,430 exposures amounting to 260 hours of total exposure time over the duration of the program. DECaPS2 comprises observations starting in March 2016 and ending in May 2019. Approximately $83\%$ of exposures were of high enough quality to be included in our photometric catalog (see Section \ref{sec:cal}). A detailed per-band breakdown is provided in Table \ref{tab:exposures}.

The survey strategy aimed for three overlapping visits (in every band) for the majority of the survey footprint, though the tiling pattern leads to some areas having a different number of visits. The tiling strategy used for DECaPS follows the strategy developed for the DECam Legacy Survey \citep[][DECaLS]{Dey:2019:AJ:}. The field centers of the three passes are offset relative to one another to fill in gaps in the DECam focal plane. The DECaLS tiling scheme had the unfortunate feature that DECam chip gaps partially overlapped in all three passes near right ascensions of 270\degree. We remedied this by adding an additional pass in this area with a small fixed offset relative to other passes in $({\rm RA} \cos({\rm DEC}), {\rm DEC})$ to cover the chip gap.

In addition, in regions with a low number of visits, there can be very small regions with no coverage following a spatial pattern matching CCD-level malfunctions on DECam. Further reductions in the coverage are attributable to some imaging not being suitable for inclusion in the photometric catalog, for example as a result of partial cloud cover (see Section \ref{sec:cal}). A representative high-resolution coverage map for the $g$-band data used to build the photometric catalog presented here is shown in Figure \ref{fig:covg} and \ref{fig:covg_small}, and similar maps for the other bands are available in Section \ref{sec:dataavil}.

The three visits at a given location could occur the same night (rarely), on adjacent nights (more typical), or during observing runs a year apart. This mixing of temporal and spatial overlap of observations improves the ability of the calibration to constrain variations in throughput as a function of time, stabilizing the calibration over the long duration of the survey. We describe the major observing runs (a range of nights with more than 50 exposures each\footnote{A small number of DECaPS exposures were taken outside of these runs as part of time trades with other programs.}) in Table \ref{tab:obsrun} and identify which filters were primarily used during that run. During each observing run, $0-6$ calibration exposures were taken of high latitude fields, which were usually 30s for $gr$-bands and 20s for $izY$-bands.

Observations include approximately 23 full nights, 29 half nights, and 12 quarter nights. We aimed to observe in the $gr$-bands when the Moon was down and $izY$-bands when the Moon was up. During run 7, $r$-band images were taken while the Moon was up and used 50-second exposure times to compensate for the brighter sky. As the survey progressed our planning software became more sophisticated, leading us to take better advantage of brief periods where the moon was down in a bright night, or vice versa, leading most runs to include observations in all of $grizY$. Runs 13 and 14 were impacted by a temporary mechanical failure of the filter wheel preventing observations in $Y$-band; run 15 skipped observations in $iz$ to focus on $Y$ to make up for the lost time.

\begin{deluxetable*}{cllll}[t]
\tablenum{2}
\tablecaption{DECaPS Observing Runs
\label{tab:obsrun}}
\tablecolumns{5}
\tablehead{
Run \# & Date Range & Filters & \# Nights & Notes
}
\startdata
1 & 2016-03-13 to 03-16 & $gr$ & $4$ & 2016-03-16 clouded out\\
2 & 2016-03-23 to 03-26 & $izY$ & $4$ & 2016-03-25 clouded out\\
3 & 2016-08-10 & $izY$ & $0.5$ & Scattered clouds\\
4 & 2016-08-14 to 08-16 & $izY$ & $2 \times 0.5$ & Scattered clouds\\
5 & 2016-08-22 & $izY$ & 0.5 & \\
6 & 2016-08-23 to 08-26 & $gr$ & $4 \times 0.5$ & 08-23 and 08-25 clouded out\\
7 & 2017-01-16 to 01-23 & $r^{\star}izY$ & $6 \times 0.5$ & 01-22 and 01-23 clouded out ($r^{\star}$ used 50s exposure time)\\
8 & 2017-04-19 to 04-20 & $grizY$ & $2$ & 04-19 marginal\\
9 & 2017-04-27 to 04-30 & $grizY$ & $2 + 2 \times 0.5$ & \\
10 & 2017-05-03 to 05-04 & $grizY$ & $2 \times 0.5$ & 05-04 cloudy\\
\midrule
11 & 2018-02-02 to 02-03 & $izY$ & $2 \times 0.5$ & 02-02 marginal\\
12 & 2018-02-25 to 02-27 & $grizY$ & $3$ & 02-27 clouded out\\
13 & 2018-05-08 to 05-11 & $griz$ & $4 \times 0.5$ & 05-08 clouded out\\
14 & 2018-05-18 to 05-20 & $griz$ & $3$ & 05-18, 05-19 marginal\\
15 & 2018-08-01 to 08-04 & $grY$ & $4 \times 0.5$ & 08-02 clouded out, 08-04 marginal\\
16 & 2019-01-09 to 01-18 & $grizY$ & $10 \times 0.25$ & 01-10, 01-13 marginal\\
17 & 2019-01-30 to 01-31 & $grizY$ & $2 \times 0.25$ & \\
18 & 2019-04-25 & $grizY$ & $0.5$ & \\
19 & 2019-04-27 to 05-02 & $grizY$ & $5$ & 04-29 clouded out\\
\enddata
\end{deluxetable*}

\section{Catalog Building} \label{sec:buildCat}
\subsection{Single-Epoch Processing} \label{sec:imageprocess}

Each exposure was processed in three serial steps by separate pipelines: the DECam Community Pipeline (CP), \crowdsourcens, and \cloudcoverrns.

\subsubsection{DECam Community Pipeline} \label{sec:CP}

The DECam Community Pipeline \citep{Valdes:2014:ASPC:} is managed and run by NOIRLab and converts \texttt{Raw} exposures to the instrument-calibrated (\texttt{InstCal}) products that serve as inputs to user-managed photometric pipelines. In addition to images, these products include a per-pixel inverse variance (weight) map and an artifact mask. In this reduction, the CP performs the following steps (among others): 
\begin{itemize}
\setlength\itemsep{-0.5em}
\item overscan subtraction (bias correction)
\item amplifier cross-talk correction
\item static bad pixel masking
\item saturation and bleed trail masking
\item nonlinearity correction
\item flat-field correction (dome and star flat)
\item large reflection pattern (``pupil ghost'') removal
\item large scale background gradient removal
\item fringe correction
\item astrometric WCS solution
\item single exposure cosmic ray masking
\end{itemize}
For more details on the processing steps, see the \href{https://noirlab.edu/science/sites/default/files/media/archives/documents/scidoc1203.pdf}{NOIRLab Data Handbook}. The CP excludes completely nonfunctional CCDs from its reduction which includes N30 for the full duration of the survey (damaged from an over-illumination event) and S30 from November 2013 to December 2016 (Runs 1-6, first half of DECaPS1). See \href{https://noirlab.edu/science/programs/ctio/instruments/Dark-Energy-Camera/Status-DECam-CCDs}{Status of DECam CCDs} from NOIRLab for more details. The CP has evolved over time, and we report the version numbers over time in Table \ref{tab:cpvers}.\footnote{While nonuniform processing might be concerning, some components of the software changes track drifts within the instrument itself; thus the software variability is in part a reflection of the true instrument variability that requires the processing to change.}

Throughout the DECaPS2 observing period there were several changes in the header keywords, such as the reference catalog used for astrometric solutions and the CCD saturation levels. We observed cases where these changed without an accompanying change in the CP version number. The CP automatically and reliably provided calibrated images shortly following each DECaPS observing run.  We note one minor flaw that had a significant impact on DECaPS2, however. In the majority of exposures (9377) taken during DECaPS2, the saturation thresholds were set slightly too high during the initial CP processing for seven CCDs (N3, N9, S13, S19, S20, S22, S26).  The \crowdsource pipeline uses the brightest 200 unsaturated stars for PSF fitting in order to limit the effects of blending, and elevated saturated limits can lead to substantial contamination of the PSF stars. This led to poor PSF fits on these CCDs and severely impacted our photometry. As such, we used the sensitivity of our PSF fits to help reset the saturation thresholds (see Appendix \ref{sec:satdet}) in the CP and reprocessed the impacted exposures (CP v5.5), resolving the issue.

\begin{deluxetable}{lll}[h]
\tablenum{4}
\tablecaption{DECam Community Pipeline Versions
\label{tab:cpvers}}
\tablecolumns{3}
\tablehead{
Version & \# of Exposures & Date Range
}
\startdata
v3.9.0 & 3780 & 2016-03-13 to 2016-03-27 \\ 
v3.9.2 & 1845 & 2016-08-10 to 2016-08-27 \\ 
v3.10.0 & 5 & 2016-03-24 to 2016-03-27 \\ 
v3.12.0 & 1350 & 2017-01-17 to 2017-01-22 \\ 
v4.1.0 & 5064 & 2017-01-25 to 2018-02-28 \\ 
v5.2.3 & 12 & 2018-02-04 to 2018-02-04 \\
v5.5 & 9377 & 2018-05-09 to 2019-05-03 \\ 
\enddata
\end{deluxetable}\vspace{-12 mm}

Fringe correction is most important for Y-band where the longer wavelengths more easily form interference fringes across the CCD that need to be modeled and removed. However, we found that for 12 exposures the CP fringe-correction algorithm failed and visibly increased the amplitude of fringing on the \texttt{InstCal} images. The impacted exposures were reprocessed using no fringe correction (CP v 5.2.3). We note that DECaPS images pose more of a challenge than typical extragalactic fields to fringe fitting algorithms because of the large numbers of stars present in the images.

\subsubsection{\emph{\crowdsource}} \label{sec:crowdsource}

The photometric pipeline \crowdsource takes in the \texttt{InstCal} products and estimates the PSF, finds the location of sources (deblending crowded fields), and estimates a variety of statistics about those detections, flux and flux uncertainty being among the most important. The \crowdsource flux uncertainties simply combine the PSF estimates with the the \texttt{InstCal}  inverse variance maps. Several improvements to \crowdsource were made since DECaPS1, including a synthetic source injection module described in Section \ref{sec:inject}. Both a detailed description of the code and new features will be described elsewhere \citep{crowdsourceunpub}.

Briefly, \crowdsource applies an additional bad pixel mask and modifies the weight map so that CP-masked pixels (except bit 7) have zero weight (see Table \ref{tab:cs_bitmask}). \crowdsource also implements special handling for the partially functional CCD S7 where the gain of amplifier B is not stable. In several special cases, \crowdsource reduces how aggressively it deblends. One case is around objects in a galaxy catalog, which is a new feature compared to DECaPS1. Another case is in regions identified as nebulous by a (band-agnostic) convolutional neural network (CNN), the ``nebulosity CNN'', which was improved relative to DECaPS1. The algorithm proceeds by iteratively estimating the sky using a masked moving median, finding peaks in the PSF-convolved residual image, jointly estimating fluxes of all sources, and refining the PSF. In addition to the native \crowdsource stopping conditions, we chose hard limits of a minimum of 4 and maximum of 10 iterations. The pipeline transitions between conservatively deblending sources and aggressively deblending sources on the third pass. A global maximum on the number of sources that can be found on a given CCD was set to $320,000$.\footnote{\crowdsource only modeled 6250 CCDs, 0.5\% of those processed, as having more than $200,000$ sources. So, the hard limit should only impact $<0.5\%$ of images, if at all.}

The PSF used was a model of the ideal-seeing instrumental response (static) from the Dark Energy Survey (DES, \citealt{Abbott:2021:ApJS:}), which includes diffraction spikes and other features in the PSF wings, convolved with a 2D, possibly anisotropic Moffat (parametric). Additionally, per-pixel residuals were fit in the central $9 \times 9$ pixels of the PSF to account for departures from a Moffat profile and added to form the final PSF model. The Moffat and core residual parameters were allowed to vary linearly across the CCD. The PSF model was refit at each iteration of source finding using up to the 200 brightest stars passing a quality cut (e.g., not saturated). Despite known variations in the PSF as a function of magnitude (i.e., the ``brighter-fatter'' effect, \citealt{Stubbs_2006_ApJ, Antilogus_2014_JInst}), no such magnitude dependence was included here and the PSF model used is most representative of the bright stars used for the PSF parameter fitting.\footnote{In Section \ref{sec:blend}, we observe a slight magnitude dependent bias in the recovered flux. However, in those injection tests, all synthetic sources (even faint stars) have exactly the PSF model \crowdsource fit to the original image. Thus, such injection tests provide no measure of magnitude dependent biases resulting from not modeling the magnitude dependence of the PSF.} 

The spatial extent of the PSF used to model a source depends on its flux, with larger extents used for larger fluxes. The intention in choosing the model extent is that the flux of a given star is captured down to a surface brightness significantly fainter than the per-pixel uncertainty in the sky. Roughly, the extents are $19 \times 19$ pix for sources with peak per-pixel fluxes less than 1000 ADU, 59 pix for sources less than 20000 ADU, 149 pix for sources that are saturated or brighter than 20000 ADU, and 299 pix for sources within 5 pixels of a source in a ``bright'' star catalog---these saturate a large number of pixels, making flux estimates challenging.

A full description of the individual-image catalogs produced by \crowdsource is available in Section \ref{sec:dataavil}. We describe three important quality-assurance quantities here since they will be discussed below. These quantities measure the overlap of a source with neighbors, the quality of the input data for the fit, and the quality of the fit.

The ``blendedness'' of an object is measured by \texttt{fracflux}, the PSF-weighted fraction of flux at the source location in the image that is coming from that source (as opposed to coming from surrounding sources). Pixels masked by the CP (zero weight in \crowdsourcens) are excluded from the average. Very blended objects have \texttt{fracflux} of 0 while isolated objects have \texttt{fracflux} of 1. The precise definition is

\begin{ceqn}
\begin{align} \label{eq:fracflux}
\texttt{fracflux} = \frac{\int du \, N(u) \, P(u) \, \mathds{1}_{\left(\sigma^{-2} > 0\right)}}{\int du \, I(u) \, P(u) \, \mathds{1}_{\left(\sigma^{-2} > 0\right)}}
\end{align}
\end{ceqn}
where the integral is over a region 5" $\times$ 5" (19 $\times$ 19 pixels) around the source (hereafter ``stamp''), $P$ is the PSF (renormalized to sum to 1 on the ``stamp''), and $\mathds{1}_{\left(\sigma^{-2} > 0\right)}$ is an indicator function requiring that the inverse variance weights be nonzero (pixels not saturated, part of a cosmic ray, etc.). In the numerator, the integral is against $N$, which is the residual image plus the source model for the source of interest only (i.e., neighbor-subtracted image). In the denominator, the integral is against $I$, the image with all sources present.

The quality factor (\texttt{QF}) is the PSF-weighted fraction of good pixels (nonzero weight) within the stamp around the source.
\begin{ceqn}
\begin{align} \label{eq:qfeq}
\texttt{QF} = \int du \, P(u) \, \mathds{1}_{\left(\sigma^{-2} > 0\right)}
\end{align}
\end{ceqn}
Stars which coincide with a masked cosmic ray, occur near the edge of the detector, or are significantly saturated will have \texttt{QF} closer to $0.0$ while clean detections have a \texttt{QF} of $1.0$

The reduced chi-squared ($\chi_r^2$) is the PSF-weighted inverse variance-weighted squared residuals, normalized by the integral of the PSF-weighting of active pixels over the stamp (which is exactly the \texttt{QF}).

\begin{ceqn}
\begin{align} \label{eq:rchi2}
\chi_r^2 = \frac{1}{\int du \, P(u) \, \mathds{1}_{\left(\sigma^{-2} > 0\right)}} \int du \, \frac{R^2(u)}{\sigma^2(u)} \, P(u) \, \mathds{1}_{\left(\sigma^{-2} > 0\right)},
\end{align}
\end{ceqn}
where $R$ is the residual (image - model). Here the quality factor plays the role of the degrees of freedom by specifying the fraction of the effective area of the PSF that is used in the fit (has nonzero weight). The $\chi_r^2$ serves as a measure of goodness of fit of the PSF model to a given source.

\subsubsection{\cloudcoverr} \label{sec:CloudCovErr}

The source locations, PSF models, and residuals for each image are reprocessed by \cloudcoverr to correct the flux and flux uncertainties in the presence of structured (nebulous) backgrounds. \cloudcoverr works by predicting the distribution of possible backgrounds masked by the star. This interpolation (known as Local Pixelwise Infilling, LPI) actually predicts the distribution of residuals of those possible backgrounds relative to the smooth background model used by \crowdsourcens, which reduces the fraction of the image that must be masked because of the presence of sources. 

\begin{deluxetable}{llll}[h]
\tablenum{5}
\tablecaption{\crowdsource Bitmask
\label{tab:cs_bitmask}}
\tablecolumns{4}
\tablehead{
\multirow{2}{2 em}{Bit} & \multirow{2}{2 em}{Description} & \multicolumn{2}{c}{Exclude Source?} \\
& & \multicolumn{1}{c}{Catalog} & \multicolumn{1}{c}{Injections}
}
\startdata
0 & No problems & No & No \\ 
1 & Bad pixel & Yes & Yes \\ 
3 & Saturated & Yes & Yes \\
4 & Bleed Trail & Yes & Yes \\
5 & Cosmic Ray & Yes & Yes \\
6 & Low Weight & Yes & No \\
7 & Difference Detection & No & Yes \\
8 & Long Streak & Yes & No \\
\midrule
20 & Additional Bad Pixel & Yes & Yes \\
21 & Nebulosity & No & No \\
22 & CCD S7 amplifier B & Yes & No \\
23 & Near Bright Star & No & Yes \\ 
24 & Near Galaxy & No & Yes \\ 
\midrule
30 & No Deblend & No & No\\
31 & Sharp & No & No\\
\enddata
\tablecomments{Bits $0 - 8$ are inherited from the CP, $20 - 24$ indicate a special region on the CCD or sky, and $30 - 31$ indicate a change in the \crowdsource source identification parameters.}
\end{deluxetable} \vspace{-11mm}

To do this interpolation, LPI trains a covariance matrix over the residuals of local data representative of the true background (unmasked). It then leverages pixels in an annulus around the star to obtain the conditional Gaussian distribution of the background residuals within the stellar footprint. The resulting flux uncertainty is quasi-independent of the \texttt{InstCal} inverse variance weights as it is estimated based on observed correlations of background pixels in the image nearby the source of interest. We refer to the background-corrected flux and flux uncertainty as \cflux and \texttt{dcflux}, respectively. 

\begin{figure*}[t]
\centering
\includegraphics[width=\linewidth]{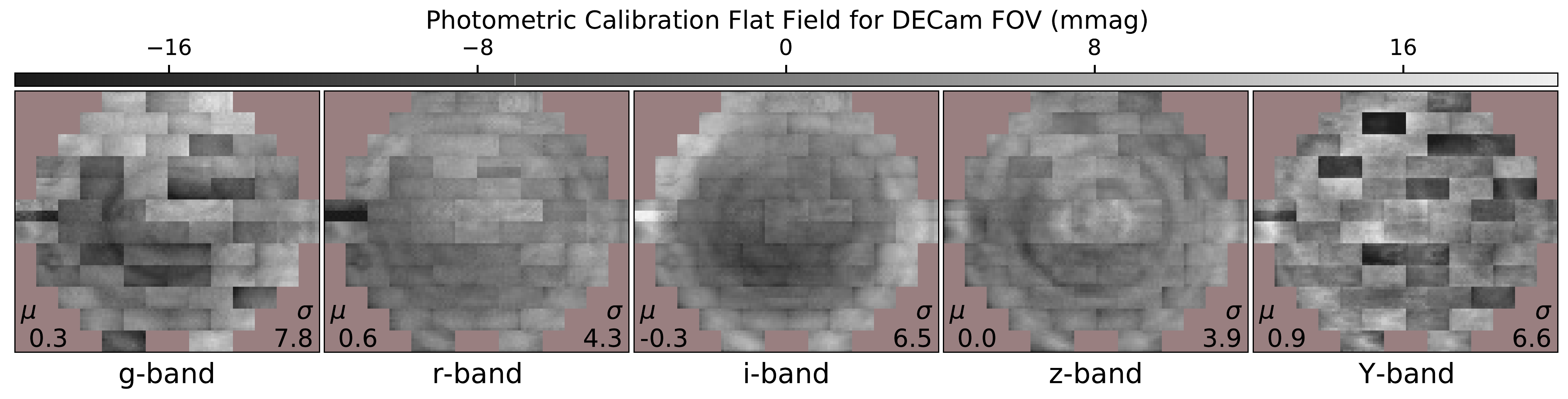}
\caption{Time-invariant flat-field across the full DECam field of view produced during photometric calibration of DECaPS2. Each panel shows a different photometric band and the median and $\sigma_{\rm IQR}$ in mmag of the flat-field are in the bottom left and right corner, respectively. Gaps between CCDs of order 200 pixels are not shown in this large-scale representation. The sky orientation shown for the focal plane is north up, west right.}
\label{fig:ffieldCal}
\end{figure*}

An additional output of \cloudcoverr is \texttt{dnt}, which is a quality flag on the correction algorithm. A table of possible nonzero values and the issues those bits indicate is available in Section \ref{sec:dataavil}. A detailed description of the method and validation of \cloudcoverr is provided by \cite{Saydjari:2022:arXiv:}. Further, the precise parameter choices used with \cloudcoverr to process DECaPS2 are detailed in Section 5.1 of \cite{Saydjari:2022:arXiv:}, so we do not repeat them here.

\subsection{Photometric Calibration} \label{sec:cal}

The DECaPS2 photometric calibration follows the same procedure as DECaPS1 \citep{Schlafly:2019:ApJS:} which is based on the photometric calibration of SDSS \citep{Padmanabhan:2008:ApJ:} and PS1 \citep{Schlafly:2012:ApJ:,Magnier:2013:ApJS:,Magnier:2020:ApJS:}. Parametric models of the instrument and observing conditions are fit to minimize exposure-to-exposure variations in the measured flux of a given source. For DECaPS2, the model consists of a time-invariant flat-field correction, a system zeropoint per night, and a time-invariant term linear in the airmass (a constant ``k-term'' in  the notation of \citealt{Padmanabhan:2008:ApJ:}). From the calibration, we obtain a per-exposure zeropoint accounting for throughput variations between exposures to put all measurements on the same (relative) scale. Zeropoints were not allowed to vary at the CCD-level meaning that all CCD-level variations were assumed to be static and corrected by the time-invariant flat-field. We apply the following cuts to obtain high-quality sources used in the calibration:
\begin{itemize}\setlength\itemsep{-0.2em}
\item \texttt{QF} $> 0.99$
\item $-14 < -2.5 \log_{10} \left(\rm{\texttt{flux}}\right) < -12$ (instr. mags)
\item $\chi^2_r / \left( 1 + \left(2 \cdot 10^{-5} \times \texttt{flux}\right)^2 \right) < 2$
\end{itemize}
Each photometric band is fit independently.

Additionally, only exposures taken in ideal photometric conditions were included in the calibration.  First, we manually inspected the photometric trends on each night and excluded observations taken in clearly non-photometric conditions. Then, as part of the determination of the photometric solution, we repeatedly solve for the calibration parameters using a linear least squares fit and apply them to all of the detections in the survey.  At each iteration, we increasingly aggressively remove individual observations of stars discrepant with their mean magnitudes, as well entire exposures or CCDs when the measured fluxes are discrepant with the rest of the night or the rest of the exposure. For more details, see \textsection~2.3 of \citet{Schlafly:2012:ApJ:}. A somewhat more relaxed photometricity cut is used for inclusion of a detection in the final calibrated flux of an object, defined below.

The flat-field correction treats each $256 \times 256$ pixel$^2$ region of each detector independently and is shown in Figure \ref{fig:ffieldCal}. In Figure \ref{fig:ffieldCal} and throughout the text we use $\sigma_{\rm IQR} = {\rm IQR}/1.34896$ as a measure of scatter. The interquartile range (IQR) is an outlier robust measure of scatter, which should be $\sim1.34896$ for the unit normal distribution. We then normalize the IQR by the normal value to provide an outlier robust measure of $\sigma$.

Offsets between adjacent CCDs are most apparent in $g$ and $Y$-bands. These offsets stem from slightly differing bandpasses for the different CCDs, and the different mean color used to construct flat fields in the CP and for the mean star in DECaPS. The edges and corners within a given CCD often differ slightly from the center of the chip, suggesting uncorrected throughput gradients across the CCDs that we are correcting. Amplifier B on S7, which has variable gain, sticks out clearly. Even with this low-resolution treatment, ``tree ring'' artifacts (concentric circles within a single CCD) from impurity migration during silicon growth are apparent, especially in the top right CCD (S29). Much of ``tree ring'' structure is at scales smaller than the $256 \times 256$ blocking used in the flat-field and remains uncorrected. 

There is a large radial gradient in $i$-band, attributed to a strong angle dependence of the $i$-filter bandpass. Radial rings are evident in most bands, especially $z$-band and are attributed to imperfect pupil ghost removal impacting initial CP flat-fields. In most cases, the photometric calibration accounts for small residuals from large corrections made by the CP. The small scatter ($< 8$ mmag) in the flat-field per band is indicative of the accuracy of the CP, especially given that most of the signal in the worst bands ($giY$) stems from color-dependent effects for which any gray corrections like those attempted here are in some way unsatisfactory.

The stability of the photometric calibration is assessed using the residuals ($r_{\star}$) between the calibrated measured fluxes per detection and the average flux over all detections for a given calibration source. In Table \ref{tab:photoparam}, we report the average scatter of $r_{\star}$ per exposure as $\sigma_{\rm{cal}\star}$. The average $r_{\star}$ per exposure can be thought of as what the calibration would predict if the system zeropoint were allowed to vary per-exposure, even though the calibration fixes a global zeropoint per-night. The scatter of these ``per-exposure zeropoints'' ($\sigma_{\rm{ZP}}$) is a measure of the goodness of fit of the fairly static photometric model used (see Table \ref{tab:photoparam}).  It describes at some level how close Cerro Tololo, the Blanco, and DECam approach the ideal of delivering perfectly repeatable fluxes night-to-night, adjusting only for an airmass and nightly throughput term; we find that it reaches this ideal at the 1\% level. We clipped samples of the residuals per exposure to within $\pm2.7\sigma_{\rm IQR}$ of the median before taking the mean and standard deviation in computing $\sigma_{\rm{cal}\star}$ and $\sigma_{\rm{ZP}}$, respectively.

\begin{deluxetable}{ccccc}[h]
\tablenum{6}
\tablecaption{Photometric Parameters
\label{tab:photoparam}}
\tablecolumns{4}
\tablehead{
\multirow{2}{2 em}{Filter} & \multirow{2}{3.5 em}{$\sigma_{\rm{cal}\star}$ (mmag)} & \multirow{2}{3.5 em}{$\sigma_{\rm{ZP}}$ (mmag)} & \multirow{2}{5.2 em}{FWHM ('')} & \multirow{2}{4.5 em}{Depth (AB mag)} \\
}
\startdata
g & 8.5 & 9.5 & 1.35 & 23.5 \\ 
r & 7.6 & 9.0 & 1.25 & 22.6 \\ 
i & 7.1 & 9.0 & 1.15 & 22.1 \\ 
z & 7.4 & 10.6 & 1.10 & 21.6 \\ 
Y & 7.0 & 9.0 & 1.07 & 20.8 \\ 
\enddata
\end{deluxetable} \vspace{-12 mm}

We apply a cut on the per exposure scatter $\sigma_{\rm{cal}\star} \leq 20$ mmag and average zeropoint offset $\langle r_{\star} \rangle < 200$ mmag to define ``photometric'' exposures ($\sim 83\%$ of the observations) that are included in the final DECaPS2 catalog. Additionally, individual CCDs are marked as excluded from the final fluxes when showing mean offsets ($\langle r_{\star} \rangle$) from the rest of the exposure more than $5\sigma$ larger than the typical offset in that band.

After applying this cut, we compute the same sigma-clipped statistics over all ``photometric'' exposures in Table \ref{tab:photoparam}. The $\sigma_{\rm{cal}\star}$ ranges from 7.0 mmag in $Y$-band to 8.5 mmag in $g$-band, and  $\sigma_{\rm{ZP}}$ ranges from 9.0 mmag in $riY$-bands to 10.6 mmag in $z$-band. Under both of these measures, the photometric calibration is good to the $\sim 10$ mmag level, or around the $1\%$ level. Given that we neglect a careful treatment of the tree-ring distortions and better modeling of the DECam PSF, which lead to errors on the order of a few mmag, this is an excellent photometric calibration.

All of the above provides a relative calibration for variations in the atmosphere plus instrument throughput exposure-to-exposure, but does nothing to set the absolute zeropoint of the survey. For DECaPS2, we tie the absolute zeropoint to PS1, which in turn ties its zeropoint to HST; see Section \ref{sec:PS1}. Various revisions of PS1 processing have altered the absolute zeropoints per band by $11 - 33$ mmag.\footnote{Here we have converted the changes in zeropoints in the PS1 photometric system to the DECam photometric system.} Thus the relative calibration of DECaPS2 is at a precision comparable to the accuracy of the absolute zeropoint.

\subsection{Constructing Objects} \label{sec:comb}

After calibration, detections from single exposures across all photometric bands are merged into objects using the Large Survey Database (see Section \ref{sec:dataavil}). Briefly, for each exposure, a k-d tree is constructed per detection to find the nearest neighbor in the list of previously known objects. If the nearest neighbor object is closer than a threshold radius (0.5"), that detection is assigned to that object. If the nearest neighbor object is further than the threshold radius, a new object is created for that detection. By this method, each detection is associated with one and only one object. Note that objects can be closer than 0.5" if they are created from detections in the same exposure where neither object had previously been found. Exposures are processed in temporal order.

Once detections are associated with objects, average properties of each object are computed.\footnote{Even though calibration is performed at the detection level, the zeropoints are not added to the database entries until the average object level.} Per object, the position is determined by a simple mean (over all photometric bands) and the position uncertainty is the standard deviation of the detection positions. The mean MJD, maximum - minimum MJD, and total number of detections (over all photometric bands) of the object are also reported. 

\begin{figure*}[ht]
\centering
\includegraphics[width=\linewidth]{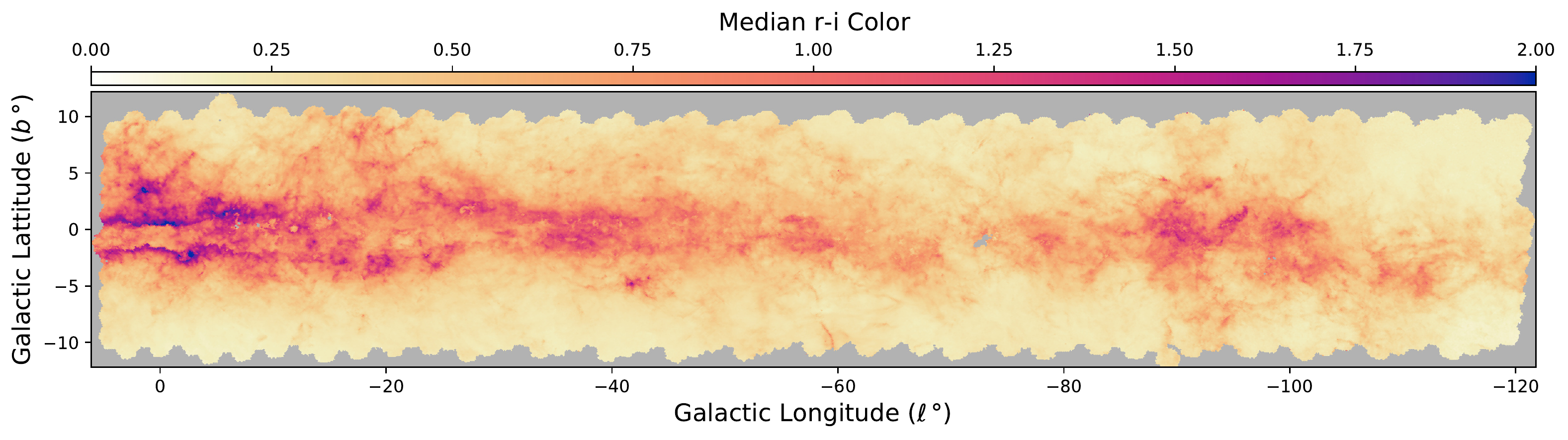}
\caption{Median $r - i$ color for sources in the DECaPS2 catalog across the survey footprint (NSide = 512). This figure employs a cut requiring sources to be brighter than 19th magnitude in the $i$-band and be detected in $r$-band.}
\label{fig:medcolorri}
\end{figure*}

In each band, we report the number of detections and the average flux, computed as a weighted mean. The weights used throughout are the inverse of the reported flux variance from the photometric pipeline, with a term added in quadrature to account for multiplicative errors in flux estimation (for example, errors due to PSF mismodeling) on the order of 0.01 mag.\footnote{We see evidence for multiplicative errors in Figure \ref{fig:low_SNR}.}

\begin{ceqn}
\begin{align} \label{eq:weight}
\rm{\texttt{weight}} = \frac{1}{\rm{\texttt{dflux}}^2 + \left(0.00921 \times \rm{\texttt{flux}}\right)^2}
\end{align}
\end{ceqn}

We report both the (weighted) standard deviation of the detection fluxes from this mean and the uncertainty associated with the mean weight. Even though there are only three visits on average, the 25th-percentile, median, and 75th-percentile flux are reported.\footnote{These values are reported as a result of the historical use of portions of the pipeline with PS1 and we do \textbf{not} recommend using the upper and lower quartiles.} These flux-related statistics are computed for both \flux and \cfluxns. 

A rough magnitude limit, which corresponds to $-2.5 \log(5 \times \rm{\texttt{dflux}) + {\rm zp}}$, where ${\rm zp}$ is the exposure zeropoint, is reported per object as a maximum over all detections. This magnitude limit is an approximate estimate of the photometric depth at which that source would be only a $5 \sigma$ detection. We discuss the spatial dependence of the magnitude limit of DECaPS2 further in Section \ref{sec:depth}.

Positional, epoch, and number of detection-related quantities are reported both over all detections and over only \texttt{OK} detections, which are deemed to be of high-quality. Flux-related quantities (and flags) are only computed for \texttt{OK} detections. Detections are \texttt{OK} if no bad flag bits were set for the center pixel (see Table \ref{tab:cs_bitmask}), the flux uncertainty is nonzero, and the \texttt{QF} and $\chi^2_r$ pass the cuts outlined in Section \ref{sec:qual}.

All flux-like quantities are reported in units of ``maggies'' (Mgy), which are equivalent to 3631 Jy and are a convenient unit such that $-2.5\log_{10}(\fluxns)$ is in AB magnitudes. Throughout the band-merged catalogs, fields that cannot be populated are replaced with zero. For example, this occurs when there are no detections in a one of the bands for an object. Thus, it is imperative that a $\texttt{nmag} > 0$  or $\texttt{nmag\_ok} > 0$ cut is applied before using the per-band fluxes. Here $\texttt{nmag}$ and $\texttt{nmag\_ok}$ are the number of detections (or \texttt{OK} detections) for a given object per band (see Section \ref{sec:dataavil} for more).

The \crowdsource quality metrics \texttt{fracflux}, \texttt{QF}, and $\chi_r^2$ as well as the \cloudcoverr quality metric \texttt{cchi2} are reported as a (weighted) average per-band. The (weighted) average \texttt{flux}/\texttt{dflux} and predicted class probability from the nebulosity CNN are also reported. The \texttt{dnt} bitmask for \cloudcoverr and the flag bitmask for both \crowdsource and the CP were propagated with both a bitwise AND and OR to show if a given bit was thrown for all or any of the detections, respectively.

A complete description of all fields in the band-merged catalogs is available in Section \ref{sec:dataavil}.
\vspace{17mm}

\section{Catalog Characterization} \label{sec:characterize}

Using the crowded-field photometry code \crowdsourcens, we created a catalog of 3.32 billion sources from 34 billion detections in the DECaPS imaging.\footnote{In terms of the number of objects, this makes DECaPS2 one of the largest photometric catalogs. The NOAO source catalog (NSC) DR2 \citep{Nidever:2021:AJ:}, which used SExtractor to uniformly reprocess public data including DECaPS1 and DECaPS2, contains 3.9 B. Pan-STARRS1 contains 2.9B sources \citep{Magnier:2020:ApJS:}, though DECaPS2 has fewer epochs and thus fewer detections. The Zwicky Transient Facility catalogs have $\sim1.5$B objects \citetext{private communication, Prince, 2022}.} We further post-processed those photometric outputs using \cloudcoverr to improve the flux and flux uncertainty estimates in the presence of structured backgrounds (such as clouds of gas and dust). We present the catalog here and its validation in the next section.

We visualize the source density in r- and z-band in Figure \ref{fig:density} (using a HEALPix grid at NSide = 512 resolution, \citealt{Gorski_2005_ApJ}). In both, reductions in source density are apparent as a result of dust clouds. However, the relative reduction is less pronounced in z-band, illustrating that our NIR photometry penetrates to greater distances through dusty regions. Source densities over the survey footprint for the other photometric bands are available in Section \ref{sec:dataavil}.

\begin{figure*}[ht]
\centering
\includegraphics[width=\linewidth]{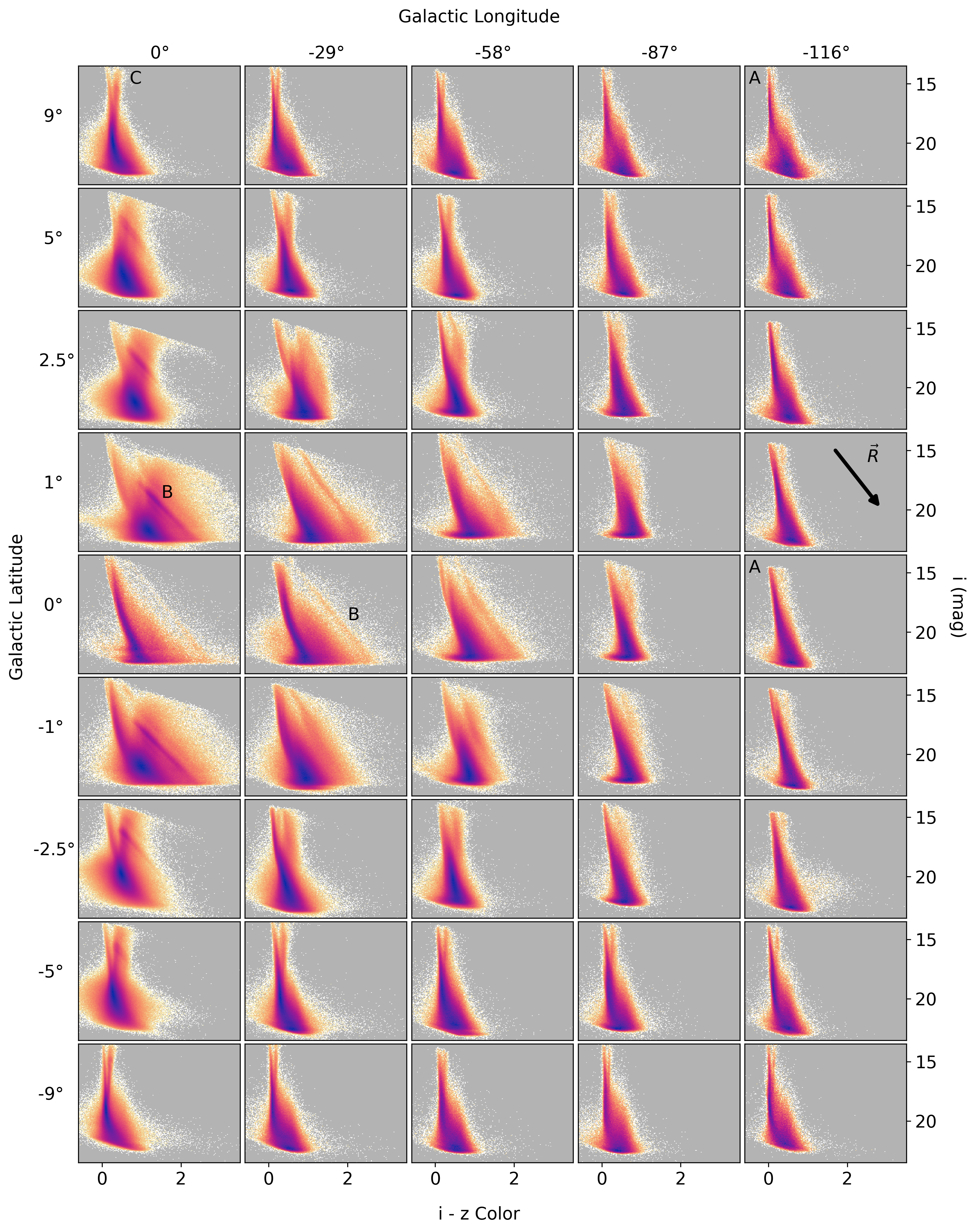}
\caption{Color-magnitude diagrams (CMDs) in $i$ versus $i-z$ for a beam of radius $0.5\degree$ on a grid of Galactic latitude and longitude over the survey footprint. Each CMD color scale is in log-density and has its own normalization (light, white, low; dark, blue, high density). The features associated with blue main sequence stars in the disk (A), red clump stars (B), and red-giant-branch stars (C) are labeled. Reddening vector showing effect of dust plotted in ($-116\degree$, $1\degree$) for reference.}
\label{fig:cmdgrid}
\end{figure*}

\begin{figure*}[ht]
\centering
\includegraphics[width=\linewidth]{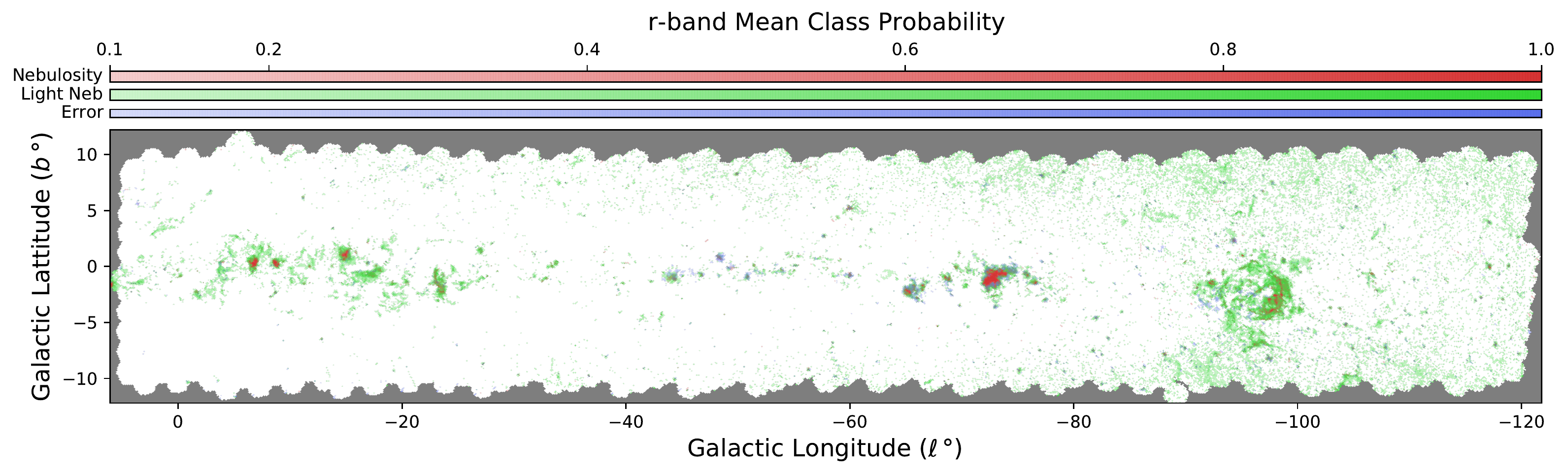}
\caption{Classification of the DECaPS2 $r$-band imaging by \crowdsource nebulosity CNN. Three of the four possible classes are shown as increasing opacity RGB color (\texttt{nebulosity}, \texttt{light neb}, and \texttt{error}) and the fourth class (\texttt{normal}) can be inferred by taking the complement. Each NSide = 512 HEALPix pixel displays the mean probability of each class over all photometric detections in the region. The probability associated with each source detection is the class probability of the pixel closest to the center of the source.}
\label{fig:nebmap}
\end{figure*}

Globular clusters (down to 9th V-band mag) are already visibly prominent in the source density map as high density points in Figure \ref{fig:density}. The homogeneous stellar populations provided by these globular and open clusters along relatively dusty lines of sight can help better constrain the stellar modeling uncertainties associated with reddening models. Of the 157 globular clusters in the 2010 edition of the Harris catalog \citep{Harris:1996:AJ:}, 47 fall in the DECaPS2 footprint (34 in DECaPS1). Of the 2858 open clusters in the Milky Way Star Clusters Catalog \citep{Scholz:2015:A&A:}, 972 fall in the DECaPS2 footprint (783 in DECaPS1).

In a few cases, high-density points are the result of spurious sources fit to massive pupil ghosts near very bright stars. Notable examples are $\gamma$ Crucis ($300.2\degree$, $5.6\degree$, V = 1.6 mag), $\lambda$ Velorum ($265.9\degree$, $2.8\degree$, V = 2.2 mag), and $\eta$ Canis Majoris ($242.6\degree$, $-6.5\degree$, V = 2.5 mag). We use these in part to develop the quality cuts designed to remove spurious sources (see Section \ref{sec:qual}).

\subsection{Dust Diversity} \label{sec:dustdiv}

In the Galactic midplane, variations in the median $r - i$ color of stars (Figure \ref{fig:medcolorri}) are dominated by reddening of the starlight from dust, and thus acts as a tracer for dust density.\footnote{The restriction to sources brighter than 19th magnitude in many all sky plots in the text is meant to focus on high confidence sources and retains 6\% to 24\% of sources in $g$ to $Y$ band.} However, for regions with very dense dust clouds, the median star lies in front of the cloud, and thus a line of sight will only appear to have less reddening (since there are no sources found behind the dust cloud). A transition to this case occurs around $|b| = \pm 1$ toward the Galactic center.

The DECaPS2 survey footprint includes famous dust clouds such as Pipe, Lupus, Circinus, the Coalsack, the Vela Molecular Ridge, and portions of Musca and Ophiuchus, all of which appear prominently in the median $r - i$ color map (Figure \ref{fig:medcolorri}). The high-quality photometry from DECaPS2 through a large range of extinction and across a diversity of structures will prove useful in probing the variation of dust properties throughout the disk, as was demonstrated already in the PS1 footprint \citep{Schlafly:2016:ApJ:, Schlafly:2017:ApJ:}.

\subsection{Stellar Diversity} \label{sec:stardiv}

To illustrate the variety of stellar populations captured by DECaPS2, we show (apparent) color-magnitude diagrams (CMDs) in $i$ versus $i-z$ for $0.5\degree$ radius beams in a grid across the survey footprint (Figure \ref{fig:cmdgrid}). Each CMD has its own log-density normalization. These sight lines sample a large range of extinction and stellar densities and capture the transition from the Galactic bulge to the Galactic disk. 

We expect detailed stellar investigations to be developed as follow-up work to this data release and point out only a few major features (and their variations) that are readily apparent. At high latitudes, ($-116\degree$, $9\degree$), we observe a sharp vertical track (A) from blue main-sequence stars in the Galactic disk. That track widens and tilts in the direction of the reddening vector (indicated by $\vec{R}$ in Figure \ref{fig:cmdgrid}) as the latitude approaches the plane and experiences more extinction from dust ($-116\degree$, $0\degree$). Moving toward that Galactic center ($0\degree$, $0\degree$), a parallel track along the reddening vector is observed (B) associated with red clump stars.

The long red clump track (B) observed in ($0\degree$, $1\degree$), resulting from variations in reddening from dust, tightens to more closely resemble the expected clump at higher latitudes where there is less differential extinction ($0\degree$, $2.5\degree$) before disappearing at the highest latitudes above the Galactic bulge ($0\degree$, $9\degree$). Variable ceilings in the maximum stellar magnitude, coming from a cut removing saturated sources in $z$-band, are apparent in several fields, including ($0\degree$, $2.5\degree$).

We also observe a track (C) with positive slope that is redder than A which we attribute to the red-giant branch (with contributions from either the disk, bulge, or both). Along certain lines of sight, such as ($-29\degree$, $-2.5\degree$) and ($-58\degree$, $2.5\degree$), there are tracks between A and C and ($-58\degree$, $2.5\degree$) which are likely associated globular/open clusters. In $g-r$ (see Section \ref{sec:dataavil}) a track bluer than A is observed along two lines of sight. The apparent gap in the main sequence, observed along several high latitude lines of sight, ($-116\degree$, $9\degree$) for example, could be a signature of probing different components (thin and thick) of the disk. We believe the high-quality photometry delivered by DECaPS2 over these diverse fields should provide a wealth of information that aids in solving the coupled problems of stellar evolution, Galactic structure, and dust density and reddening variations.

\begin{figure*}[ht]
\centering
\includegraphics[width=\linewidth]{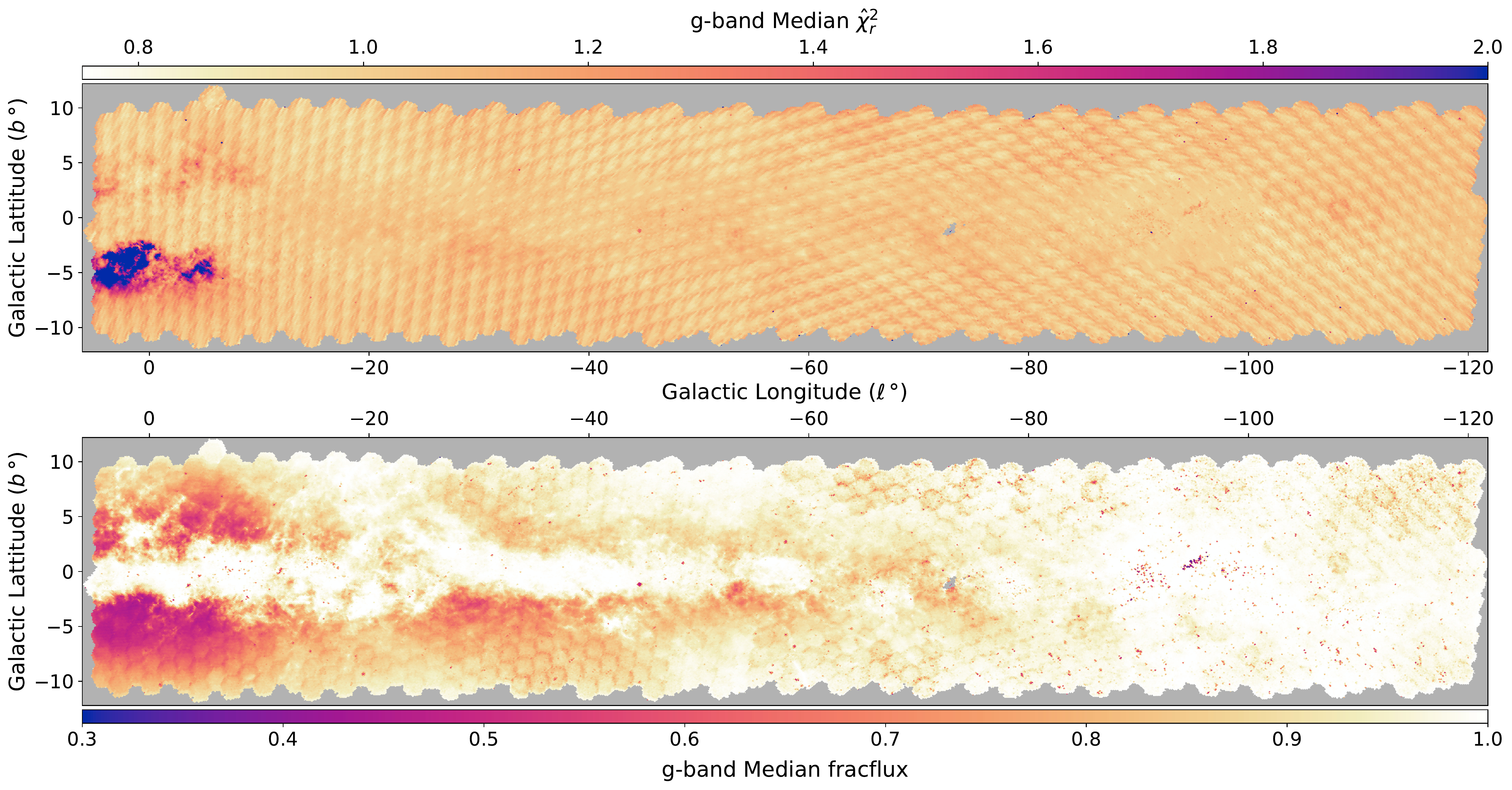}
\caption{Median $g$-band $\chi^2_r$ (top) and \texttt{fracflux} (bottom) over all objects in the survey footprint.}
\label{fig:mean_rchi2_fracflux}
\end{figure*}

\subsection{Nebulosity} \label{sec:neb}

As discussed in Section \ref{sec:crowdsource}, \crowdsource uses a CNN to identify regions of nebulosity and reduces the degree of targeted deblending in those regions to avoid attributing diffuse emission to the sum of many small point sources. To do so, the CNN reports a probability that a given image sub-region (512 $\times$ 512 pixels) is in one of four classes: 

\begin{itemize}
\setlength\itemsep{-0.5em}
\item \texttt{nebulosity}, containing significant contamination by nebulosity
\item \texttt{light neb}, containing faint nebulosity
\item \texttt{normal}, no contamination
\item \texttt{error}, containing artifacts from bright star pupil ghosts or spurious sky-level fluctuations introduced by the CP 
\end{itemize}

These classes were trained on human-sorted representative images and are, in part, subjective. To elucidate what features each class actually corresponds to, we show the response of the CNN to $r$-band images across the survey footprint (Figure \ref{fig:nebmap}). The map encodes the class probabilities as the transparency of RGB channels for \texttt{nebulosity}, \texttt{light neb}, and \texttt{error} classes, respectively. The \texttt{normal} class probability is the complement of the sum of the other three.

The \texttt{nebulosity} class appears to dominate along ridges or cores surrounded by regions where \texttt{light neb} dominates. This nested behavior validates \texttt{light neb} and \texttt{nebulosity} as characterizing different degrees of the same physical feature. The features with large \texttt{nebulosity} or \texttt{light neb} probability strongly resemble the main features in H$\alpha$ maps, tracing emission nebulae in the Gum catalog, for example. Notable nebulae visible in the probability map include Lobster ($353.2\degree$, $0.9\degree$, NGC 6357), Cat's Paw ($351.1\degree$, $0.5\degree$, NGC 6334), Prawn ($344.8\degree$, $1.6\degree$, IC 4628), and Carina ($287.6\degree$, $-0.6\degree$, NGC 3372). In addition, a large shell associated with the Vela supernova remnant is seen. Since H$\alpha$ emission is the primary source of nebulosity in $r$-band, this correlation validates the accuracy of the CNN in identifying the features of interest in practice.

The \texttt{error} class has high probability in a few points scattered across the footprint. These correspond to several extremely bright stars such as, $\lambda$ Velorum ($265.9\degree$, $2.8\degree$, V = 2.2 mag), $\alpha$ Crucis ($300.1\degree$, $-0.4\degree$, H = 1.3 mag), and $\beta$ Crucis ($302.5\degree$, $3.2\degree$, V = 1.3 mag). This validates the response of the \texttt{error} class to artifacts, such as those from pupil ghosts of bright stars. Further the \texttt{error} class probability is elevated around bright nebula, such as Carina, where spurious sky-level fluctuations are most common.

\begin{deluxetable*}{lrrrrrrrr}[t]
\tablenum{7}
\tablecaption{Quality Cut Counts (in Millions)
\label{tab:qacut}}
\tablecolumns{9}
\tablehead{
Cuts & Any(grizY) & g & r & i & z & Y & All(gr) & All(izY)
}
\startdata
\textbf{Detections} & 34,032 & 4,939 & 6,783 & 7,538 & 8,417 & 6,355 & - & - \\
+flags & 33,476 & 4,850 & 6,674 & 7,414 & 8,277 & 6,262 & - & - \\
+flags+\texttt{QF} & 33,216 & 4,808 & 6,622 & 7,357 & 8,211 & 6,218 & - & - \\
+flags+$\hat{\chi}^2_r$ & 33,138 & 4,792 & 6,616 & 7,334 & 8,178 & 6,218 & - & - \\
+flags+\texttt{QF}+$\hat{\chi}^2_r$ & 32,896 & 4,754 & 6,567 & 7,282 & 8,117
 & 6,177 & - & - \\
\textbf{Objects} & 3,319 & 1,405 & 1,911 & 2,330 & 2,588 & 2,092 & 1,288 & 1,804 \\
+\texttt{fracflux} & 1,558 & 784 & 1,000 & 1,203 & 1,285 & 1,090 & 722 & 929 \\
\enddata
\tablecomments{Detection counts (for all bands and per band) and how they are modified by cuts on \crowdsource quality metrics. These detection level cuts are used to define \texttt{OK} detections at the object level. Similar counts for objects before and after applying a cut at \texttt{fracflux} of 0.75. Counts for objects with detections in both $g$ and $r$-band and all three of $i$, $z$, and, $Y$-band shown in last two columns.}
\end{deluxetable*} 

The final nebulosity mask used to change the deblending in \crowdsource affects an even smaller area than those highlighted in Figure \ref{fig:nebmap}. These regions are indicated by sources with bit 21 set and per-CCD mask images saved in the single-exposure catalog files (see Section \ref{sec:dataavil}). The decision boundary for the nebulosity mask is

\begin{ceqn}
\begin{align} \label{eq:nebdecide}
\frac{p(\texttt{nebulosity}) + 0.5 \, p(\texttt{light neb}))}{p(\texttt{light neb}) + p(\texttt{normal}) + p(\texttt{error})} > 2
\end{align}
\end{ceqn}
This boundary was selected conservatively, to mask only the most nebulous regions in order to apply the full deblending power of \crowdsource to the vast majority of the survey footprint. However, irrespective of any boundary selection, it is clear from Figure \ref{fig:nebmap} that most of the sky is seen to be \texttt{normal} by the CNN, as desired.

\subsection{Fit Quality} \label{sec:fitqa}

We use $\chi^2_r$ and \texttt{fracflux} for all objects over the survey footprint as indicators of variations in the quality of fit. When analyzing $\chi^2_r$ throughout this work, we rescale $\chi^2_r$ to give $\hat{\chi}^2_r$, using the \texttt{flux} of the source to prevent multiplicative systematics from dominating in the bright limit. 

\begin{ceqn}
\begin{align} \label{eq:normchi2}
\hat{\chi}^2_r = \frac{\chi^2_r}{1 + \left(10^{-5.5} \times \texttt{flux}\right)^2}
\end{align}
\end{ceqn}

The choice of constant in Equation \ref{eq:normchi2} is described in Appendix \ref{sec:satdet} and becomes important for sources brighter than $-13.75$ instrumental mags ($\sim 16.2$ g-band mag).

The median $g$-band $\hat{\chi}^2_r$ is $\sim 1$ for most of the survey footprint, indicating an excellent goodness of fit in most cases. The $\hat{\chi}^2_r$ is elevated in the most crowded regions, specifically the southern Galactic bulge, where PSF estimation can be difficult in the presence of such extreme blending.

In addition to variations resulting from Galactic structure, there is residual hexagonal pattern noise, especially near the higher latitudes taken later in the observing program. Specifically, the $\hat{\chi}^2_r$ increases toward the edge of the focal plane. We attribute this pattern to variability in the quality of PSF estimation for the different CCDs in the focal plane. In addition to PSF variations as a function of position in the field of view (see Appendix \ref{sec:fov_fwhm}), each of the DECam detectors has a slightly different nonlinear onset and saturation level, which can vary over time. As discussed in Appendix \ref{sec:satdet}, the impact of underestimated saturation levels on the $\hat{\chi}^2_r$ was so severe that we reprocessed most of the DECaPS2 observations, which significantly reduced the amplitude of the hexagonal pattern noise seen here. Some of the discrete steps in $\hat{\chi}^2_r$ may be explained by version changes in the CP, including an improvement in weight estimation between version 3 and 4---that is, the steps may reflect changes in the CP uncertainty rather than actual differences in the match of the model to the data.

The median $g$-band \texttt{fracflux} over the survey footprint appears to predominantly track source density. In regions of high source density, such as the southern Galactic bulge, the median \texttt{fracflux} is $\sim 0.5$. At higher latitudes, or in the presence of strong foreground dust extinction preventing the detection of most stars, we see the median \texttt{fracflux} approaches $\sim 1.0$, as it should for isolated sources. Small fluctuations of order $\sim 0.05$ tracking the hexagonal tiling pattern can be seen in some regions of the footprint, most notably at ($-75\degree$, $10\degree$) and ($-35\degree$, $-10\degree$). 

Specifically, the median \texttt{fracflux} decreases toward the edge of the focal plane. This appears related to variations in the PSF FWHM in pixels over the field of view shown in Appendix \ref{sec:fov_fwhm}, though we have not identified the exact connection. The regions where this pattern is most notable are regions with unusually poor seeing or highly variable seeing between different visits. However, the overall smooth variation of \texttt{fracflux} throughout the survey footprint is a testament to the photometric uniformity of the survey, measured here through the deblending stability.


\begin{figure}[t]
\centering
\includegraphics[width=\linewidth]{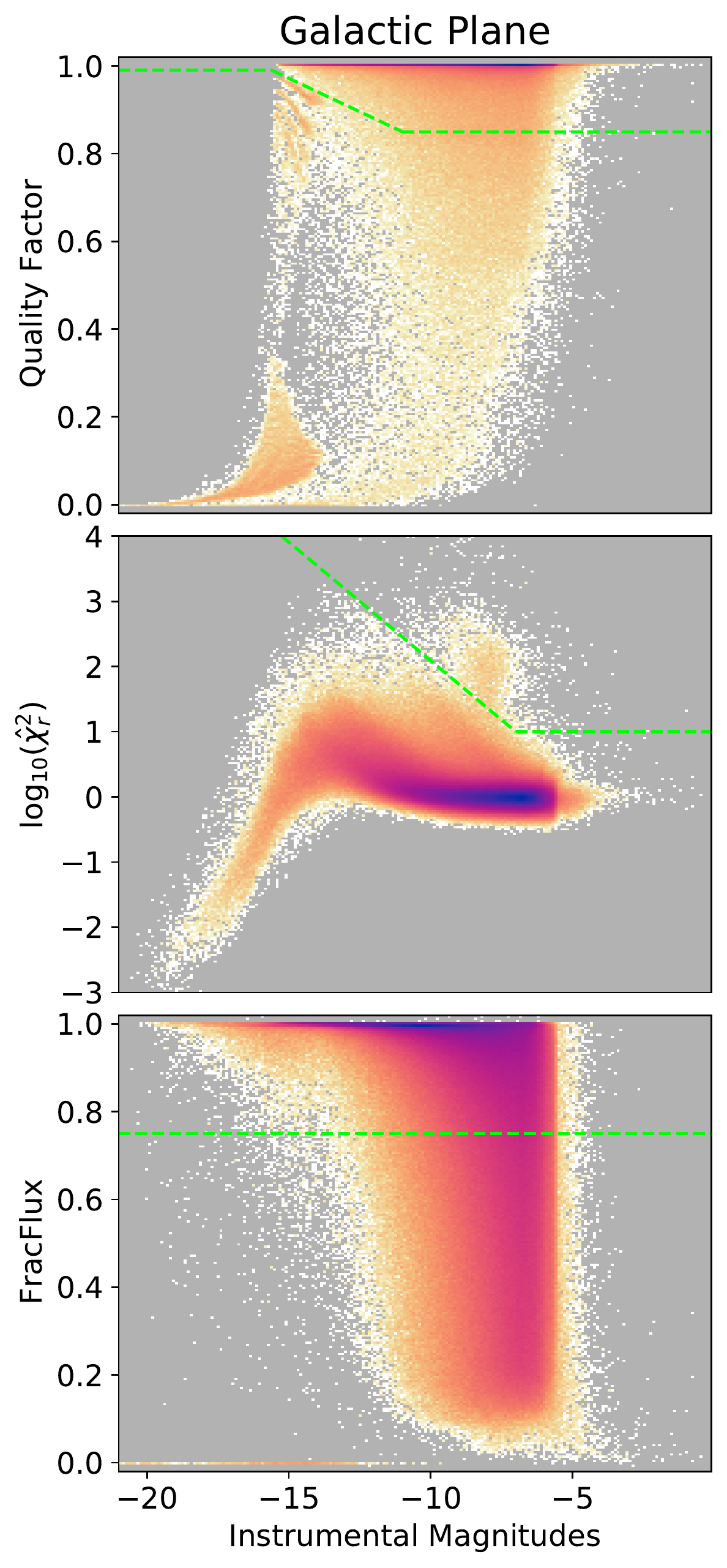}
\caption{Distribution (2D histogram) of source detection as a function of instrumental magnitude and \texttt{QF} (top), $\log_{10} \hat{\chi}^2_r$ (middle), \texttt{fracflux} (bottom). Detections shown are in $r$-band for a typical line of sight ($0.5\degree$ radius) toward the Galactic plane ($\ell$,$b$) = ($-65\degree$, $0\degree$), with no cut on photometricity. Edge case sources with negative \flux are excluded. The color scale is in log-density and each panel has its own normalization (light, white, low; dark, blue, high density). Quality cuts are shown as green-dashed lines.}
\label{fig:det_qa}
\end{figure}

\section{Suggested Quality Cuts} \label{sec:qual}

We discuss different populations of photometric outputs as they appear under metrics of the photometric fit at the detection level. We then describe the detection-level cuts we apply to define \texttt{OK} detections that are used in constructing the object-level catalog. We conclude by evaluating the distribution of objects in terms of the bands in which they are detected and how an object-level cut impacts that distribution.

At the per-band object level, we describe a cut requiring:
\begin{itemize}
\setlength\itemsep{-0.5em}
\item \texttt{nmag\_ok} $> 0$ (mandatory)
\item \texttt{fracflux} $> 0.75$ (optional)
\end{itemize}
The first cut is mandatory as a result of the catalog construction (see Section \ref{sec:buildCat}) and the second provides a possible cut to yield an extremely conservative ``high-quality'' catalog.

Further cuts at the object level on 
\begin{itemize}
\setlength\itemsep{-0.5em}
\item the average $\hat{\chi}^2_r$,
\item the total number of \texttt{OK} detections, or
\item requiring \texttt{OK} detections in multiple bands
\end{itemize}
could be explored for various applications.

In Figure \ref{fig:det_qa}, we show the distribution of $r$-band detections for a typical line of sight ($0.5\degree$ radius) toward the Galactic plane ($\ell$,$b$) = ($-65\degree$, $0\degree$), where no cut on photometricity is applied. Rare sources with negative \flux are excluded. In the top panel, the distribution is \texttt{QF} as a function of instrumental magnitude. At the far left, ridges $\texttt{QF} \sim 0.7 - 0.95$ and a large foot $\texttt{QF} \sim 0 - 0.3$ are associated with sources at or near saturation. The highest ridge remains even after a cut on CP flags. For moderately faint sources ($-12$ to $-6$ inst. mag), most sources have \texttt{QF} $\sim 1$, but a broad distribution all the way to $0$ is observed. Sources in this distribution with lower quality factor tend to be closer to the chip edge or are more likely to be spurious detections associated with artifacts (bleed trails, cosmic rays). There is no sharp transition between ``good'' and spurious detections in this broad distribution, so we cut at $0.85$ to be consistent with DECaPS1. The final detection-level cut on \texttt{QF} (green-dashed boundary) linearly connects the faint $0.85$ cut and a cut at $0.99$ to eliminate sources exhibiting characteristics of saturation.

\begin{figure}[t]
\centering
\includegraphics[width=0.99\linewidth]{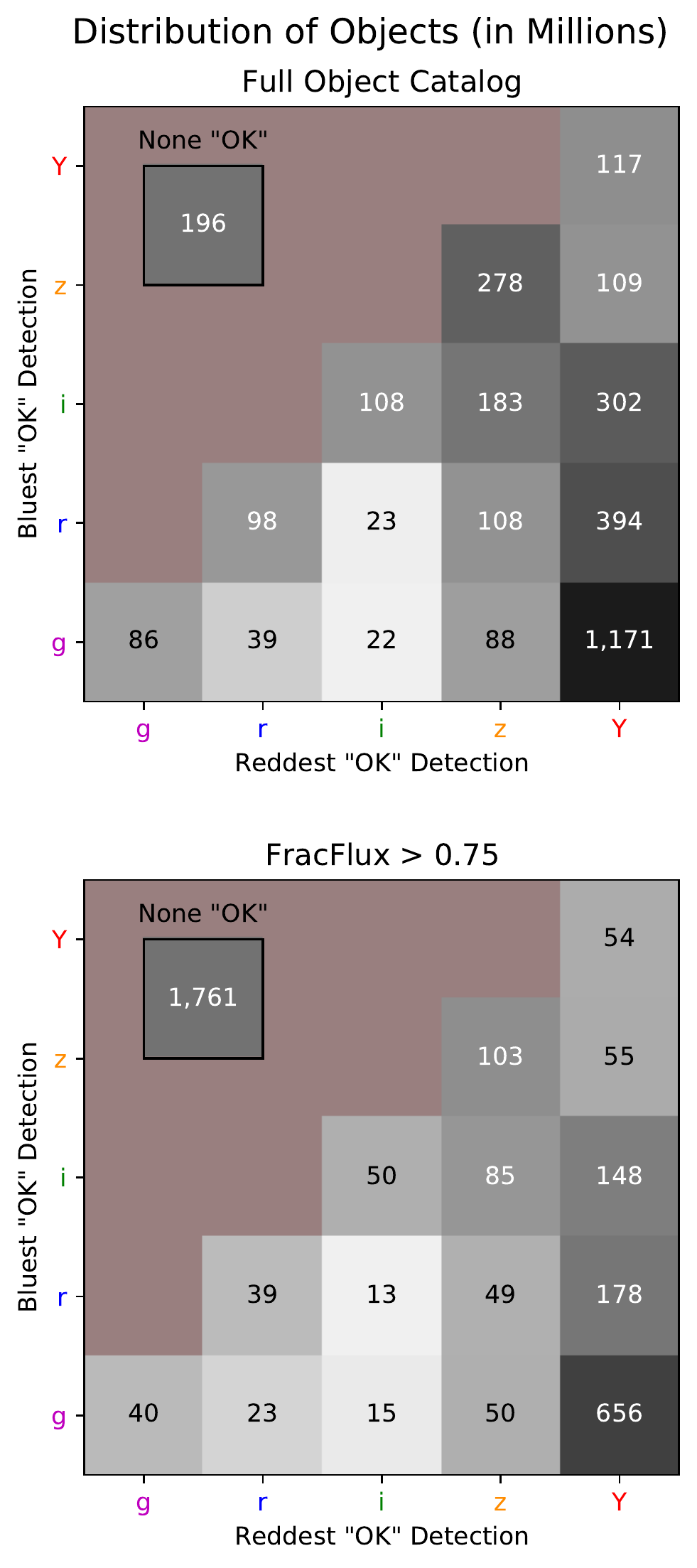}
\caption{DECaPS2 objects grouped by the bluest and reddest bands in which that object has an \texttt{OK} detection. The number of objects (in millions) is shown and each entry is colored based on its relative proportion of the sample (logarithmic color scale, white low, black high). Sources with no \texttt{OK} detections are shown as an inset. The top panel uses the usual definition of \texttt{OK} detection while the bottom panel additionally requires \texttt{fracflux}$>0.75$.}
\label{fig:objtable}
\end{figure}

The middle panel shows $\log_{10} \hat{\chi}^2_r$ as a function of instrumental magnitude. For moderately faint sources ($-12$ to $-6$ inst. mag), the $\hat{\chi}^2_r$ is centered at 1, indicating a good fit. The center of the $\hat{\chi}^2_r$ distribution increases to $\sim 10$ before turning over and rapidly decreasing toward $0$. The exact values here are strongly dependent on the regularization applied to convert $\chi^2_r$ to $\hat{\chi}^2_r$, but the sources after turnover are all saturated and removed by cuts on CP \texttt{flags}. There are two main populations of larger $\hat{\chi}^2_r$ sources at the faint end. An approximately vertical track around $-7$ with $\hat{\chi}^2_r \sim 100 - 1000$ is primarily associated with spurious sources around artifacts, particularly unmasked cosmic rays, which sometimes escape the CP cosmic ray finder especially in crowded fields. Sources in a diagonal track increasing in $\hat{\chi}^2_r$ with increasingly source flux are more likely to be: (1) one or more sources used to approximate a galaxy or (2) either real or spurious sources in the wings of bright sources. In these cases, it is more difficult to distinguish spurious from real populations. Thus, we suggest a detection-level $\hat{\chi}^2_r$ cut that only eliminates the first subpopulation (green-dashed boundary).

\begin{figure*}[ht]
\centering
\includegraphics[width=\linewidth]{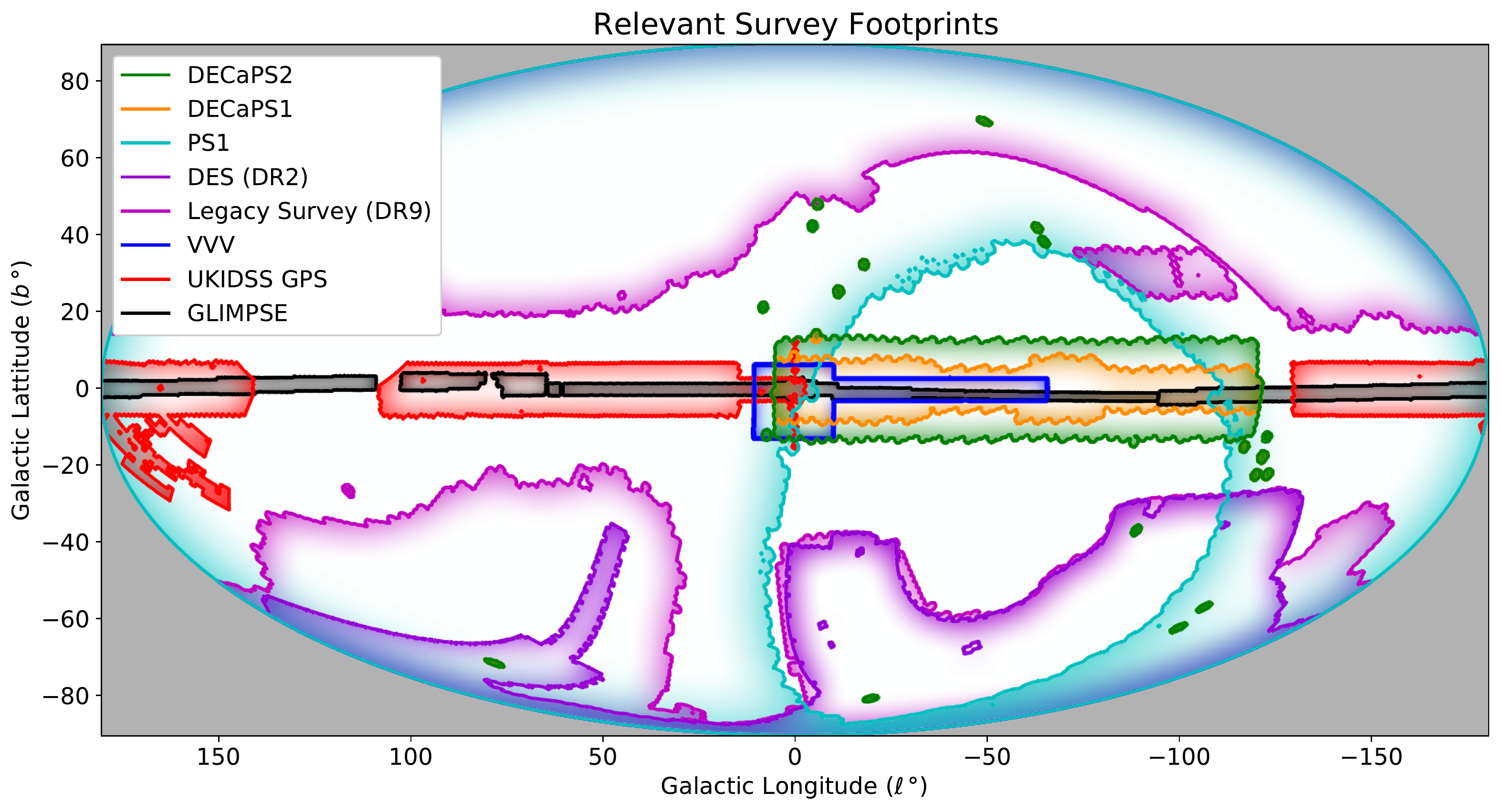}
\caption{Mollweide projection in Galactic coordinates of large survey footprints with spatial or photometric coverage overlapping DECaPS2. The boundaries of each footprint are shown with shading on the side the survey covers. Gaia, 2MASS, and WISE are not shown because they are all sky surveys.}
\label{fig:suroverlap}
\end{figure*}

The bottom panel shows \texttt{fracflux} as a function of instrumental magnitude. For faint sources ($-12$ inst. mag and fainter), the \texttt{fracflux} distribution peaks at $1$, but has a broad distribution extending to $\sim 0.1$. Presumably, below $\sim 0.1$ the faint source is simply not deblended from the nearby brighter source(s). In Section \ref{sec:iqr} we find a minimum \texttt{fracflux} of 0.3 for SNR $9 - 10$ sources. There is a second mode to the \texttt{fracflux} distribution for bright sources ($< -15$) and a population of bright, \texttt{fracflux} 0 sources. Both are eliminated by CP flags and relate to saturation effects. While we do not impose a quality cut at the detection level for \texttt{fracflux}, we find evidence in Section \ref{sec:iqr} via injection tests that \texttt{fracflux} $ = 0.75$ provides a conservative cut to eliminate sources for which deblending the source and its neighbor into one or two sources is uncertain. Since this cut is independent of magnitude, it can be applied on the object catalog as desired in a given analysis using the DECaPS2 products.

Because the error modes caught by cuts on \texttt{QF} and $\hat{\chi}^2_r$ are best represented in instrumental magnitudes, we apply these cuts prior to merging detections into objects. We confirmed the generality of these cuts across all five filters and for several pointings with different stellar densities. The equivalent distributions for Figure \ref{fig:det_qa} at the object level have all CP flagged populations (i.e., saturated sources) and sources within the detection level cuts above removed. At the object level, magnitudes are calibrated and on the AB system; thus there is slight broadening of the object distributions due to variable zeropoints between exposures. Table \ref{tab:qacut} provides the number of detections and objects before and after these cuts. Tighter cuts imposed on $\log_{10}(\hat{\chi}^2_r)$ or a cut on the probability that the region of the image around the source was of class \texttt{error} could provide even more conservative catalogs.

It is also important to apply these cuts in defining \texttt{OK} detections at the detection level because of failure modes in detection-object association in the catalog construction. Detections are either correctly assigned to an object or are subject to object-object, object-spurious, spurious-spurious confusion. In the first confusion case, the detection belongs to a real object (which may not exist in the catalog), but is assigned to another real object. In the second, a spurious detection (from a cosmic ray or diffraction spike residual) is incorrectly assigned to a real object. In the last case, a spurious detection is assigned to an object which was created off of another spurious detection. Since every detection is associated with an object and there are limits for the separation between objects after the first exposure ingested into the catalog, we know all of these failure modes likely occur, though to different extents. 

In Figure \ref{fig:objtable} we group objects by the bluest and reddest band for which that object has a detection which is \texttt{OK} (see Section \ref{sec:comb}). The diagonal of Figure \ref{fig:objtable} (top) represents the number of objects with \texttt{OK} detections in only one band. To the right of the diagonal, objects have \texttt{OK} detections spanning a larger range of photometric bands. The object counts in Figure \ref{fig:objtable} do \emph{not} check that all intervening bands, between the bluest and reddest \texttt{OK} bands, have \texttt{OK} detections. However, our expectation for real objects is that they should be observed through a contiguous range of photometric bands. To ensure that the sum of object counts in Figure \ref{fig:objtable} is the total number of objects in the catalog, we also included a count of objects which have no \texttt{OK} detections. Given that \texttt{OK} detections require no bad flags from the CP at the central pixel of the source (see Table \ref{tab:cs_bitmask}), saturated sources are an example of a real source expected to have no \texttt{OK} detections under this definition.

In the bottom panel of Figure \ref{fig:objtable}, the same grouping of the DECaPS2 objects are shown with the additional constraint that \texttt{fracflux}$> 0.75$ for a detection to be \texttt{OK}. Multi-band detections in $gri$-bands are less impacted by the \texttt{fracflux} cut as compared to $zY$-bands. This cut decreases the number of objects with \texttt{OK} detections in a single band by a factor of $\sim 2$. This reduction agrees with the intuition that faint spurious sources associated with transient artifacts (i.e., cosmic rays) are unlikely to be detected in multiple exposures.


\section{Relative Validation} \label{sec:cross}

\begin{figure*}[ht]
\centering
\includegraphics[width=\linewidth]{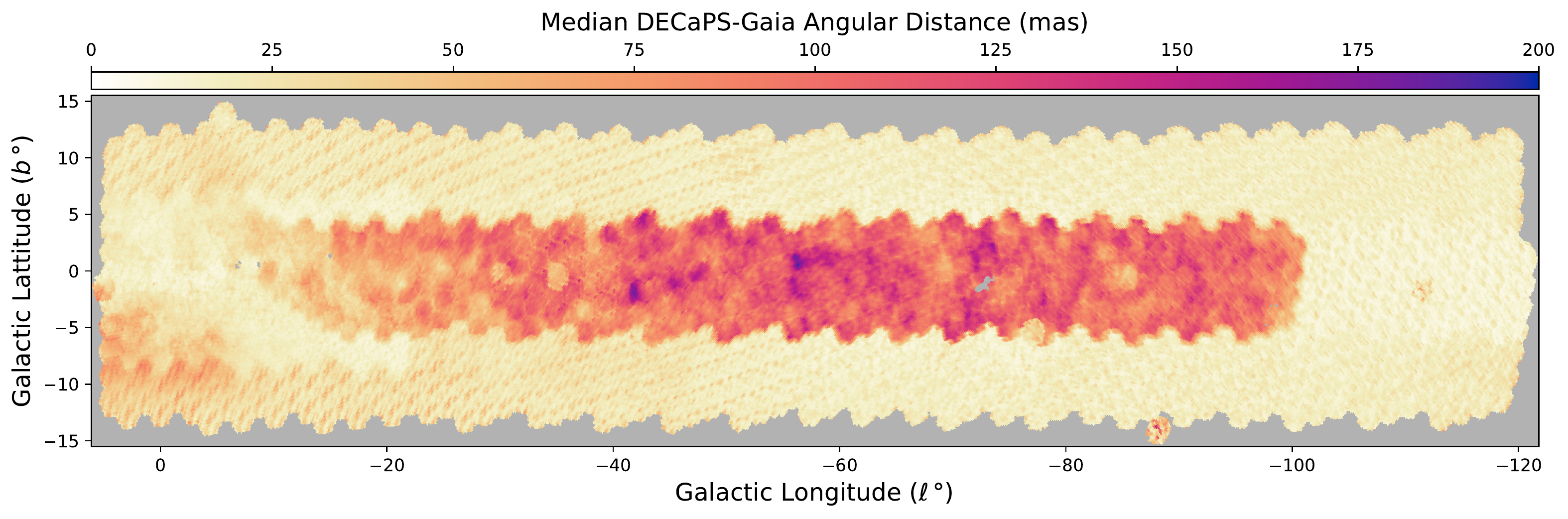}
\caption{Median angular distance (NSide = 512) between source location in DECaPS2 and the crossmatched source location in Gaia eDR3. This figure employs a cut requiring that sources be brighter than $19$th magnitude in $g$-band and be detected three times (total, regardless of band).}
\label{fig:gaia_gcdist}
\end{figure*}

To better contextualize DECaPS2, we show the spatial extent and overlap of other large-sky-coverage surveys (Figure \ref{fig:suroverlap}). DECaPS2 fills a hole in coverage of the Galactic plane with deep ($r \sim 23^{\rm{rd}}$ mag) arcsecond resolution optical-NIR photometry.

The most comparable survey to DECaPS2 is Pan-STARRS1 (PS1) in the equatorial North \citep{Chambers:2016:arXiv:}. PS1 uses similar filters ($grizy$) and reaches similar photometric depths (23.3, 23.2, 23.1, 22.3, 21.3 mag). On single-exposures, DECaPS2 is $\sim 1$ mag deeper than PS1, but contains fewer visits (3 versus 12). We engineered overlap with PS1 on both ends of the survey footprint (Figure \ref{fig:suroverlap}) in order to cross-calibrate the surveys. Other optical surveys targeting the Milky Way disk include IPHAS (r, i, H$\alpha$) and VPHAS+ (u, g, r, i, H$\alpha$), which are roughly two magnitudes shallower than DECaPS2 in the overlapping photometric bands.

Other large programs on DECam include the Dark Energy Camera Legacy Surveys (DECaLS, \citealt{Burleigh:2020:arXiv:}) and Dark Energy Survey (DES, \citealt{Abbott:2021:ApJS:}) which provide $grz$ (24.7, 23.9, 23.0 mag) and $grizY$-bands (24.7, 24.4, 23.8, 23.1, 21.7 mag), respectively, at higher Galactic latitudes. The Legacy Survey footprint shown in Figure~\ref{fig:suroverlap} includes both observations from DECam (declination less than +32\degree) and related programs observed from Kitt Peak National Observatory in the north. Smaller individual programs (including DECaPS1 data) on DECam were reprocessed as the NOAO Source Catalog (NSC) which fills the remaining optical-NIR hole in the equatorial South (see Figure 1 in \citealt{Nidever:2021:AJ:}).

The 2 Micron All-Sky Survey (2MASS, \citealt{Skrutskie:2006:AJ:}) can be used to probe further into the Galactic disk as a result of lower extinction from dust in redder wavelengths ($\sim 15.8, 15.1, 14.3$ mag depth in $J, H, K_s$). Targeted infrared plane surveys reach far fainter magnitudes ($K_s \sim 18^{\rm{th}}$ mag), such as the UKIDSS \citep{Lawrence:2007:MNRAS:} Galactic Plane Survey (GPS, \citealt{Lucas:2008:MNRAS:}) and Vista Variables in the Via Lactea (VVV, \citealt{Minniti:2010:NewA:,Saito:2012:A&A:,Alonso-Garcia:2018:A&A:}). In terms of longer wavelength space-based infrared astronomy, the Spitzer survey GLIMPSE \citep{Benjamin:2003:PASP:,Churchwell:2009:PASP:} focused specifically on the Galactic plane ($\sim 18-17$ mag depth in bands 1-4) and the all-sky Wide-Field Infrared Survey Explorer (WISE, \citealt{Wright:2010:AJ:, Schlafly:2019:ApJS:}) have imaged the Galactic plane at even longer wavelengths (20.7, 20.0 mag depth in W1, W2) where the effect of dust is even further suppressed.

\subsection{Gaia} \label{sec:gaia}
While DECaPS2 is primarily a photometric survey, it is important to have accurate astrometry in order to crossmatch between surveys. To evaluate the DECaPS2 astrometry, we match Gaia eDR3 sources (with a matching radius of 0.5") to DECaPS2 and view the source locations in Gaia as ground-truth (Figure \ref{fig:gaia_gcdist}). We further require that the DECaPS2 sources be brighter than $19$th magnitude in $g$-band and be detected three times (total, regardless of band).

There is a clear, discontinuous change in the astrometry from a median error of $\sim 100$ mas in the center of the survey footprint to $\sim 18$ mas in the outer portion of the survey footprint. This is the result of changes in the astrometric reference catalog used by the DECam Community Pipeline (CP) as the survey progressed. At different points in the survey, the CP used 2MASS, Gaia DR1, or Gaia eDR3 to obtain the WCS solutions for a given CCD in an exposure.\footnote{These changes in the astrometric reference catalog are not indicated by changes in header keywords prior to CP v5 (indicated by \texttt{ASTRMREF} thereafter) and are not necessarily consistent within a given CP version.} Since this astrometry is sufficient to match DECaPS2 to other photometric surveys, we take the heterogeneous astrometry from the CP without any further modifications. We do not attempt to resolve variations in the astrometry on the field-of-view scale visible at high Galactic latitudes because the magnitude of these variations are much smaller than the average astrometric precision over much of the DECaPS1 footprint. See Appendix \ref{sec:locprec} for more on the astrometric performance of \crowdsource alone.

\subsection{Pan-STARRS1} \label{sec:PS1}

\begin{figure*}[ht]
\centering
\includegraphics[width=\linewidth]{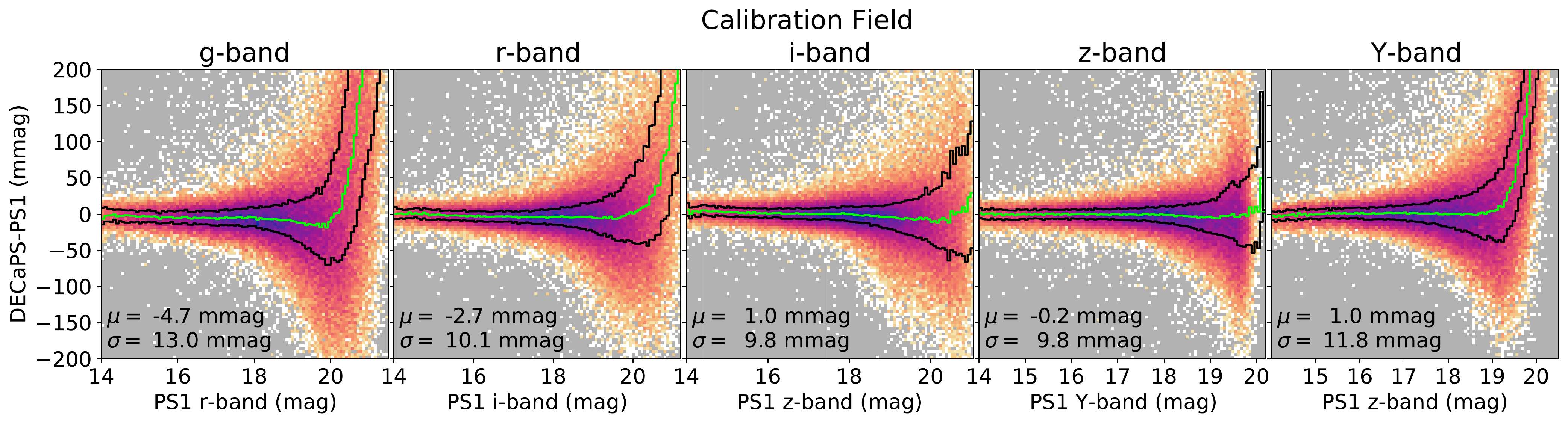}
\caption{Comparison of DECaPS2 and PS1 catalogs on a low-extinction calibration field at ($\ell,b,\Delta\theta$) = ($236\degree$, $-14\degree$, $3\degree$), where $\Delta\theta$ is the angular radius of the circular field. The color scale is in log-density and each panel has its own normalization (light, white, low; dark, blue, high density). Lines indicate quartiles in the y-axis per x-axis bin with 25\% and 75\% in black and 50\% (the median) in green. Outlier robust center (median) and scatter ($\sigma_{\rm IQR}$) metrics are shown for stars between $15^{\rm{th}}$ and $17^{\rm{th}}$ mag.}
\label{fig:PS1cal}
\end{figure*}

The absolute calibration of DECaPS2 was tied to PS1 by comparing a low-extinction calibration field at $(\ell,b)=(236\degree,-14\degree)$ in the overlap of both survey footprints (see Figure \ref{fig:suroverlap} and \ref{fig:PS1cal}). The crossmatch for the figure required sources in PS1 to be within 0.5" of the source location in DECaPS2 and that there be no more than 1 match for a given source.\footnote{A 1'' crossmatch radius was used in practice to set the zeropoints, but does not change the results because of the low stellar density in the calibration field.} We use a private reduction of PS1, available upon request.  This catalog uses the original DR1 PS1 single-epoch detections, plus a somewhat more aggressive flagging of non-photometric and problematic detections than the public PS1 data. This reflects the catalog's historical ties to the photometric calibration of PS1 \citep{Schlafly:2012:ApJ:}, and is not expected to have any meaningful differences with respect to the public PS1 catalog for the kind of broad population-wide flux comparisons most relevant to DECaPS2.

We use the color transformation derived in the DECaPS1 paper on this calibration field to convert PS1 to DECaPS filters (Equation 2 in \citealt{Schlafly:2018:ApJS:}). This transformation was obtained as a cubic polynomial fit to the color difference between DECaPS and PS1 bands as a function of the PS1 $g-i$ color. The zeropoint of that transformation was fixed by integrating HST-derived spectral energy distributions over the DECam and PS1 filter bandpasses, and thus ultimately derives from \citep{Bohlin_2014_AJ}. We then adjust the absolute zeropoint of the DECaPS2 catalog to bring the fluxes measured in DECaPS2 into agreement with the color-transformed PS1 fluxes, for unsaturated point sources brighter than 17th mag in DECaPS. We show the spatial variation of this offset over the intersection of the DECaPS2 and PS1 footprints in Appendix \ref{sec:colorTvary}.

The median offset for stars $15^{\rm{th}}$ to $17^{\rm{th}}$ mag in PS1 is $< 5$ mmag by construction of the absolute calibration. The median DECaPS2-PS1 magnitude difference is approximately flat down to $20^{\rm{th}}$ mag ($19^{\rm{th}}$ mag in $Y$-band). A sharp positive rise toward the faintest stars indicates that faint stars are estimated to be brighter in PS1 than DECaPS2. We attribute this to a selection effect given that the single exposure depth of DECaPS2 is $\sim 1$ mag deeper than PS1 (see Appendix \ref{sec:threshbias}). Faint sources near the detection limit of PS1 are only detected if their flux fluctuates high via Poisson noise, but are not detected if they fluctuate low. Thus the PS1 flux will be overestimated relative to the true flux and the flux estimated by DECaPS2, since these sources are further from the DECaPS2 detection limit than the PS1 detection limit. In the bright limit, the scatter ($\sigma_{\rm IQR}$) between DECaPS2 and PS1 is $10 - 13$ mmag. This is a similar order of magnitude to the $10$ mmag uncertainty in the absolute calibration of PS1 and relative calibration of DECaPS2.

\subsection{DECaPS1} \label{sec:DECaPS1}

\begin{figure*}[ht!]
\centering

\includegraphics[width=0.87\linewidth]{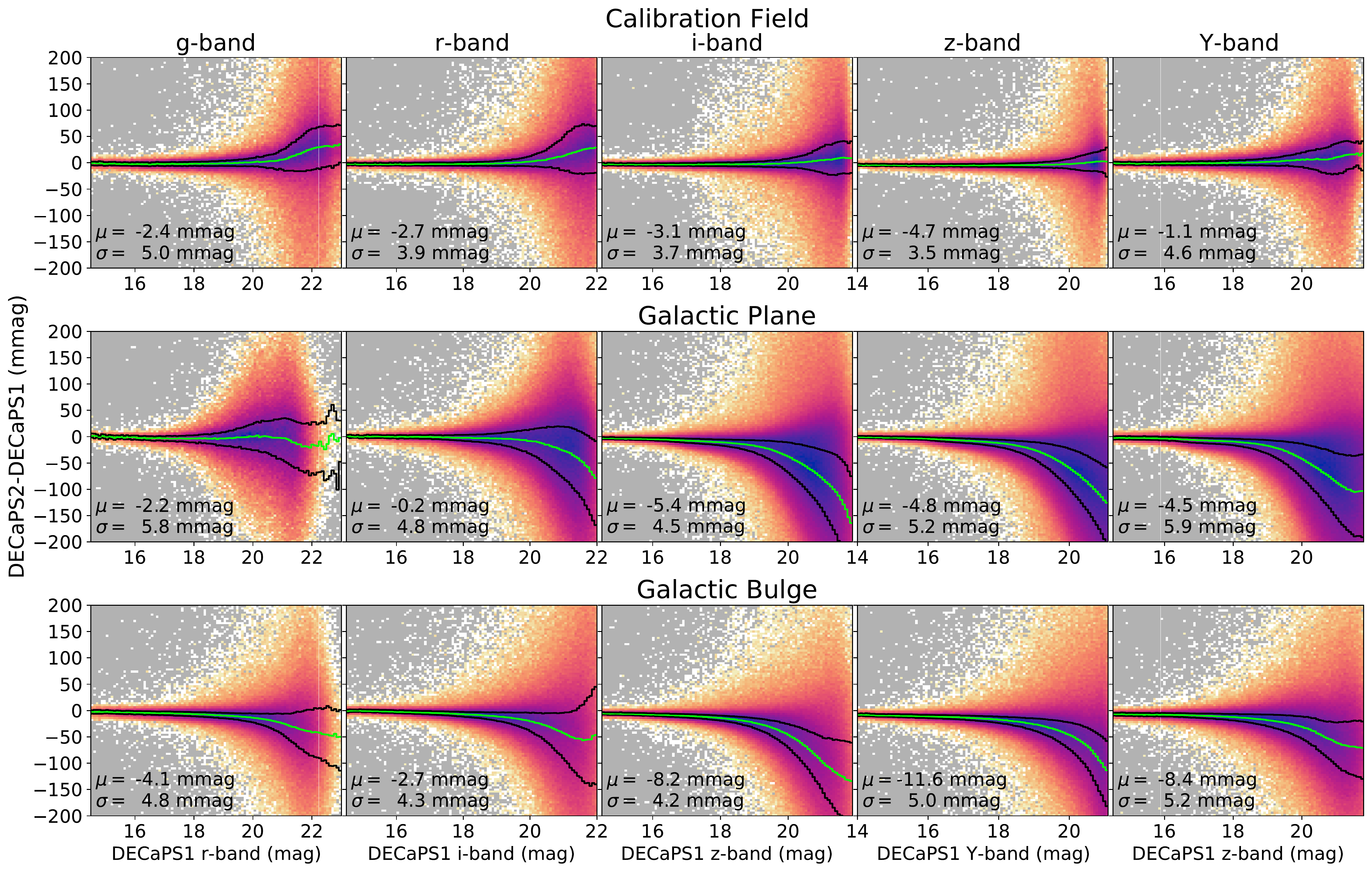}
\caption{Comparison of DECaPS2 and DECaPS1 catalogs on three representative fields: in a low-reddening calibration field ($\ell,b,\Delta\theta$) = ($236\degree$, $-14\degree$, $3\degree$), in the Galactic plane ($\ell$, $b$, $\Delta\theta$) = ($-65\degree$, $0\degree$, $0.5\degree$), and in the Galactic bulge ($\ell$, $b$, $\Delta\theta$) = ($-5\degree$, $2\degree$, $0.5\degree$). $\Delta\theta$ is the angular radius of the circular field. The color scale is in log-density, and each panel has its own normalization (light, white, low; dark, blue, high density). Lines indicate quartiles in the y-axis per x-axis bin with 25\% and 75\% in black and 50\% (the median) in green. Outlier robust center (median) and scatter ($\sigma_{\rm IQR}$) metrics are shown for stars between $15^{\rm{th}}$ and $17^{\rm{th}}$ mag.}
\label{fig:DECaPS1cal}

\includegraphics[width=0.87\linewidth]{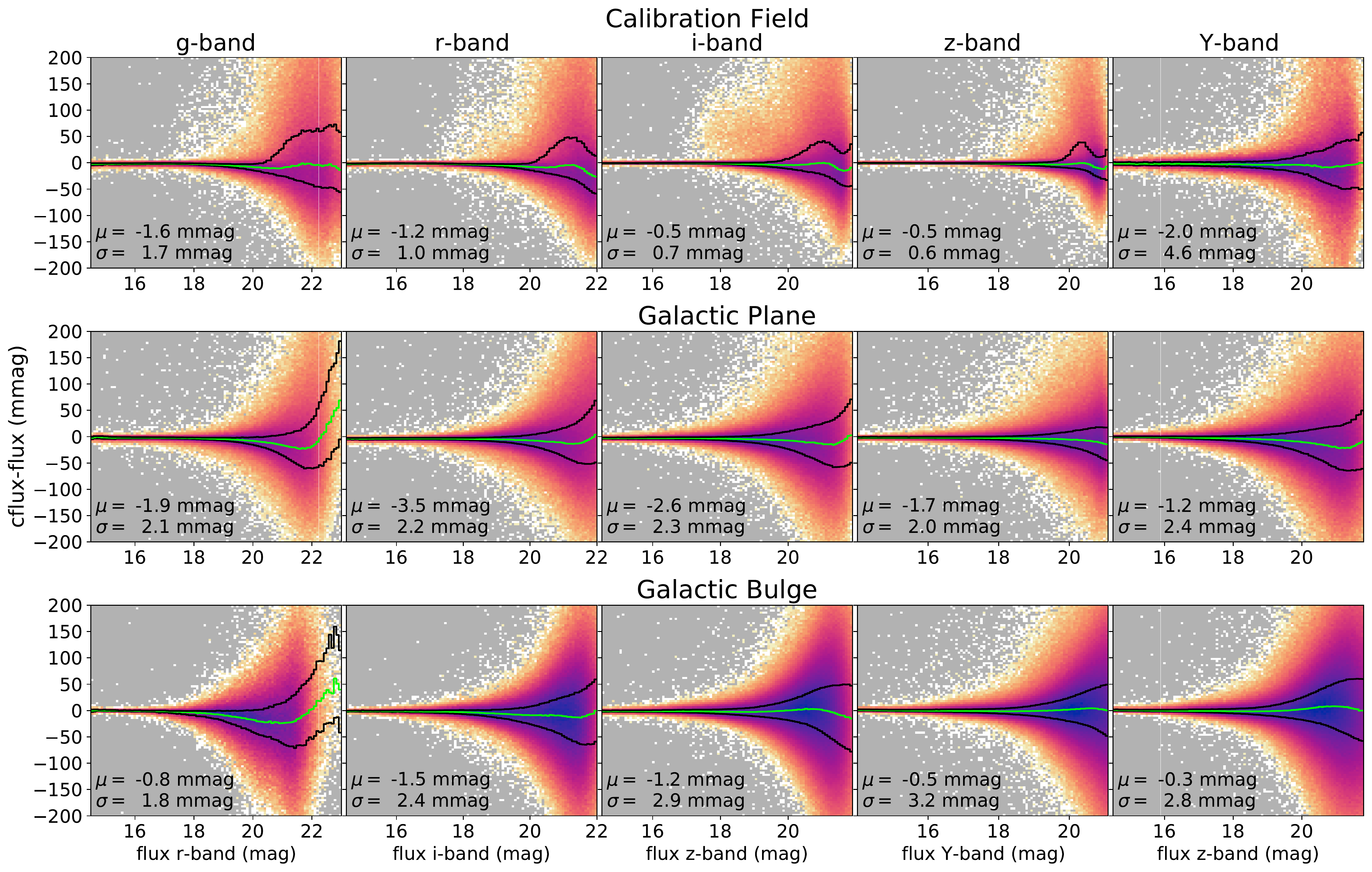}
\caption{Same as Figure \ref{fig:DECaPS1cal} except comparing \cflux - \flux within the DECaPS2 catalog.}
\label{fig:cflux_flux_cal}
\end{figure*}

We also compare the new reduction to DECaPS1 on the overlapping footprint. This provides an internal consistency check on the relative calibration (which was performed independently for the DECaPS1 and DECaPS2 processing). The crossmatch required sources in DECaPS2 to be within 0.5" of the source location in DECaPS1 and that there be no more than one match for a given source. We show a comparison in all five bands for three representative fields in the Galactic plane, Galactic bulge, and at high Galactic latitude in Figure \ref{fig:DECaPS1cal}. The median offsets between DECaPS1 and DECaPS2 ($-5$ to $0$ mmag) computed using the bright stars ($15^{\rm{th}}$ to $17^{\rm{th}}$ mag) on the calibration field are artificially fixed by both surveys being calibrated to PS1 on that field. 

However, the offsets for the Galactic plane and bulge fields are a quasi-independent check on the relative calibration for DECaPS data releases. The offsets observed ($-12$ to $0$ mmag) are again less than or equal to the absolute and relative calibration uncertainties ($\sim10$ mmag). We attribute these offsets to the improved background estimation in the most recent version of \crowdsource (see \citealt{crowdsourceunpub}), which enables sources to be more completely separated from the background. Since \crowdsource iteratively finds sources, removes their flux, and recomputes a masked, moving-median background, it has a bias toward incomplete deblending (underestimating flux) for the faintest sources, which is exacerbated by worse background/PSF modeling. These offsets are typically at the $\sim0.5-1\sigma$ level using the \texttt{dflux} per-object uncertainties. The scatter around the median for the DECaPS2-DECaPS1 crossmatch is at least a factor of two smaller than the DECaPS2-PS1 crossmatch. As expected, the scatter introduced by different processing and more data on the same instrument is smaller than the scatter comparing to a completely different photons measured by a different pipeline and instrument. 

\subsection{cflux} \label{sec:cfluxcompare}

We also perform a consistency check on the background-corrected flux (Section \ref{sec:CloudCovErr}) by examining \cflux - \flux as a function of magnitude on the same three representative fields in DECaPS2 (Figure \ref{fig:cflux_flux_cal}). Systematic offsets in all plots are $<1\sigma$ using the \texttt{dflux} per-object uncertainties, and are more typically $\sim0.25\sigma$. The median offsets on bright ($15^{\rm{th}}$ to $17^{\rm{th}}$ mag) stars are small, only $-3.5$ to $0$ mmag, but tend negative, indicating \crowdsource is still not completely separating sources from the background. As expected, the largest fractional changes occur for the faint sources, which have comparable flux to diffuse background emission. These large relative changes lead to a larger scatter at faint magnitudes, as indicated by the larger separation between the 25\% and 75\% quartile lines. The faint interquartile range is much larger for the Galactic bulge and Galactic plane compared to the calibration field, which is expected given the much more complex structure of the background residuals which \cflux corrects. Unlike the DECaPS2-DECaPS1 comparison, offsets for faint stars are in general small ($\sim 20$ mmag) and for the Galactic bulge and calibration field, have both signs. For the Galactic plane, all five bands dip slightly negative, indicating that the faint sources are estimated to be slightly brighter by \cflux as compared to \fluxns.


\section{Injection Tests} \label{sec:inject}

While comparisons to other surveys provide helpful context and confirm consistency between different instruments or pipelines, it is important to evaluate the performance of a pipeline at recovering known sources-- ``synthetic injection tests.'' We use these injection tests at the single-visit level to model the biases at low SNR. The single free parameter of our low SNR model is $s_d$, and we show in Appendix \ref{sec:threshbias} that $s_d$ is related to the usual definition of photometric depth in terms of recovering 50\% of sources at a given magnitude. Thus, we measure the photometric depth, including crowding effects and other complications via the observed low SNR bias. We further use the injections to evaluate the deblending performance of \crowdsource and the accuracy of the reported flux uncertainties.

\begin{figure*}[t]
\centering
\includegraphics[width=\linewidth]{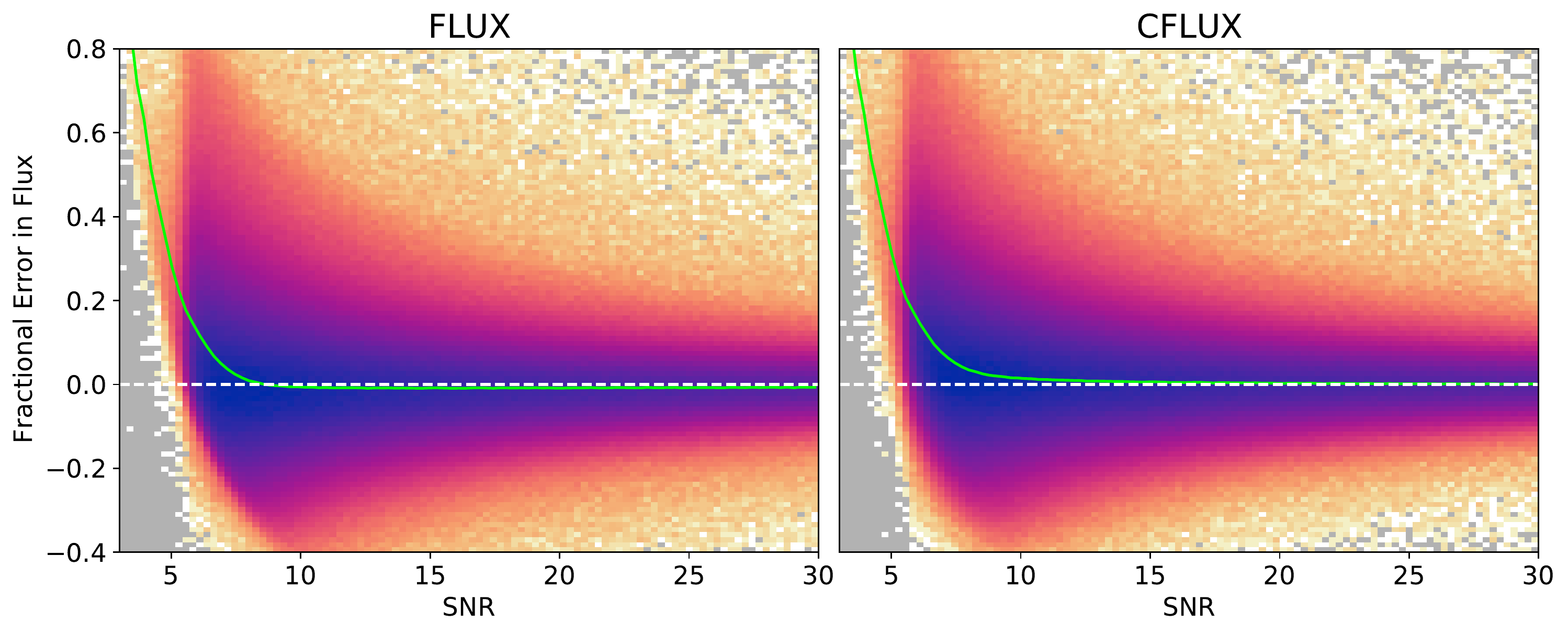}
\caption{Histogram (2D) for injected source tests as a function of the ``true'' injected SNR ($s_0$) and fractional error in recovered flux using either \flux (left) or \cflux (right). Color scale is logarithmic density (light, white, low; dark, blue, high density) with gray indicating zero density. All panels include a cut selecting for unblended sources (see text). The green line shows the binned trimmed (excluding the highest and lowest 10\%) mean as a function of SNR and the white-dashed line provides the zero-error reference.
}
\label{fig:low_SNR}
\end{figure*}

To perform synthetic injection tests, we first obtain the photometric outputs for a given image, inject sources back into that image, and then solve the image with injections as if it were an additional CCD observed during that exposure. For each exposure in the survey, we select at random one CCD out of the $60-61$ observed CCDs per exposure on which to perform these injection tests. The density of injected sources is chosen to be $10\%$ of the source density in the original image, which we found to not significantly perturb the original solution.

We draw the flux of injected sources from the distribution of sources found in the original image, after excluding some sources. To prevent injecting very bright sources that will impact a significant fraction of the CCD area, we apply a strict cut on flux (in ADU) which corresponds to $\sim 17^{\rm{th}}$ magnitude in $g$-band. We also exclude sources from the seed distribution that have a ``bad'' flag set at their central pixel (saturation, broken pixel, etc.; see Table \ref{tab:cs_bitmask}). We then sample sources from that flux distribution by uniform sampling of the linearly interpolated CDF. The source positions are drawn from a uniform (float, not integer) spatial distribution across the CCD, with a 33-pixel exclusion zone from the edges of the image.

The test sources are injected with the position-dependent PSF model obtained during the solution of the initial image. Injecting each source involves evaluating the PSF model at the injected location, an independent Poisson draw (consistent with the \crowdsource gain) for each pixel in a stamp of pixels impacted by the star, and adjusting the weight image to account for the injected counts (again, consistent with the gain). We choose a stamp size of 511 pixels, which is much larger than the stamp size used to model most sources in \crowdsource (19 or 59 pixels). 

The image, weight image, and data mask (unchanged) after the injections are saved with RICE (lossy) compression to mimic the outputs of the CP.\footnote{We explicitly choose a dither seed for the RICE compression, which differs from the one originally used to save the images in order to minimize possible systematics between the injected and original sources. However, tests using the same dither seed for the compression suggest that the quantization noise is sufficiently subdominant to not perturb the photometric solutions.} All random steps in our injection module use a PCG64 random generator seeded on the date-time in the filename of the exposure for reproduciblility. We save the locations and fluxes of the injected sources as an additional field in the catalog files (see Section \ref{sec:dataavil}).

One limitation of these injection tests is that the empirical flux distribution used for the flux draws may differ from the true flux distribution, especially near the faint end of the distribution. That is to say, the \crowdsource outputs are incomplete for the faintest stars detected and thus the injections underestimate the number of faint stars to be injected. Another limitation is that injections use the \crowdsource model PSF which may not match the true PSF for the image. However, these tests act as a consistency check on the model and provide a valuable measure of sensitivity in crowded fields.

\subsection{Low SNR Bias} \label{sec:lowSNR}

All survey pipelines have a limit below which they are unable to differentiate faint sources from noise. Using injection tests, we can characterize this limit and the biases resulting from the handling of faint sources. To first avoid the complexities of blending, we restrict our sample to sources injected at least 2 FWHM away from a source in the original image and found within 0.25 FWHM of the injected location. In rare instances, \crowdsource can model the background as having negative counts or a source as having negative flux. We exclude both cases here.

We show in Figure \ref{fig:low_SNR} a (logarithmic) histogram of the fractional error in the flux recovered for injected sources over the entire survey footprint as a function of their injected signal to noise ratio (SNR). The green line shows the binned, trimmed (excluding the highest and lowest 10\%) mean as a function of SNR. The white-dashed line provides the zero-error reference. The SNR is computed in the background-noise-dominated limit, assuming the PSF effective area is that of a Gaussian ($4\pi\sigma^2$), with the same FWHM as the source PSF.

\begin{ceqn}
\begin{align} \label{eq:SNR}
{\rm SNR} = \frac{f \times g}{\sqrt{4\pi\sigma^2 \times b \times g}}
\end{align}
\end{ceqn}
where $g$ is the \crowdsourcens-estimated gain, $b$ is the background (in ADU) at the center of the source, and $f$ is the ground-truth flux of the injected source.

\begin{figure}[t]
\centering
\includegraphics[width=\linewidth]{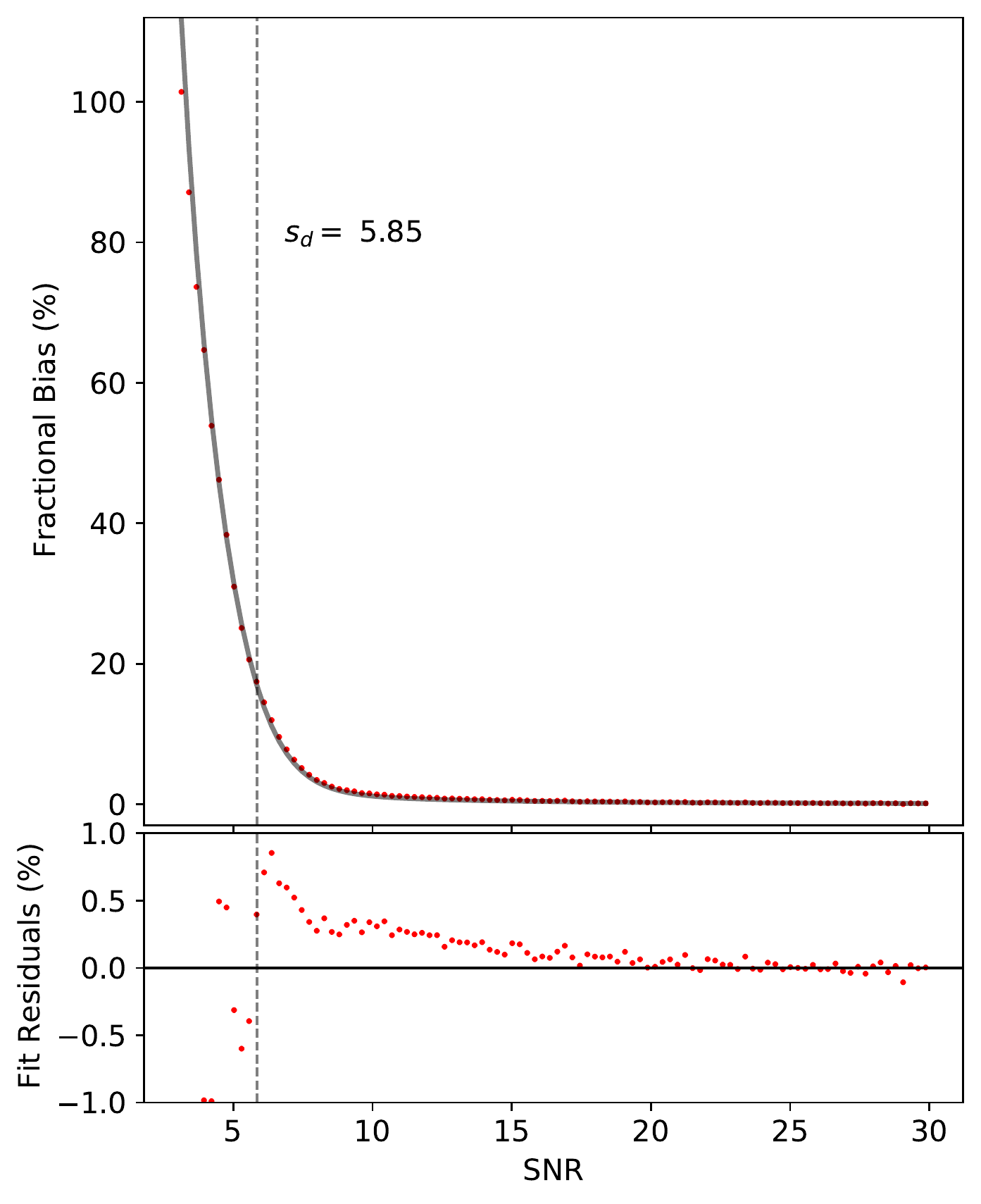}
\caption{Average low SNR bias of \cflux (red points) and least-squares fit to Equation \ref{eq:fitfunc} (gray solid line). The effective detection threshold of \crowdsource obtained from the fit is shown (vertical gray-dashed line). Fit residuals are shown in the bottom panel.
}
\label{fig:snrfit}
\end{figure}

When ambiguous, we denote this SNR by $s_0$ to distinguish it from SNR computed using the recovered flux $s_r$, the injected flux including the realization of Poisson noise $s_i$, and the SNR detection threshold of the photometric pipeline $s_d$.

The left and right panels of Figure \ref{fig:low_SNR} compare the recovered flux from \crowdsource (\fluxns) and the background-corrected flux from \cloudcoverr (\cfluxns). The true flux is underestimated on average in the high SNR limit by \flux at $-0.7\%$ (median for SNR 25-30). In contrast, \cflux approaches zero bias from above (see discussion below) and is a $0.1\%$ overestimate over the same SNR range. A multiplicative underestimate of flux is in part expected for \crowdsource since it sequentially detects and separates sources from the background. That is, until a source has been identified, the flux from that source contributes to the average that determines the background model. However, with each iteration (of which there are $4-10$), the flux estimate should converge to the truth from below. The background-corrected \cflux likely detects this incomplete deblending relative to regions of pure sky and reapportions the flux from the background to the star.

The recovered flux is overestimated (positive bias) in the low SNR limit as shown by the positively diverging average (red line) in both panels. We explain (and model) this bias as resulting from two effects. For moderate SNR (10-20), a positive bias derived in \cite{Portillo:2020:AJ:} dominates, which results from using the maximum likelihood position for a source. Intuitively, this occurs because the maximum likelihood solution seeks to model as much flux as possible, even if that flux is partly background noise. This bias depends only on the true SNR of the source and an expansion to 4$^{\rm{th}}$ order gives the last two terms in Equation \ref{eq:fitfunc}.

At low SNR, a detection bias dominates. In order to be found by \crowdsourcens, the peak in the PSF-convolved image (after background subtraction) must be $5\sigma$ above the noise. Faint sources with noise that causes them to fluctuate high are found, but those that fluctuate low are not. This bias is reflected in the lack of sources in the lower left compared to the upper right of Figure \ref{fig:low_SNR}. 

In Appendix \ref{sec:threshbias}, we derive a form of this detection bias which depends only on the true SNR of the injected source ($s_0$) and the source-detection threshold ($s_d$). Further, we apply an approximation to obtain the first term in Equation \ref{eq:fitfunc}. This approximation is useful because it allows us to easily fit to the average bias from the injection tests and measure the effective $s_d$ of \crowdsourcens.

\begin{ceqn}
\begin{align} \label{eq:fitfunc}
    \frac{\left< f - f_0 \right>}{f_0} = \frac{B\left(\frac{s_d}{s_0} - 1\right)}{1 - e^{-\frac{A}{\sqrt{2}}\left(s_d-s_0\right)}} + \frac{1}{s_0^2} + \frac{1}{s_0^4}
\end{align}
\end{ceqn}
Here (A,B) = (1.98, 1.135) are constants fixed by the approximation of the complementary error function ($\erfc$), $f$ is the measured flux, $f_0$ is the true flux, and $s_d$ is the single free parameter. 

\begin{figure}[b]
\centering
\includegraphics[width=\linewidth]{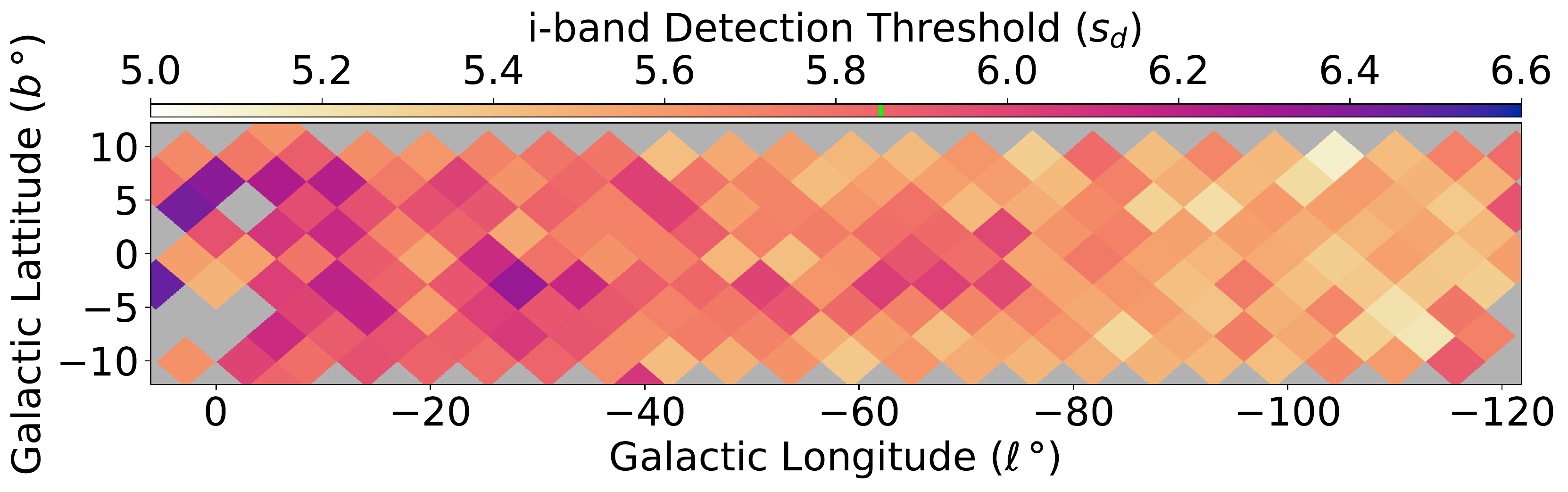}
\caption{Spatial variation of the effective detection threshold $s_d$ in $i$-band, which tracks source density. The average value of $s_d$ obtained from the fit in Figure \ref{fig:snrfit} is indicated on the color bar by a green line.
}
\label{fig:i_var_s}
\end{figure}

Figure \ref{fig:snrfit} shows the weighted, least-squares fit of the average bias (red line in Figure \ref{fig:low_SNR}) for \cflux to Equation \ref{eq:fitfunc}, using the square root of the number of sources per bin as weights. The optimal value for the effective source-detection threshold of \crowdsource is $s_d = 5.85$, which is higher than the formal $5\sigma$ detection threshold. The observation of $s_d > 5$ is expected given the relatively local sky modeling and the fact that the local sky is biased high until a source is identified. The residuals are not trendless, but are $\sim 0.1\%$ in the high SNR limit and $\sim 1\%$ near the detection threshold. The remaining trend could result from the analytical approximations made to arrive at Equation \ref{eq:fitfunc}, the use of a binned outlier-rejected average, the contribution of other unmodeled biases, or the breakdown of the background-noise dominated limit and the need to include the Poisson noise from the source.\footnote{Contributions from read noise and dark current are small for DECam, but could also contribute here.}

\begin{figure}[hb]
\centering
\includegraphics[width=\linewidth]{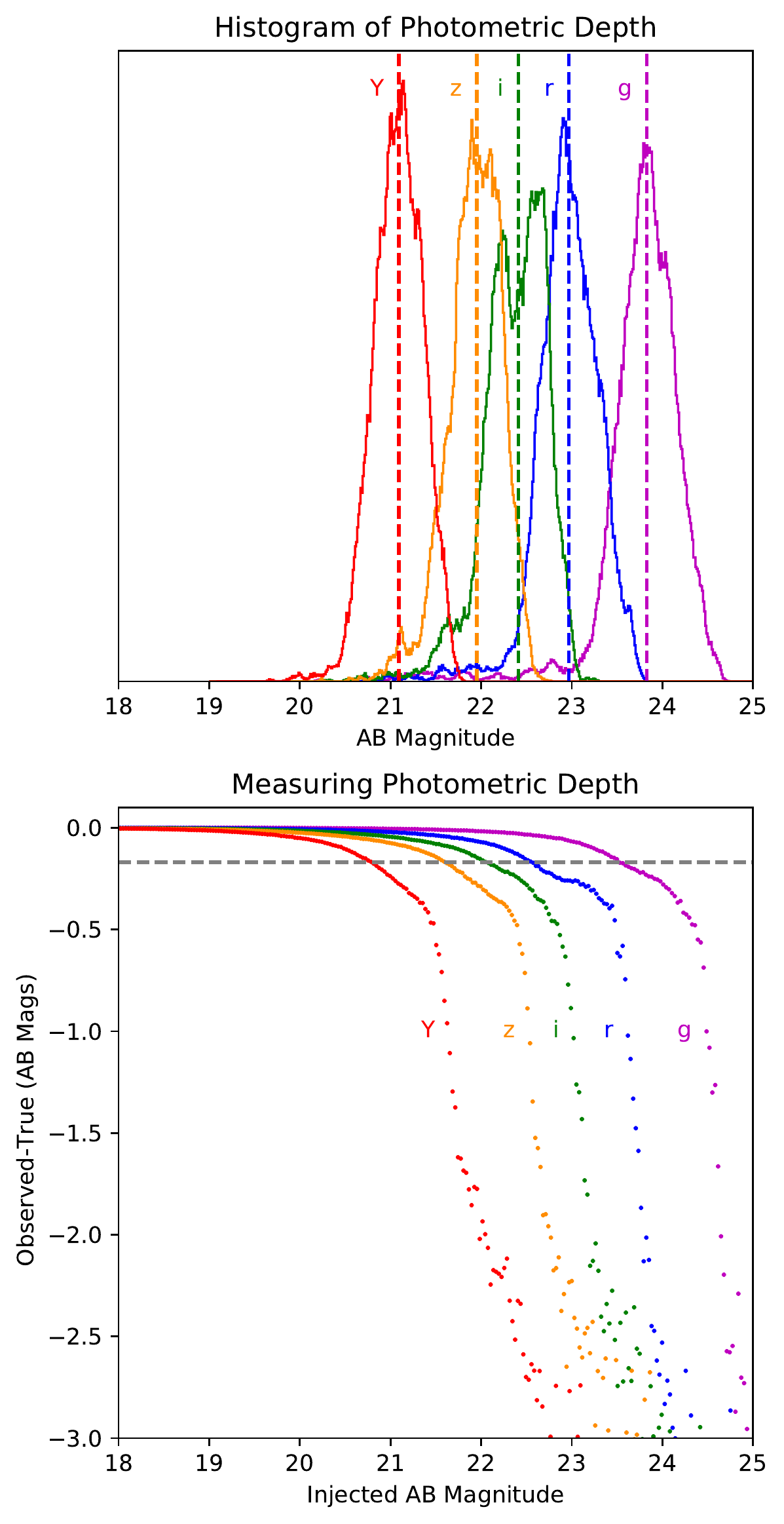}
\caption{(Top) Histogram of imaging depth assuming 5.85$\sigma$-detection limit for each photometric CCD image included in the DECaPS2 catalog, separated by band. Dashed vertical line indicates the median. (Bottom) Median bias for recovered injected sources that are unblended as a function of magnitude, again separated by band. Horizontal dashed line indicates $-169$ mmag bias, which is the bias at which only $50\%$ of sources with that magnitude would be detected.
}
\label{fig:hist_depth}
\end{figure}

In the above fit, we have pooled over all injection tests over the survey footprint and over all photometric bands. However, variability in the background, source density, and PSF can change the effective detection threshold. In Figure \ref{fig:i_var_s}, we restrict to $i$-band and perform the same fit as in Figure \ref{fig:snrfit} to sources subdivided into NSide = 16 HEALPix pixels (about 2 degrees on a side). Only pixels with more than 900 sources passing the aforementioned cuts (to limit blending) are fit, which is why several pixels in the Galactic bulge are not treated by this analysis. 

Despite the low resolution, there is a clear trend with source density--a lower detection threshold with lower source density, approaching the formal $5\sigma$ threshold. At higher source densities, $s_d$ approaches $\sim 6.5$. This could be a result of a larger density of real, below-threshold sources or the onset of blending effects. We exclude, at least partially, the effects of blending by a cut on sources in this plot, but have not accounted for the impact of blending on the bias model. Despite the variability shown in Figure \ref{fig:snrfit}, the average $s_d = 5.85$ appears to adequately describe most of the survey footprint; we use this single value in what follows to avoid handling the Galactic bulge pixels which did not have enough isolated sources to estimate $s_d$.

These low SNR biases could be partially suppressed downstream of the single-exposure stage by using forced photometry at the average location of objects identified over different exposures and/or photometric bands. In that case, only sources with noise that fluctuates negative in every observation would be missed. However, no such post-processing is performed in DECaPS2. The catalog simply takes an average of all individual detections of an object (ignoring non-detections), which bakes in a positive detection bias. In order to implement a ``correction'' for the low SNR biases outlined above on a photometric catalog, we would need to use a prior on the true source-flux distribution. Choosing a prior would allow a statistical correction as a function of the observed SNR. We leave individual use cases to decide on the appropriate prior and here simply use the bias to inform the quality of our photometry. 

\subsection{Photometric Depth} \label{sec:depth}

One of the most important metrics of a photometric survey is its depth, often defined as the magnitude such that $50\%$ of sources with that flux are recovered. One way to estimate this metric is to assume a recovery limit in SNR for the photometric pipeline, which we will choose to be 5.8 as informed by the fit above. Then, using the gain, average PSF width, and average sky counts for a CCD, we can use Equation \ref{eq:SNR} to solve for the equivalent flux. Accounting for the variable zeropoints per exposure and only including photometric exposures that are used in creating the catalog, we obtain Figure \ref{fig:hist_depth} (top). The histogram of depths spans roughly $\pm1$ mag. The strong bimodality in $i$-band is the result of brighter-than-usual sky during the beginning of the DECaPS1 observing run.

In light of Figure \ref{fig:low_SNR}, calculating photometric depth by fixing an SNR detection limit (without using injection tests to determine $s_d$) is only marginally more accurate than choosing the depth based on the turnover in the magnitude histogram for recovered sources. While it might be attractive to use injection tests to determine at what magnitude $50\%$ of injections are recovered and define that as the depth, such a definition is strongly sensitive to how one defines a source as being correctly recovered. How close is the source to the injected position and, even more importantly, how close is the recovered flux to the true injected flux? These questions are significantly complicated by blending in crowded fields and the bias shown in Figure \ref{fig:low_SNR}. 

Instead, we can use the injections to measure the achieved survey depth based on the observed low SNR bias. As shown in Appendix \ref{sec:threshbias}, it turns out that $s_d$ is exactly the SNR ratio such that a $s_d$ source has a $50\%$ chance of recovery in a given exposure. Then, once $s_d$ is known, we can find the fractional bias at $s_d$, 16.9\% for $s_d = 5.85$ (see Appendix \ref{sec:threshbias}). Since we often consider differences in magnitudes, we can convert from fractional error to obtain a bias of $-169$ mmag. By measuring at what calibrated magnitude the recovered injection tests are biased by $-169$ mmag, we measure the effective photometric depth of the survey, including crowding effects and other complications.

\begin{figure}[t]
\centering
\includegraphics[width=\linewidth]{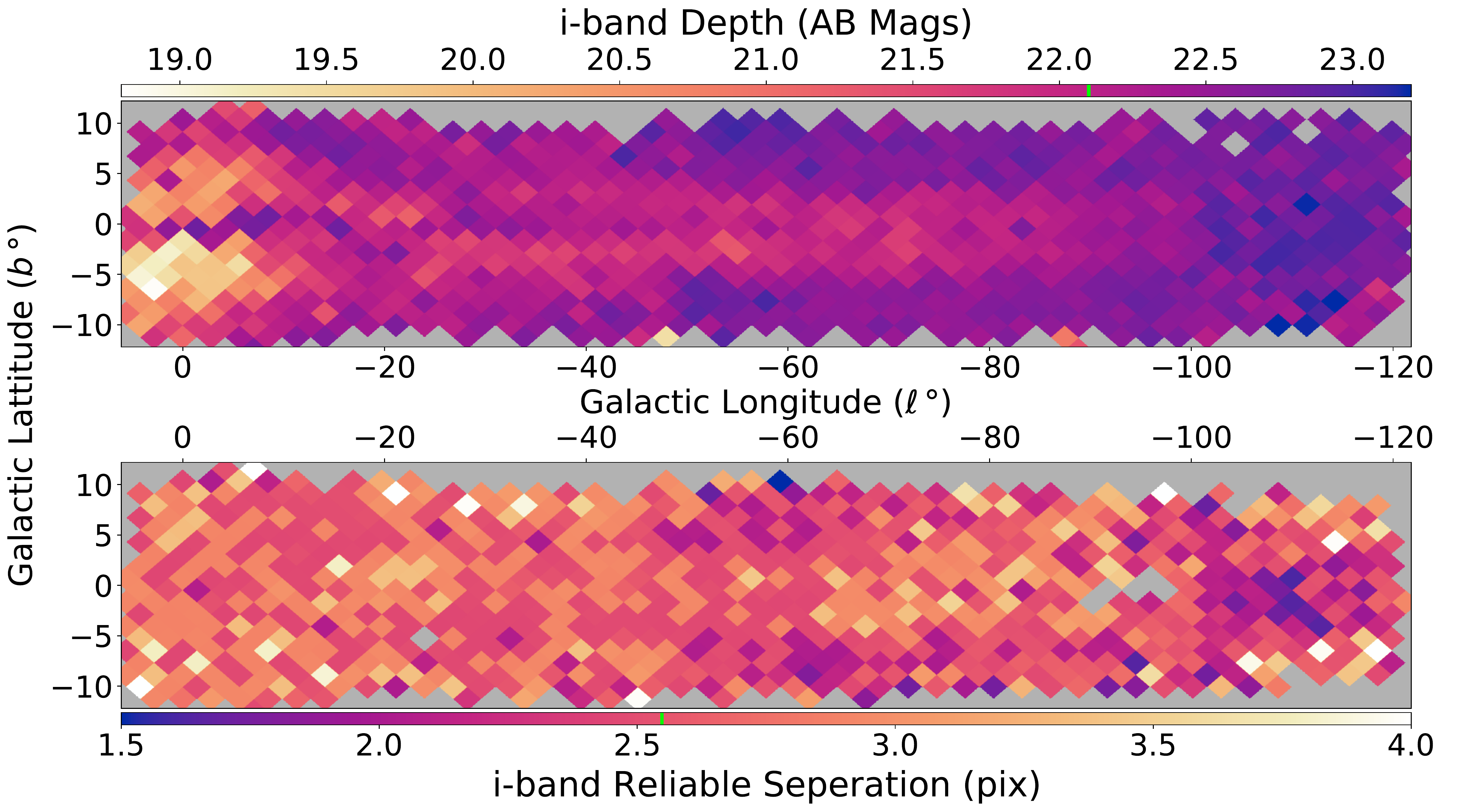}
\caption{Variation of the photometric depth (top) and reliable separation (bottom) in $i$-band across the survey footprint. Both are determined by a threshold of $-169$ mmag bias on recovered injected sources. The average depth from Figure \ref{fig:hist_depth} and the median reliable separation over the survey footprint are indicated on the respective color bar by a green line.
}
\label{fig:i_var}
\end{figure}

Using the same cuts to obtain only well-separated sources as in Figure \ref{fig:low_SNR}, we obtain Figure \ref{fig:hist_depth} (bottom). For each photometric band, we show the median bias for the recovered flux in a narrow magnitude range as a function of the magnitude of the injected source. The onset of a positive bias is observed toward the faint limit and the intersection of the median curve with the $-169$ mmag threshold (gray, dashed) determines the limiting magnitude in that band.

Using the bias method, we find the photometric depths are 23.5, 22.6, 22.1, 21.6, 20.8 mag in grizY bands, respectively. These depths are usually slightly brighter ($\sim300$ mmag) than using the median of the histograms assuming a sharp SNR cutoff which give depths of 23.8, 23.0, 22.4, 22.0, and 21.1 mag. Using either method, these depths surpass the targets (24.1, 22.3, 21.2, 20.6, 20.3 mag; see Section \ref{sec:Observe}) in all bands but $g$-band, which is a 0.5 mag shallow. However, we have three images in each band, so even in $g$-band, a future processing that coadds images would reach our target depth.

Another benefit of the bias method on injections is that we can map spatial variations in the depth (available in Section \ref{sec:dataavil}) at a spatial resolution set by the injected source density (Figure \ref{fig:i_var}, top). To do so, we no longer apply the cuts restricting to unblended sources in order to measure the depth in the Galactic bulge. We observe spatially correlated variations across the survey footprint for $i$-band, which clearly track source density. The survey is $\sim1$ magnitude deeper at high latitude and in regions of low source density and $\sim1-2$ magnitudes shallower toward regions of high source density. In the Galactic bulge, some of this bias likely derives from blending, but in the crowded field limit where depth is difficult to define, it is desirable to have the depth track the reliability (bias) of sources as a function of magnitude.

\subsection{Blending} \label{sec:blend}

We can also use injection tests to provide a measure of a reliable separation at which sources are correctly deblended by \crowdsourcens. We can measure the maximum separation between the injected source and a source in the original image before the source is biased more than a threshold. Here we again choose a $-169$ mmag bias threshold, but here it just serves as a well-motivated measure of when the blending bias becomes ``significant,'' as measured relative to the detection bias at threshold. The variation of this ``reliable distance'' in pixels (which are 0.26'' for DECam) is shown over the survey footprint in $i$-band in Figure \ref{fig:i_var}, bottom. We consider only injections within 8 pixels of an original source and only show HEALPix NSide = 32 pixels with more than 200 such sources.

There is a sharp cut preventing source deblending below one pixel in \crowdsourcens. At low source densities, \crowdsource appears to approach this limit with reliable deblending distances of $\sim 1.5-1.7$ pixels. However, for much of the survey footprint, the reliable separation is closer to 3 pixels. Taking a median over the survey footprint, we find an average reliable separation of 3.0, 2.9, 2.5, 2.5, 2.5 pixels and median FWHM of 5.2, 4.8, 4.4, 4.2, 4.1 pixels for $grizY$, respectively. This suggests the median performance is deblending sources separated by $\sim60\%$ of the FWHM, which is better than the Rayleigh resolution limit. These measures are complicated in part by the variations in the underlying source-density distribution which can modify the observed bias.

\begin{figure}[t]
\centering
\includegraphics[width=\linewidth]{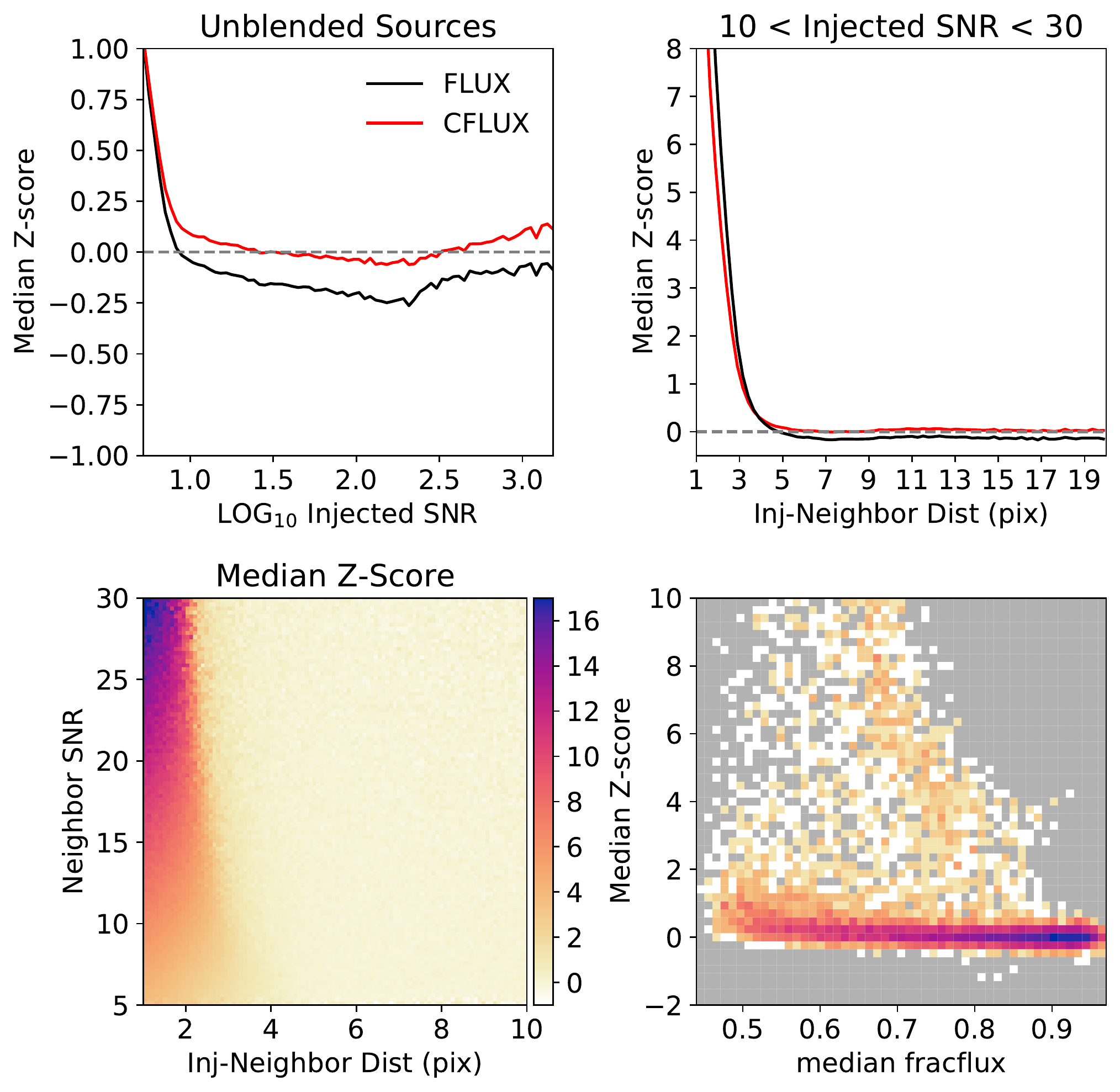}
\caption{(Top Left): Median Z-score for unblended, injected $z$-band sources as a function of the injected SNR for both \flux and \cfluxns. (Top Right): Median Z-score for injected sources with moderate SNR as a function of distance to a source in the original image. (Bottom Left): Additional dependence on the neighbor SNR is shown. (Bottom Right): Converts the bottom left panel into a 2D histogram of the median Z-score bias versus the median \texttt{fracflux} for the given injected-neighbor source distance. Color scale is logarithmic density (light, white, low; dark, blue, high density) with gray indicating zero density. 
}
\label{fig:nb_bias}
\end{figure}

We further investigate the onset of bias from blending using the injection tests in $z$-band in Figure \ref{fig:nb_bias}. In the top left, we apply the cut to consider only unblended sources and show the median Z-score as a function of the (log$_{10}$) injected SNR for both \flux and \cfluxns. For sources over 10 SNR, the median Z-score is approximately flat with respect to injected SNR. The remaining trends observed could arise from the discrete changes in size of the PSF stamp used by \crowdsource to model sources. For \cflux the median Z-score is centered around zero, while \flux is biased low by $-0.1\sigma$, in agreement with Figure \ref{fig:low_SNR}. Below 10 SNR, sources are biased high for reasons described in Section \ref{sec:lowSNR}.

\begin{figure}[t]
\centering
\includegraphics[width=\linewidth]{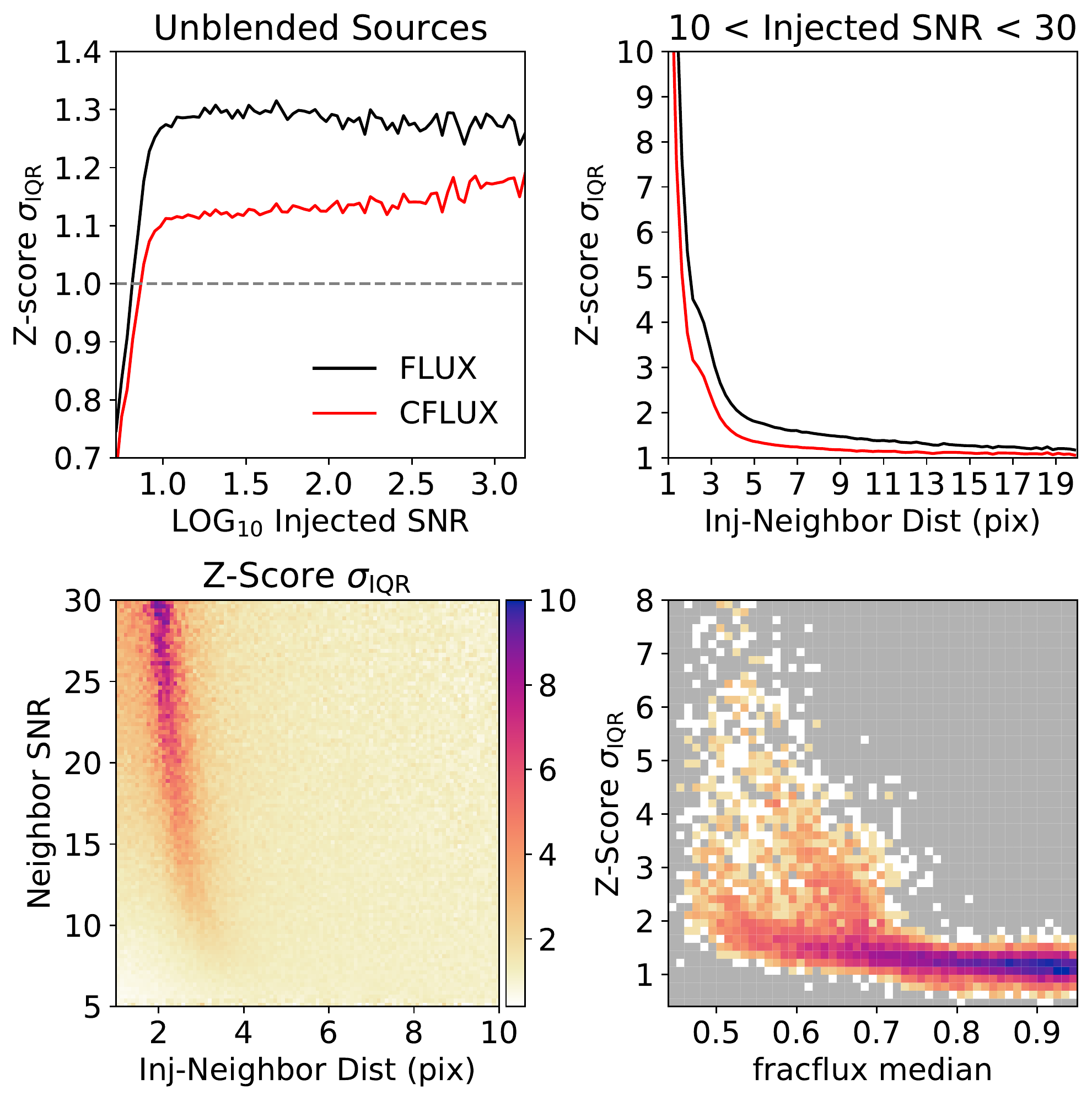}
\caption{Same as Figure \ref{fig:nb_bias} for Z-score $\sigma_{\rm IQR}$ instead of the median Z-score.
}
\label{fig:nb_iqr}
\end{figure}

\begin{figure*}[t]
\centering
\includegraphics[width=\linewidth]{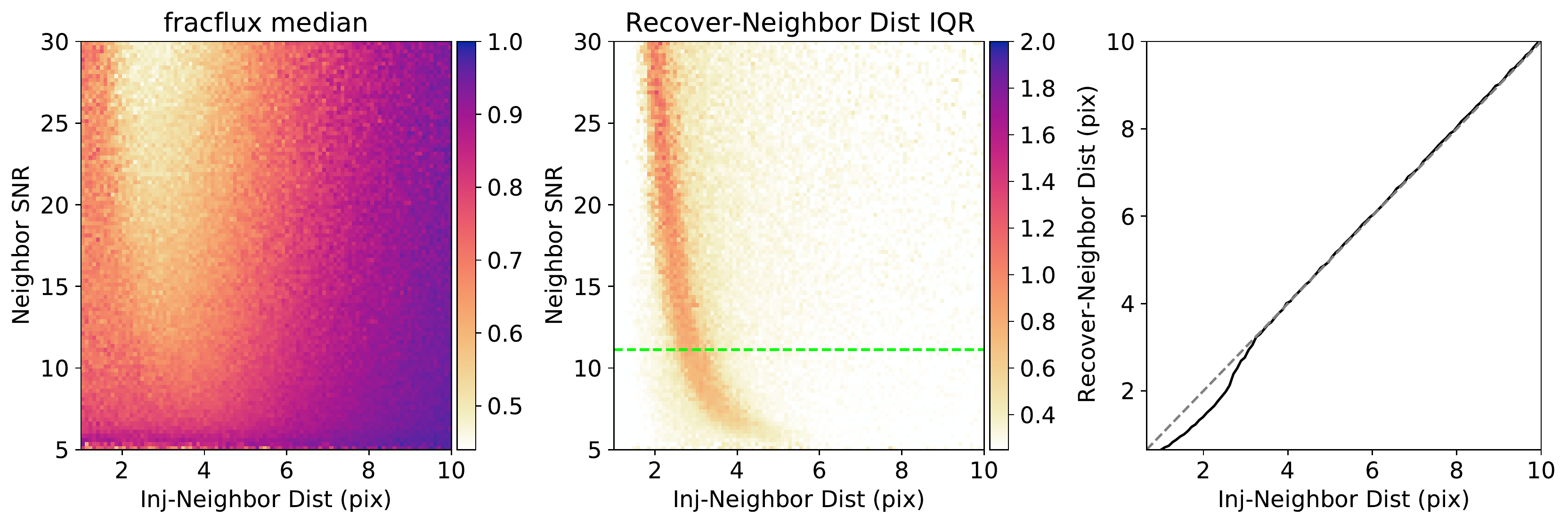}
\caption{(Left): Median \texttt{fracflux} as a function of the neighbor SNR and distance between the injected source and neighbor. (Middle): IQR of the distance between the recovered location for an injection and the nearest neighbor on the same axes. (Right): Line cut along the green-dashed line in the middle panel showing the median distance between the recovered location for an injection and the nearest neighbor versus the between the true injected location and the nearest neighbor. Perfect location recovery would fall along the 1:1 line (gray, dashed).
}
\label{fig:nb_loc}
\end{figure*}

The additional bias introduced by blending is shown in Figure \ref{fig:nb_bias}, top right, where the $z$-band sources are such that \flux is positive, the background model is positive, and the source is found within $0.25$ FWHM of the true injected location. The injected sources are further limited to moderate SNR injections, those with injected SNR between $10$ and $30$. The median Z-score diverges as the distance to a source in the original image decreases, making 1$\sigma$ errors at $\sim 3$ pixel separations and $>5\sigma$ errors at $\sim2$ pixel separations and below. At the smallest separations, we expect the injected and original source are confused and merged into a single detection. While the \cflux Z-scores diverge more slowly, \cloudcoverr makes no attempt to account for blending. 

The impact of blending also depends on the flux of the blended neighbor. Continuing to restrict to moderate SNR injections, the bottom left of Figure \ref{fig:nb_bias} shows the median Z-score as a function of both the distance to the nearest source in the original image and the SNR of that neighboring source (which is proportional to flux). If the neighbor SNR is a similar order of magnitude to the injected source, a bias is introduced at small separations that increases with increasing neighbor flux. There is a sharp onset of this bias around $2-3$ pixel separations with a trend that will be discussed below. If the neighbor SNR is very large relative to the injected source, the injected source will not be found if it is too close (further up the vertical axis, not shown). However, in the limit of the smallest separations, the injected source will be recovered regardless of the neighboring SNR, just as a very biased source which has a flux that is the sum of the two sources (far left edge of the plot).

In the bottom right of Figure \ref{fig:nb_bias} converts the bottom left panel into a 2D (logarithmic) histogram of the median Z-score bias versus the median \texttt{fracflux} for the given injected-neighbor source distance. For all \texttt{fracflux} $> 0.5$, the vast majority of Z-scores are $\leq 1$, with a trend of increasing bias toward more blended sources (lower \texttt{fracflux}). There is an onset of $>5 \sigma$ errors at $\texttt{fracflux} = 0.75$. The outliers above $\texttt{fracflux} = 0.75$ are the result of close ($<3$ pixel separation) sources with fainter SNR than the injected source ($<10$). Since the recovery fraction of such sources near the detection threshold is low, similar levels of bias from undetected sources is possible. In the bottom panels of Figure \ref{fig:nb_bias} and \ref{fig:nb_iqr}, we show results for \cfluxns. The equivalent plots for \flux are qualitatively the same, but have biases $1.5-2$ times larger.

In addition to measuring bias, injection tests can be used to measure if the reported photometric uncertainties accurately capture the errors made by the photometric pipeline. Figure \ref{fig:nb_iqr} shows this analysis in the context of blending, with cuts mirroring those in Figure \ref{fig:nb_bias}. If the error bars were exactly correct, the Z-score distribution for recovered sources would be a unit normal distribution with $\sigma_{\rm IQR} = 1$.

The $\sigma_{\rm IQR}$ of the Z-scores for unblended sources as a function of the injected source SNR is shown in the top left of Figure \ref{fig:nb_iqr}. Above SNR 10, the $\sigma_{\rm IQR}$ is approximately flat with respect to SNR. Using \flux (and its associated uncertainty), the $\sigma_{\rm IQR}$ is $\sim1.3$ times larger than the normal value, indicating that the flux uncertainties are underestimated. For \cfluxns, the uncertainties are closer to correct and lead to an $\sigma_{\rm IQR}$ only $\sim1.1$ times larger than the normal value. This is in part expected since the associated uncertainties with \cflux account for the off-diagonal correlations in the background model, in additional to the usual diagonal Poisson contributions, when computing uncertainties.\footnote{The \texttt{dflux} from \crowdsource depends both on the \crowdsource PSF model and the inverse variance weights coming from the CP. However, the \texttt{dcflux} from \cloudcoverr is only sensitive to the CP inverse variance weights via the gain estimated by \crowdsourcens. The background covariance matrix used in estimating \texttt{dcflux} is determined by the observed correlations in the local background around a given source and is thus quasi-independent of the CP inverse variance weights.}

Blending complicates the estimation of uncertainties but is not explicitly accounted for by either \flux or \cfluxns. This is illustrated in Figure \ref{fig:nb_iqr}, top right, where the Z-score $\sigma_{\rm IQR}$ diverges as the distance to a source in the original image decreases, underestimating the error bars by a factor $3-5$ at the separations on the order of a two pixels. 

The bottom left of Figure \ref{fig:nb_iqr} shows that if the neighbor SNR is small relative to the injected source, there is almost no measurable impact on the recovered Z-score $\sigma_{\rm IQR}$ of the injected source at any distance. In the limit where the sources are close together, they are not deblended and the injected source is simply recovered as having the sum of the fluxes for both stars. This merging of the sources shows up as a bias (see Figure \ref{fig:nb_bias}), but does not impact the $\sigma_{\rm IQR}$. In between the well-separated and merged limit, there is up to an order of magnitude underestimation of the uncertainty. This ``deblending'' uncertainty derives primarily from the variability in the number of sources used to model the injection and neighbor (1 or 2 sources) and the source location, as shown in Figure \ref{fig:nb_loc}. Accounting for ``deblending'' uncertainty correctly in large-scale photometric pipelines is an important remaining challenge for the field, for which tests similar to the one above can be diagnostic.

The bottom right of Figure \ref{fig:nb_iqr} converts the bottom left panel into a 2D (logarithmic) histogram of the Z-score $\sigma_{\rm IQR}$ versus the median \texttt{fracflux} for the given injected-neighbor source distance. Above $\texttt{fracflux} = 0.75$, the distribution of $\sigma_{\rm IQR}$ is tight and flat near 1. Below $\texttt{fracflux} = 0.75$, outliers with $\sigma_{\rm IQR}$ $> 3$ appear and the overall tilt toward increasing $\sigma_{\rm IQR}$ with decreasing \texttt{fracflux} becomes more pronounced. The left panel of Figure \ref{fig:nb_loc} shows the median \texttt{fracflux} as a function of the neighbor SNR and injection-neighbor distance. This reaffirms that sources past the peak in Z-score $\sigma_{\rm IQR}$ are often merging with the neighboring source, leading to an increase in the \texttt{fracflux} with decreasing separation. While the lowest \texttt{fracflux} for a given neighbor SNR does coincide with the peak in Z-score $\sigma_{\rm IQR}$, the shape of the change in \texttt{fracflux} is much broader.

The middle panel of Figure \ref{fig:nb_loc} shows the variability (IQR) in the distance between the location at which the injected source is recovered and the location of the neighbor in the original image. In the ideal case, this variability would be zero with the injected source location being recovered at the true location every time. However, there is a sharp peak in the variability of the distance between the recovered location and the neighbor, which matches the peak in Z-score $\sigma_{\rm IQR}$. 

\begin{figure}[b]
\centering
\includegraphics[width=\linewidth]{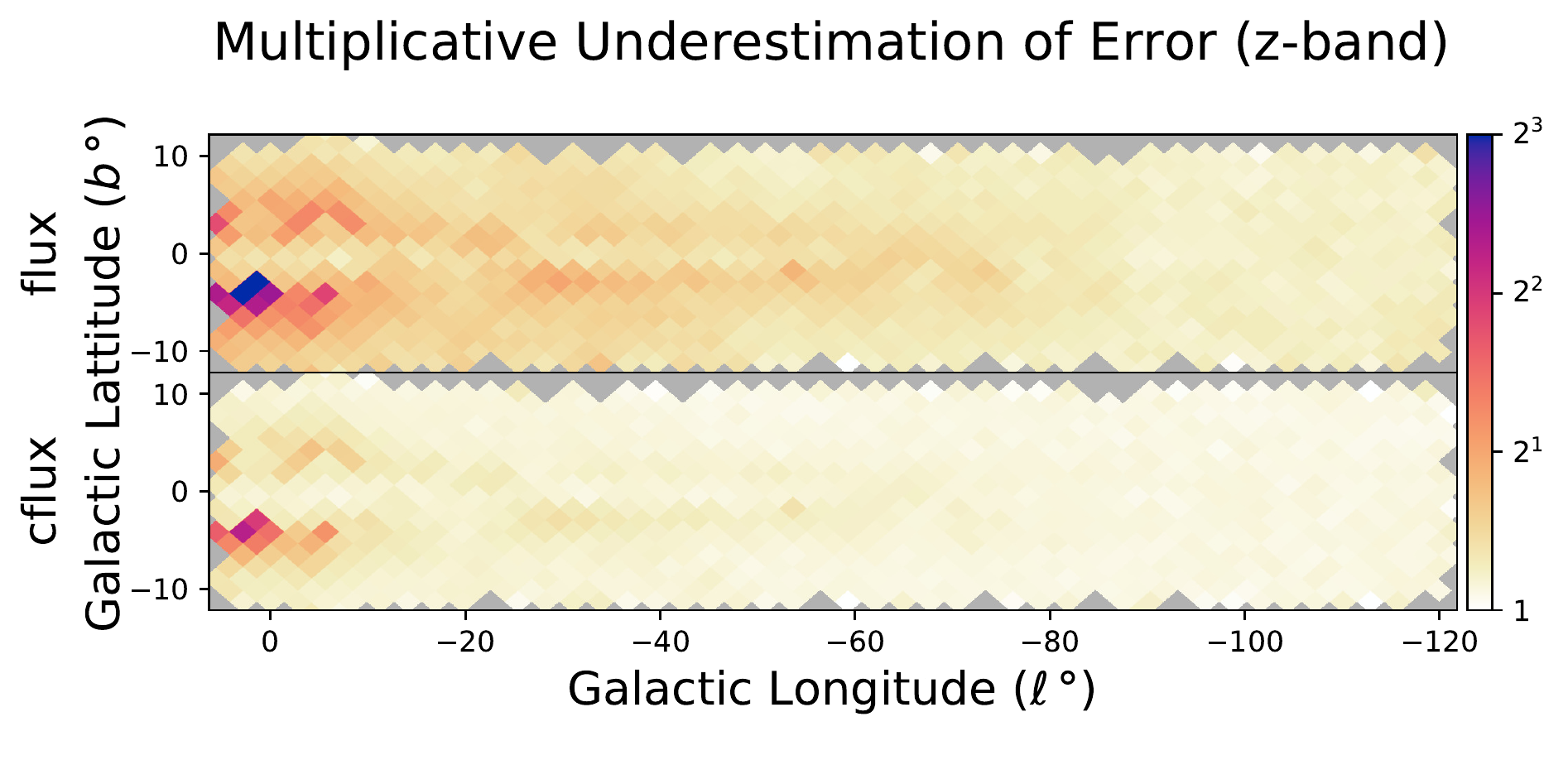}
\caption{Color map of the multiplicative underestimation of error as measured by $\sigma_{\rm IQR}$ for $z$-band \flux and \cflux over the survey footprint. The color scale is $\log_2$ to emphasize the large range of values while maintaining interpretability. Top and bottom compare results for \flux and \cfluxns, respectively.}
\label{fig:IQR_vary}
\end{figure}

\begin{figure*}[t]
\centering
\includegraphics[width=\linewidth]{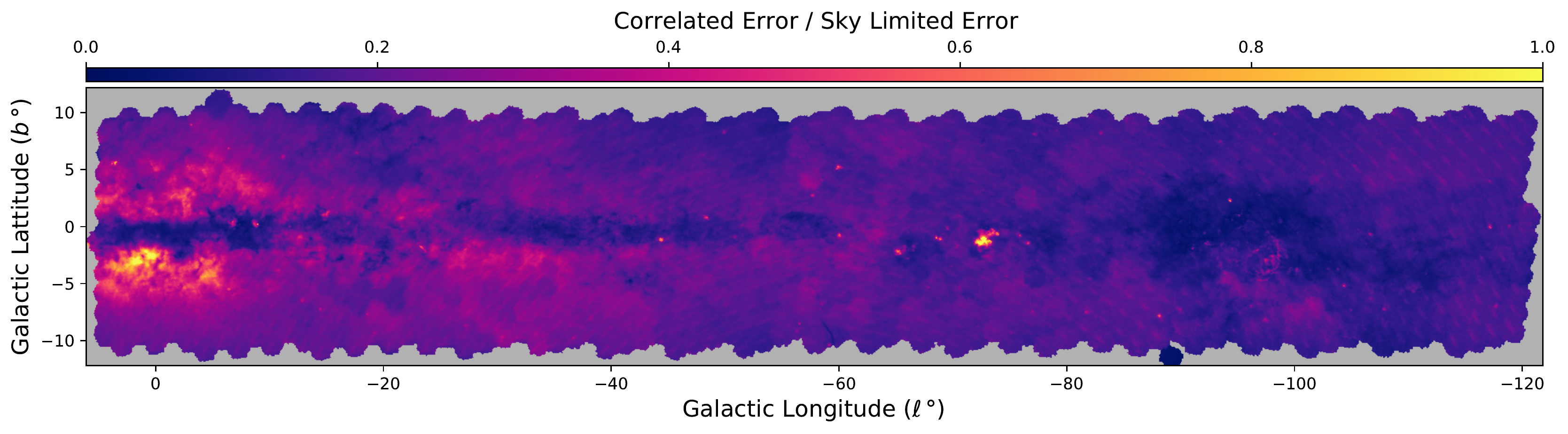}
\caption{The ratio of the additional uncertainty \cloudcoverr introduces as the result of correlations to diagonal Poisson uncertainty from the background sky counts (Equation \ref{eq:correrr}), shown over the survey footprint. Each HEALPix pixel shows an outlier-clipped mean with a further requirement that sources have the \textsc{CloudCovErr.jl} flag \texttt{dnt} == 0.}
\label{fig:corr_err}
\end{figure*}

To further clarify the nature of this peak, the right panel of Figure \ref{fig:nb_loc} shows the median distance between the recovered location and the neighbor versus the distance between the true injection location and the neighbor, for fixed neighbor SNR (slice along green-dashed line). In the ideal case, these two distances are the same and the data would follow the 1:1 line (diagonal dashed gray) as they do for all separations above $3.5$ pixels. At the onset of the peak in distance and Z-score $\sigma_{\rm IQR}$, the distance between the recovered location and the original source shows a large decrease below the true value, which persists at lower separations.\footnote{This shift in location could occur because only the brighter of the two sources is found, the two sources are modeled as a single source in between the true location of both sources, or both sources are found with the location of the injected source biased toward the location of the original source. We suspect the second case dominates, but an analysis on synthetic images with only two sources would better characterize the failure mode.} Thus, the uncertainty estimates for the source flux fail most catastrophically when the source location is being pulled toward its neighbor and the pipeline finds the number of sources ambiguous.

A tilt in the peak in the distance IQR illustrates the intuition that bright sources must be closer together before it is difficult to determine if there is one source or two sources close together. Below an SNR of $10$, this tilt gains curvature, suggesting that faint sources near the detection threshold are absorbed into the injected source at much larger separations. However, there are some selection effects at play in the sample in Figures \ref{fig:nb_bias},  \ref{fig:nb_iqr}, and \ref{fig:nb_loc}. The figures are not sensitive to the case where the injected source is not detected within $0.25$ FWHM from the injected location, because its flux is either totally absorbed by the original source (without shifting the original source location significantly) or its location is shifted by more than $0.25$ FWHM by the presence of the original source.

\subsection{Validating Uncertainties} \label{sec:iqr}

To avoid these complications associated with blending, but still make a map of the spatial variation (available in Section \ref{sec:dataavil}) in the quality of our flux uncertainties, we again limit to unblended injected sources and further limit to $\log_{10}$ injected source SNR between $1.1$ and $3.2$, where Figure \ref{fig:nb_iqr} (top left) is approximately invariant to the injected source SNR (Figure \ref{fig:IQR_vary}). For \fluxns, the median multiplicative underestimation of error over the survey footprint is $1.3$, and tracks source density as illustrated by the large $\sigma_{\rm IQR}$ in the Galactic bulge and just off the Galactic plane. For \cfluxns, the median multiplicative underestimation of error is only $1.1$ and improves the $\sigma_{\rm IQR}$ in bulge by factor of two. This comparison illustrates that while \cflux does deliver the intended improvements to background modeling, neither \flux nor \cflux handle blended-source uncertainties well. This should be a focus of future development for pipelines in crowded fields. For now, the map in Figure \ref{fig:IQR_vary} can act as a suggested multiplicative correction to the reported uncertainties in DECaPS2.

\section{Astrophysical Tests} \label{sec:astro}

Another useful set of tests is to evaluate the performance of our photometric reduction on astrophysical targets where we have prior knowledge that informs an expected result.

\subsection{Nebulous Uncertainties} \label{sec:neberr}

The main purpose of the \cloudcoverr processing is to improve the uncertainty estimates in regions with structured backgrounds by including off-diagonal terms in the covariance matrix. Thus, we want to confirm that \cloudcoverr is increasing the uncertainties predominately in regions of the Galaxy with known nebulosity. Figure \ref{fig:corr_err} shows the ratio of the additional uncertainty that \cloudcoverr introduces as the result of correlations to diagonal Poisson uncertainty from the background sky counts. 
\begin{ceqn}
\begin{align} \label{eq:correrr}
\frac{\rm{Correlated \, Error}}{\rm{Sky \, Limited \, Error}} = \frac{\sqrt{\texttt{dcflux}^2-\texttt{dflux}^2}}{\sqrt{4\pi\sigma^2 \times b \times g}}
\end{align}
\end{ceqn}
We report an outlier-clipped mean value of the ratio for stars within each HEALPix pixel. The outliers are values beyond 10*$\sigma_{\rm IQR}$ and are computed per CCD. In addition, we require the \textsc{CloudCovErr.jl} flag \texttt{dnt} = 0.

This measure provides a relative scale for the magnitude of these correlations and is robust to variations in the sky brightness, seeing, and the distribution of source fluxes. However, the qualitative features are similar to those observed by simply taking a ratio of the \cflux and \flux uncertainties (\texttt{dcflux} and \texttt{dflux}), which was shown in Figure 13 of \cite{Saydjari:2022:arXiv:}.

Overall, the correlated uncertainties track source density, are $\sim 0.5$ times the sky-limited flux uncertainty just off the plane $\ell < -40\degree$, and can be larger than the sky uncertainty in the (Southern) Galactic bulge. The peak associated with the Carina Nebula ($b = -0.8\degree, \ell = -73\degree$) and filaments associated with the Vela supernova remnant ($b = -2.8\degree, \ell = -96\degree$) show that the correlated uncertainties are sensitive to structured backgrounds as intended. Other point sources correspond to known nebulae, globular clusters, or artifacts in the background near very bright stars. Thus, the behavior of \texttt{dcflux} on the structures present in the survey footprint agrees with our prior knowledge.

\begin{figure}[h]
\centering
\includegraphics[width=\linewidth]{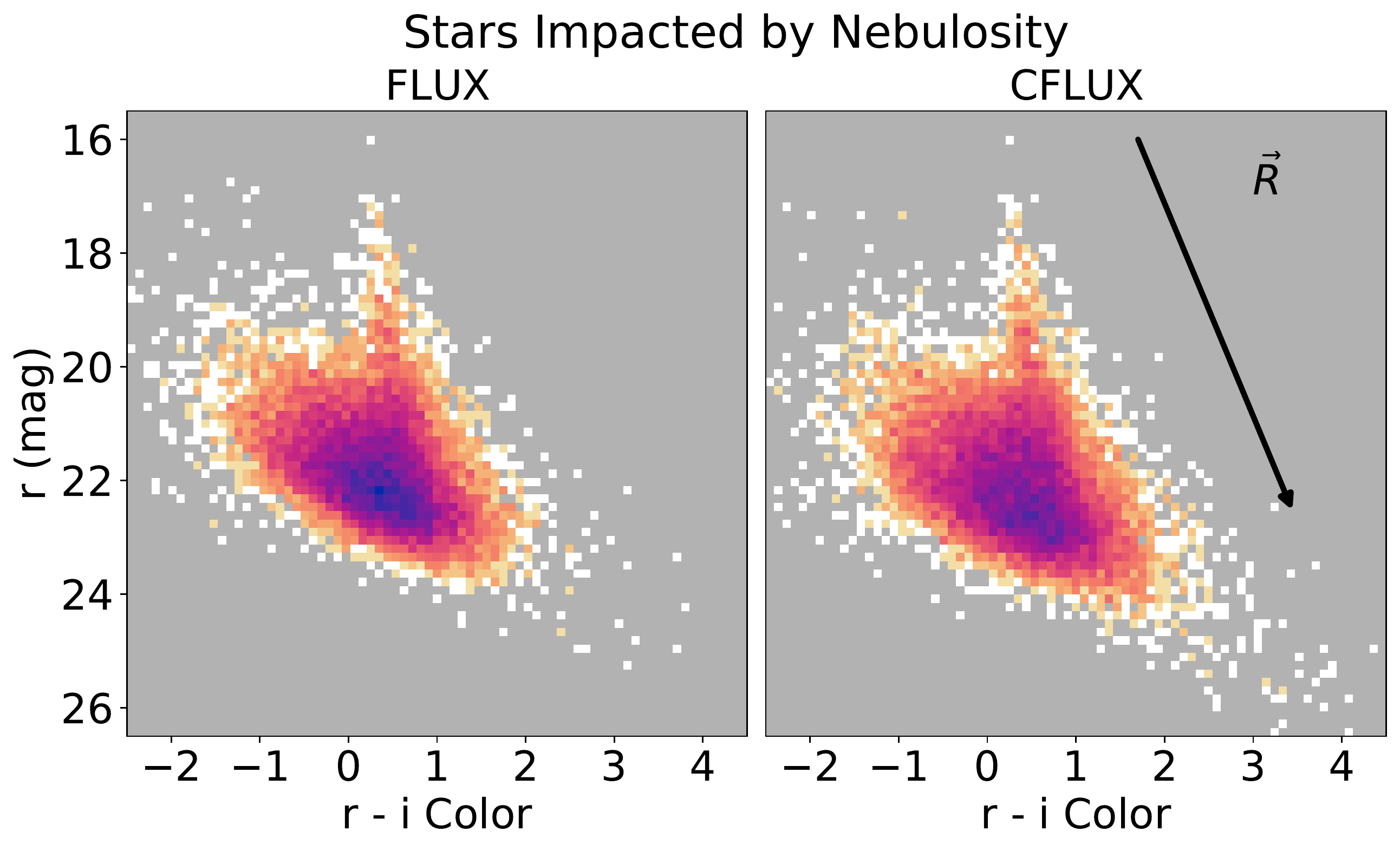}
\caption{CMD in $r$ versus $r-i$ using \flux (left) and \cflux (right) for selected sources near the Vela supernova remnant impacted by nebulosity (see text for exact cuts). Color scale shows log density (light, white, low; dark, blue, high) on a common scale. Reddening vector (right) showing effect of dust plotted for reference. \cflux shifts stars with overestimated $r$-band flux toward a simple reddened main sequence.}
\label{fig:neb_cmd}
\end{figure}

\subsection{Nebulous CMD} \label{sec:nebCMD}

To test performance of the photometry in the presence of nebulosity, we target a radius $4\degree$ region ($264.6\degree$, $-4.8\degree$) near the Vela supernova remnant which has significant filamentary H$\alpha$ emission (appearing in $r$-band). We further limit to the stars most impacted by the sparse filaments by requiring in both $r$ and $i$-bands that:
\begin{itemize}
\setlength\itemsep{-0.5em}
\item the off-diagonal contributions to the uncertainty be large, \texttt{dcflux}-\texttt{dflux} $>10^{-10.5}$ Mgy
\item the nebulosity correction to flux be large (and negative)\footnote{This asymmetric choice leads to a purer selection of stars on the edges of filaments because of the on-average positive correction of \flux by \cloudcoverr (see Figure \ref{fig:cflux_flux_cal} and \ref{fig:low_SNR})}, \flux - \cflux $>10^{-10.5}$ Mgy
\item no \texttt{dnt} flags are thrown for any detections
\end{itemize}

We show the CMD in $r$ versus $r-i$ of this sample in Figure \ref{fig:neb_cmd} using both \flux (left) and \cflux (right) for comparison. Given our knowledge of stellar evolution, we have an expectation that the CMD (for stars in this region of the disk) is dominated by the main sequence (straight vertical), with a tilt as a result of reddening from dust. The \cflux CMD more closely resembles that expectation. The \cflux CMD shifts weight from the blueward tilted population at fainter magnitudes, caused by overestimation of the $r$-band flux, toward the main sequence and along the reddening vector. The sample size here is limited by the density of sources on the edge of a filament in an evolutionarily simple region, which is why repeated injection tests in filamentary regions are valuable (see Figure 10, \citealt{Saydjari:2022:arXiv:}).

\begin{figure}[t]
\centering
\includegraphics[width=\linewidth]{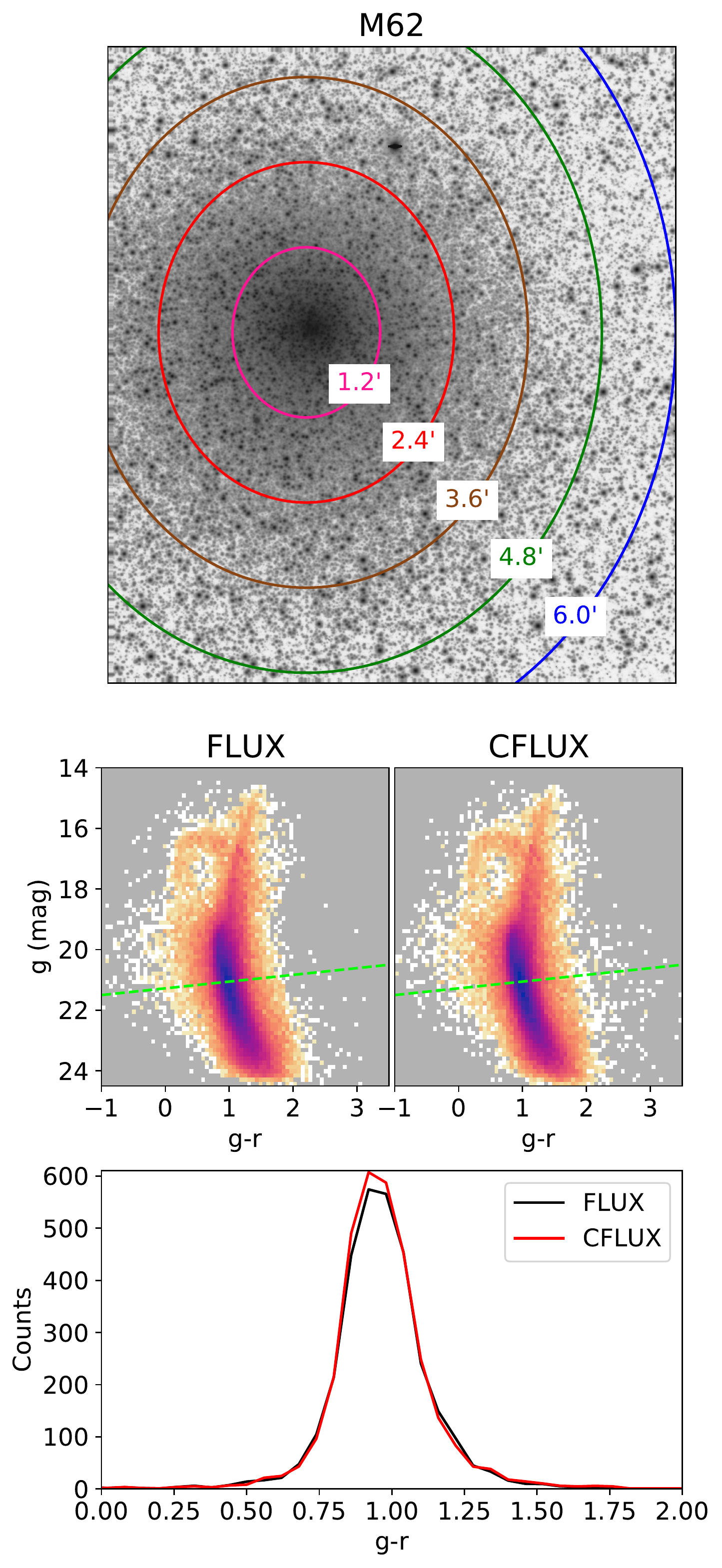}
\caption{(Top) Single-exposure $g$-band image of M62 with labeled radii in steps of $1.2'$. Color scale is $\arcsinh$ stretched around the median with high and low counts in black and white, respectively. (Middle) CMD in $g$ versus $g-r$ for all sources within $6'$ of the center of M62 using \flux (left) and \cflux (right). Color scale shows log density (light, white, low; dark, blue, high) on a common scale. (Bottom) Line cut along the green-dashed line in middle panels showing the sharper main sequence given by \cfluxns.}
\label{fig:M62_cflux}
\end{figure}

\subsection{Globular Cluster} \label{sec:globular}

Globular clusters, which are formed from only one (or a few) star-forming episodes, are useful reference objects because they act as snapshots at fixed stellar age. In addition, they also provide a gradient of source densities, with some of the densest, hardest to deblend regions at their centers.

\begin{figure*}[t]
\centering
\includegraphics[width=\linewidth]{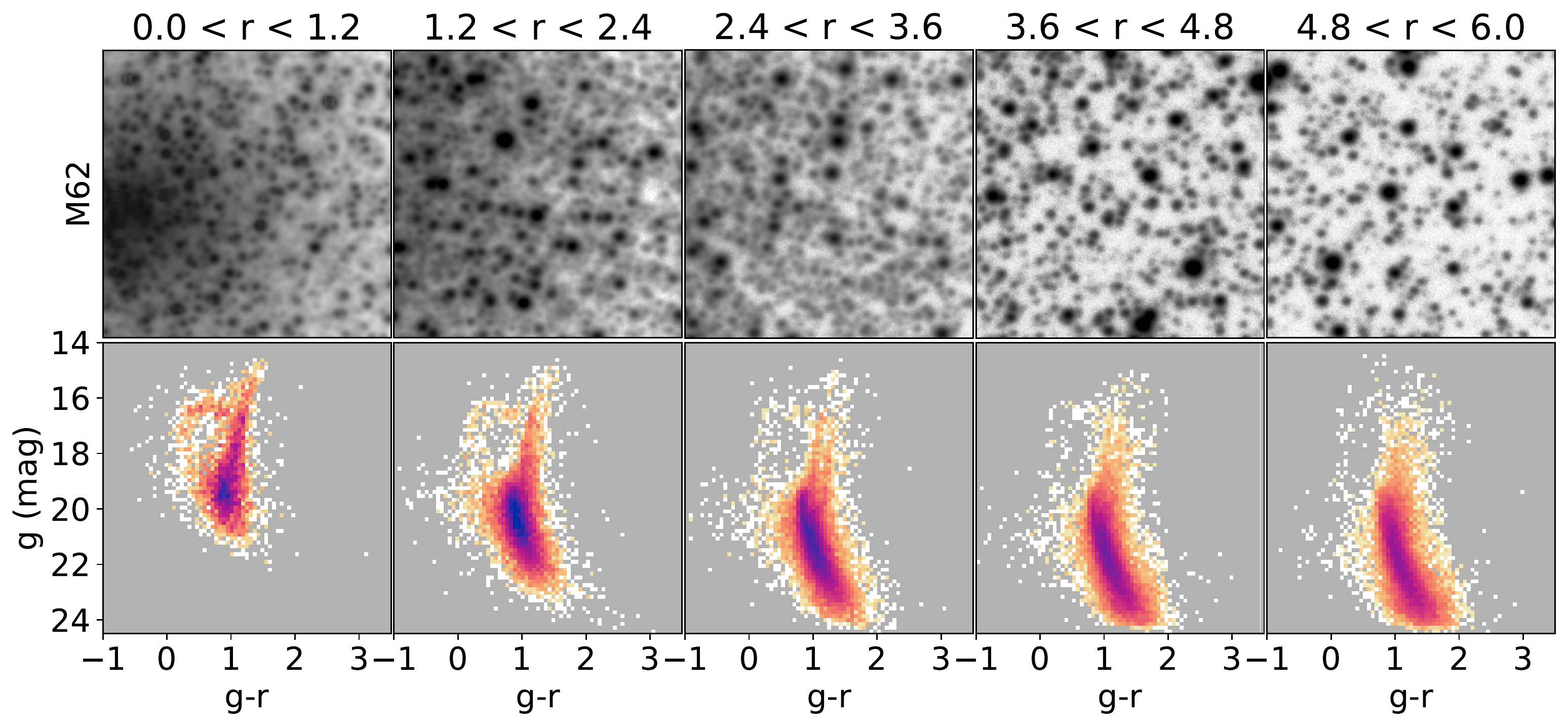}
\caption{(Top) Image ($g$-band) stamps cut from annuli around M62 between the indicated radii. Each image is $\arcsinh$ stretched around the median with its own color scale (black high and white low counts) to emphasize the increasing degree of blending toward the center of the globular cluster. (Bottom) CMDs in $g$ versus $g-r$ for sources found in the indicated annuli. Color scale shows log density (light, white, low; dark, blue, high) on a common scale in units of sources per square-degree.}
\label{fig:M62_rad}
\end{figure*}

We consider M62, the brightest globular cluster (6.45 V mag) in the DECaPS2 survey footprint, as an example (see Figure \ref{fig:M62_cflux}, top). We show the CMD using both \flux (middle left) and \cflux (middle right) in $g$ versus $g-r$ for all sources within $6'$, which is about 60\% of the tidal radius. The CMD using \cflux is slightly sharper around the main sequence track which agrees with our expectation that there should be little to no metallicity scatter. Line cuts transverse to the main sequence are shown in the bottom panel for emphasis. Importantly, the bias-variance trade-off inherent in any bias correction such as that used for \cflux does not add so much scatter as to obscure finer features visible for the globular cluster, such as the blue horizontal branch (curved feature near 16$^{\rm{th}}$ magnitude).

Globular clusters also act as a test bed for the performance of deblending algorithms. We subdivide M62 by radius in steps of $1.2'$, as indicated in Figure \ref{fig:M62_cflux}. In Figure \ref{fig:M62_rad}, we show the CMD in $g$ versus $g-r$ for sources in each annulus and an example image cutout to demonstrate the degree of blending in that field. In the largest annulus, we recover sources brighter than $24^{\rm{th}}$ mag. Closest to the center, the crowding is so significant that only sources brighter than $\sim21^{\rm{st}}$ are recovered. In all cases, the recovered CMDs appear to be differently weighted samplings of the same underlying CMD.\footnote{An apparent split in the red-giant branch observed in the third panel of Figure \ref{fig:M62_rad} (bottom) results from a dust lane passing over half of the annulus.} This is the desired behavior for when a deblending algorithm encounters a field more blended than it can deconstruct; the detection limit moves to a brighter level, but the sources identified are still of science quality.


\section{Data/Code Availability} \label{sec:dataavil}

Data products associated with DECaPS2 are publicly available on the survey website \href{http://decaps.skymaps.info/}{http://decaps.skymaps.info/}. We provide a \href{http://decaps.skymaps.info/release/datamodel/index.html}{data model} (inspired by SDSS) that documents the contents and structures of all released files. We provide single-epoch catalog outputs from both \crowdsource and \cloudcoverrns, the final object-merged catalog as subdivided FITS files for flexible database ingestion, and files that specify the zeropoints and photometric cuts from the calibration. We build and store the merged catalog using the \href{http://research.majuric.org/public/project/lsd/}{Large Survey Database} and make those tables available.

Raw images in addition to the \texttt{InstCal} processed images used here are available from the \href{https://astroarchive.noirlab.edu/portal/search/}{NOIRLab Archive}. Since there are multiple versions in the NOIRLab Archive for a subset of the images which required reprocessing by the DECam Community Pipeline, we also provide the exact list of \texttt{InstCal} processed images that we used to create DECaPS2 for the sake of reproducibility.

Work on the DECam Community Pipeline is led by F. Valdes and source code is available by request, but documentation and support for use is limited. \crowdsource is publicly available on \href{https://github.com/schlafly/crowdsource}{Github} and is distributed via \href{https://anaconda.org/conda-forge/crowdsourcephoto}{condaforge}. \cloudcoverr is publicly available on \href{https://github.com/andrew-saydjari/CloudCovErr.jl}{Github} and is distributed through the Julia package \href{https://github.com/JuliaRegistries/General/tree/master/C/CloudCovErr}{registry}. A Jupyter notebook with the code and data compilations necessary to reproduce all figures in the text are available on Zenodo at doi:\href{https://doi.org/10.5281/zenodo.6677609}{10.5281/zenodo.6677609}.

\section{Conclusion} \label{sec:conc}

With DECaPS2, we provide deep optical-NIR ($g \sim 24$th to $Y \sim 21$st mag) coverage of the southern Galactic plane (6.5\% of the sky) with approximately 1" seeing. Combined with PS1, this completes comparable imaging of the entire Galactic plane essential for probing our Galaxy's stars, gas, and dust. We use \crowdsource to process images and create a catalog with 3.3 billion objects built from 34 billion detections. The relative photometric calibration is uniform and has a precision of $\sim 7$ mmag, with an absolute calibration tied to PS1.

A detailed analysis of detection quality metrics and \crowdsource failure modes led to improved quality cuts, which were applied in creating the object catalog. Given that \texttt{fracflux} is a common quality cut used on the object catalog, we provide guidance for a threshold on \texttt{fracflux} (which is extremely conservative). We compare the DECaPS2 astrometry to Gaia and DECaPS2 photometry to both PS1 and DECaPS1, showing strong agreement and internal consistency. A similar comparison validates our new background-corrected flux, \cfluxns, produced by \cloudcoverrns. The performance of the convolutional neural network (CNN) that identifies nebulosity and modifies the deblending in \crowdsource was validated by correct identification of Galactic nebulosity over the survey footprint.

We performed synthetic injection tests over the entire survey footprint and used them to measure and model the biases for low SNR sources. This bias was leveraged to empirically measure the photometric depth of the survey. Through injection tests, we also quantify the deblending performance of \crowdsource and find uncertainties to be underestimated when the number of sources is ambiguous. We make a map of the multiplicative underestimation of error bars across the survey footprint, which can be used in downstream statistical analyses and which illustrates the improvement of \cflux over \fluxns. Further, we use astrophysical priors on nebulous regions of the Galaxy and globular clusters to further validate the deblending performance of \crowdsource and the correction introduced by \cloudcoverrns.

All images, single-visit catalogs and models, and combined source catalogs are publicly available and well-documented by a data model. These well-validated data products with many quality flags (even beyond those explored in detail here) should provide a rich, adaptable resource for the community, facilitating a variety of studies of the Milky Way galaxy. We look forward to using DECaPS2 to refine dust reddening models and as a photometric input for the statistical inference of the 3D distribution of dust in the Milky Way, probing further into the Galactic plane.

\acknowledgments
A.K.S. gratefully acknowledges support by a National Science Foundation Graduate Research Fellowship (DGE-1745303). D.P.F. acknowledges support by NSF grant AST-1614941, “Exploring the Galaxy: 3-Dimensional Structure and Stellar Streams.” D.P.F. acknowledges support by NASA ADAP grant 80NSSC21K0634 “Knitting Together the Milky Way: An Integrated Model of the Galaxy’s Stars, Gas, and Dust.”

This work was supported by the National Science Foundation under Cooperative Agreement PHY-2019786 (The NSF AI Institute for Artificial Intelligence and Fundamental Interactions). We acknowledge helpful discussions with Paul Edmon, Joel Brownstein, Charlie Conroy, Tom Prince, Christian I. Johnson, Eugene Magnier, Lucas Janson, Justina R. Yang, and Nayantara Mudur. A.K.S. acknowledges Sophia S\'{a}nchez-Maes for helpful discussions and much support.

The photometric cataloging codes \crowdsource and \cloudcoverr were run on the FASRC Cannon cluster supported by the FAS Division of Science Research Computing Group at Harvard University. Computations for the visualization of DECaPS2 images by the Legacy Survey Viewer were performed in part at the National Energy Research Scientific Computing Center, a DOE Office of Science User Facility supported by the Office of Science of the U.S. Department of Energy under Contract No. DE-AC02-05CH11231.

DECaPS2 is based on observations at Cerro Tololo Inter-American Observatory, National Optical Astronomy Observatory, which is operated by the Association of Universities for Research in Astronomy (AURA) under a cooperative agreement with the National Science Foundation. Observations were obtained across six proposals for the same program ``Mapping Dust in 3D with DECam: A Galactic Plane Survey'', PI: Finkbeiner, NOAO Prop. ID: 2014A-0429 (3 nights), 2016A-0327 (8 nights), 2016B-0279 (6 nights),	2018A-0251 (9 nights), 2018B-0271 (8 nights), 2019A-0265 (8 nights).

This project used data obtained with the Dark Energy Camera (DECam), which was constructed by the Dark Energy Survey (DES) collaboration. Funding for the DES Projects has been provided by the U.S. Department of Energy, the U.S. National Science Foundation, the Ministry of Science and Education of Spain, the Science and Technology Facilities Council of the United Kingdom, the Higher Education Funding Council for England, the National Center for Supercomputing Applications at the University of Illinois at Urbana-Champaign, the Kavli Institute of Cosmological Physics at the University of Chicago, the Center for Cosmology and Astro-Particle Physics at the Ohio State University, the Mitchell Institute for Fundamental Physics and Astronomy at Texas A\&M University, Financiadora de Estudos e Projetos, Fundação Carlos Chagas Filho de Amparo à Pesquisa do Estado do Rio de Janeiro, Conselho Nacional de Desenvolvimento Científico e Tecnológico and the Ministério da Ciência, Tecnologia e Inovacão, the Deutsche Forschungsgemeinschaft, and the Collaborating Institutions in the Dark Energy Survey. The Collaborating Institutions are Argonne National Laboratory, the University of California at Santa Cruz, the University of Cambridge, Centro de Investigaciones Enérgeticas, Medioambientales y Tecnológicas-Madrid, the University of Chicago, University College London, the DES-Brazil Consortium, the University of Edinburgh, the Eidgenössische Technische Hochschule (ETH) Zürich, Fermi National Accelerator Laboratory, the University of Illinois at Urbana-Champaign, the Institut de Ciències de l’Espai (IEEC/CSIC), the Institut de Física d’Altes Energies, Lawrence Berkeley National Laboratory, the Ludwig-Maximilians Universität München and the associated Excellence Cluster Universe, the University of Michigan, the NOIRLab (formerly known as the National Optical Astronomy Observatory), the University of Nottingham, the Ohio State University, the University of Pennsylvania, the University of Portsmouth, SLAC National Accelerator Laboratory, Stanford University, the University of Sussex, and Texas A\&M University.

The DECam Community Pipeline is a service of the Community Science Data Center of NSF's National Optical/Infrared Research Laboratory.

The Pan-STARRS1 Surveys (PS1) and the PS1 public science archive have been made possible through contributions by the Institute for Astronomy, the University of Hawaii, the Pan-STARRS Project Office, the Max-Planck Society and its participating institutes, the Max Planck Institute for Astronomy, Heidelberg and the Max Planck Institute for Extraterrestrial Physics, Garching, The Johns Hopkins University, Durham University, the University of Edinburgh, the Queen's University Belfast, the Harvard-Smithsonian Center for Astrophysics, the Las Cumbres Observatory Global Telescope Network Incorporated, the National Central University of Taiwan, the Space Telescope Science Institute, the National Aeronautics and Space Administration under Grant No. NNX08AR22G issued through the Planetary Science Division of the NASA Science Mission Directorate, the National Science Foundation Grant No. AST-1238877, the University of Maryland, Eotvos Lorand University (ELTE), the Los Alamos National Laboratory, and the Gordon and Betty Moore Foundation.

\facility{Blanco (DECam)}

\software{
GNU Parallel \citep{Tange:2020:zndo:},
\textsc{astropy} \citep{AstropyCollaboration:2013:A&A:},
\textsc{ipython} \citep{Perez:2007:CSE:}, 
\textsc{matplotlib} \citep{Hunter:2007:CSE:}, 
\textsc{numpy} \citep{vanderWalt:2011:CSE:}.
}

\onecolumngrid
\newpage
\appendix

\begin{appendices}


\section{PSF FOV Variation} \label{sec:fov_fwhm}

\begin{figure}[h]
\centering
\includegraphics[width=\linewidth]{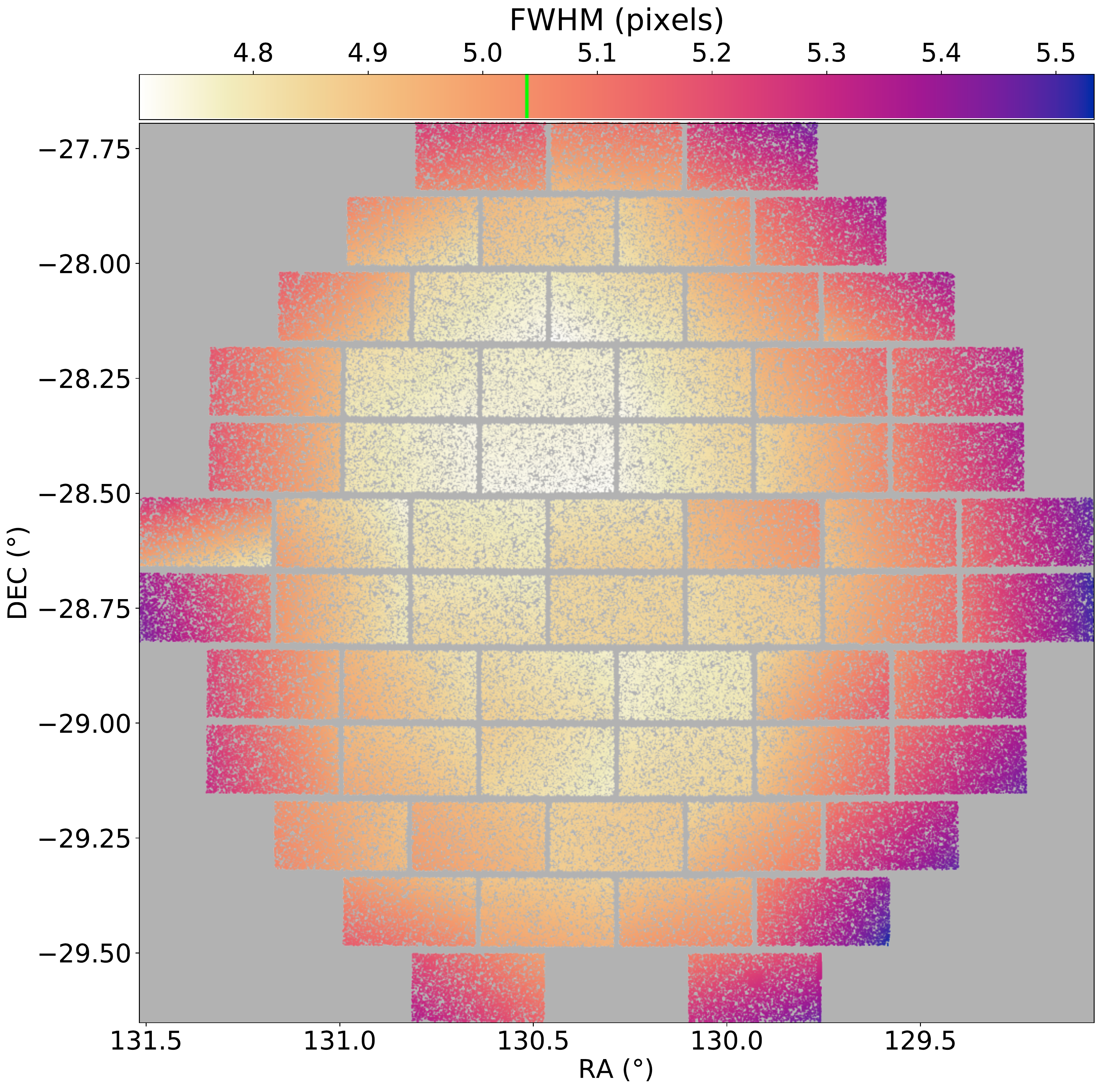}
\caption{The FWHM for sources from \crowdsource as a function of position on the focal plane of a typical $g$-band exposure. The median $g$-band FWHM reported for the whole survey is indicated on the color bar by a green line. The sky orientation shown for the focal plane is north up, west right.
}
\label{fig:fov_fwhm}
\end{figure}

DECam, as with any optical system, has a variable pixel scale across the wide field of view resulting in a variation of the PSF. A variable pixel scale would result in a photometric variation if not for the flat fielding corrections. In Figure \ref{fig:fov_fwhm} we show the FWHM in pixels for sources from \crowdsource as a function of position on the focal plane of a typical $g$-band exposure. There is a clear, asymmetric increase in the FWHM toward the edge of the field of view, larger toward the western edge (right side of plot). 

We have not isolated that source of this variation to either an instrumental effect or a stage in the CP-\crowdsource pipeline and the origin of this variation remains an open question. However, the pattern of PSF FWHM variation shown in Figure \ref{fig:fov_fwhm} appears related to the pattern noise in the average \texttt{fracflux} (Figure \ref{fig:mean_rchi2_fracflux}).

\section{CCD Saturation Thresholds} \label{sec:satdet}

At some point between Run 12 and 13 (see Table \ref{tab:obsrun}) the (software) saturation thresholds in the CP changed. This change left pixels at or near saturation unmasked for several CCDs. Since the 200 brightest ``unsaturated'' stars, stars without pixels flagged by the CP as saturated, are used for PSF fitting in \crowdsourcens, the PSF \crowdsource learned and used to measure the photometry of the affected CCDs was distorted. This unfortunately impacted all of the photometry for a given CCD, not just the measured fluxes of the stars near saturation. In most cases, making a logarithmic histogram of the per pixel counts revealed a ``pile-up'', the beginning of which could be used to define the onset of saturation. However, a few cases required even lower thresholds.

\begin{figure}[h]
\centering
\includegraphics[width=\linewidth]{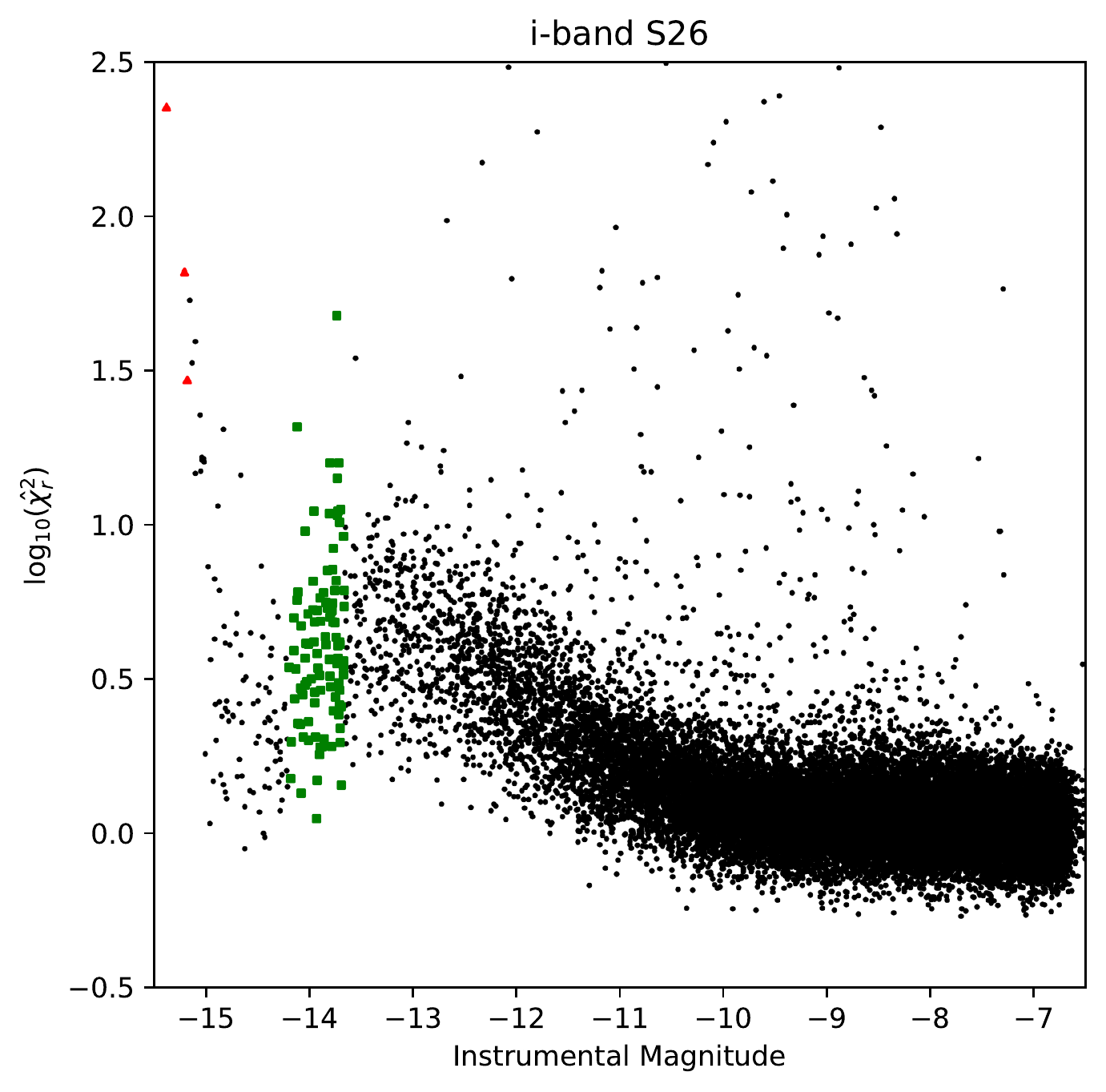}
\caption{Scatter plot of sources not flagged by the CP as saturated when the saturation threshold was set too low. The plot shows the $\hat{\chi}^2_r$ as a function of the estimated source flux (in instrumental magnitudes). There is a sharp linear increase in the $\log_{10}(\hat{\chi}^2_r)$ at the brightest magnitudes ($<-15$), indicative of sources at/near saturation being poorly modeled by the average PSF. The three brightest unflagged sources are shown as red triangles and a set of 100 bright, but unsaturated reference sources are shown as green squares. This example is for an $i$-band image taken by CCD S26.
}
\label{fig:sat_det}
\end{figure}

When the saturation threshold for a given CCD is set too low, the $\hat{\chi}^2_r$ from \crowdsource of the few brightest stars will be anomalously high because while the stars are unmasked, they are actually at/near saturation and thus distorted relative to the true PSF. We then used the ratio of the average $\hat{\chi}^2_r$ for the few brightest stars and the average $\hat{\chi}^2_r$ of bright stars far from saturation as an indicator for CCDs needing a saturation threshold adjustment (Figure \ref{fig:sat_det}). Only the saturation threshold for CCDs that were consistently flagged across multiple exposures and bands were adjusted (list in main text).

\section{Computational Details} \label{sec:compcost}

The overwhelming majority of computational time was spent using \crowdsource and \cloudcoverr in processing individual exposures. These runs were executed on the FASRC Cannon cluster at Harvard University on compute nodes with water-cooled Intel 24-core Platinum 8268 Cascade Lake CPUs with 192GB RAM and dual AVX-512 fused multiply-add (FMA) units running 64-bit CentOS 7. We used GNU Parallel \citep{Tange:2020:zndo:} to spawn, manage restarts, and track the computational time of processing the 21,430 exposures (each containing 60-61 CCDs of $2046 \times 4094$ pixels) across many multi-cpu SLURM batch allocations.

\begin{figure}[t]
\centering
\includegraphics[width=\linewidth]{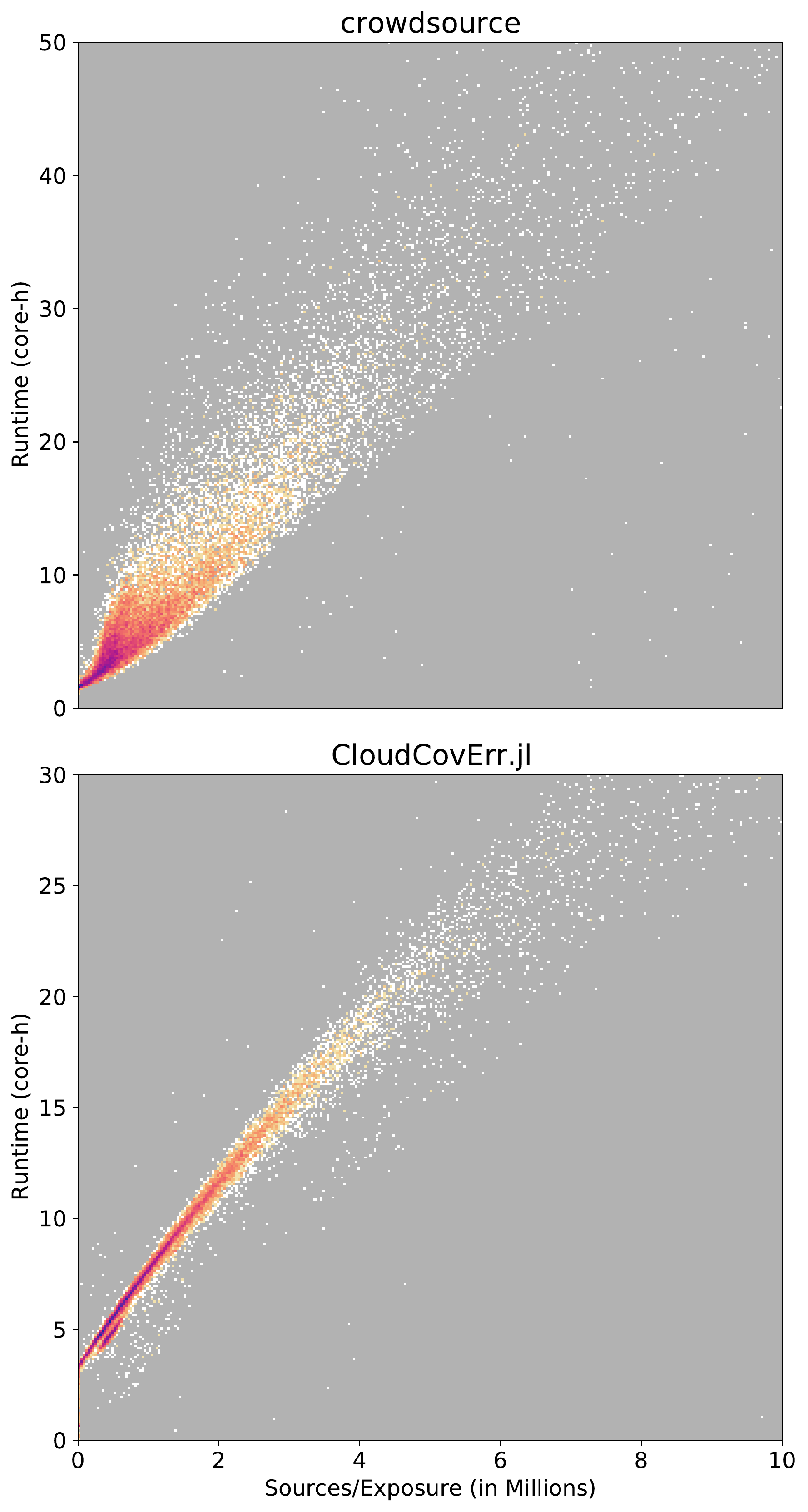}
\caption{Histogram showing the runtime of an exposure versus the number of sources found in that exposure for \crowdsource (top) and \cloudcoverr (bottom) processing of DECaPS2. The color scale is in log-density, and each panel has its own normalization (light, white, low; dark, blue, high density). 
}
\label{fig:runtime}
\end{figure}

\begin{figure*}[t]
\centering
\includegraphics[width=\linewidth]{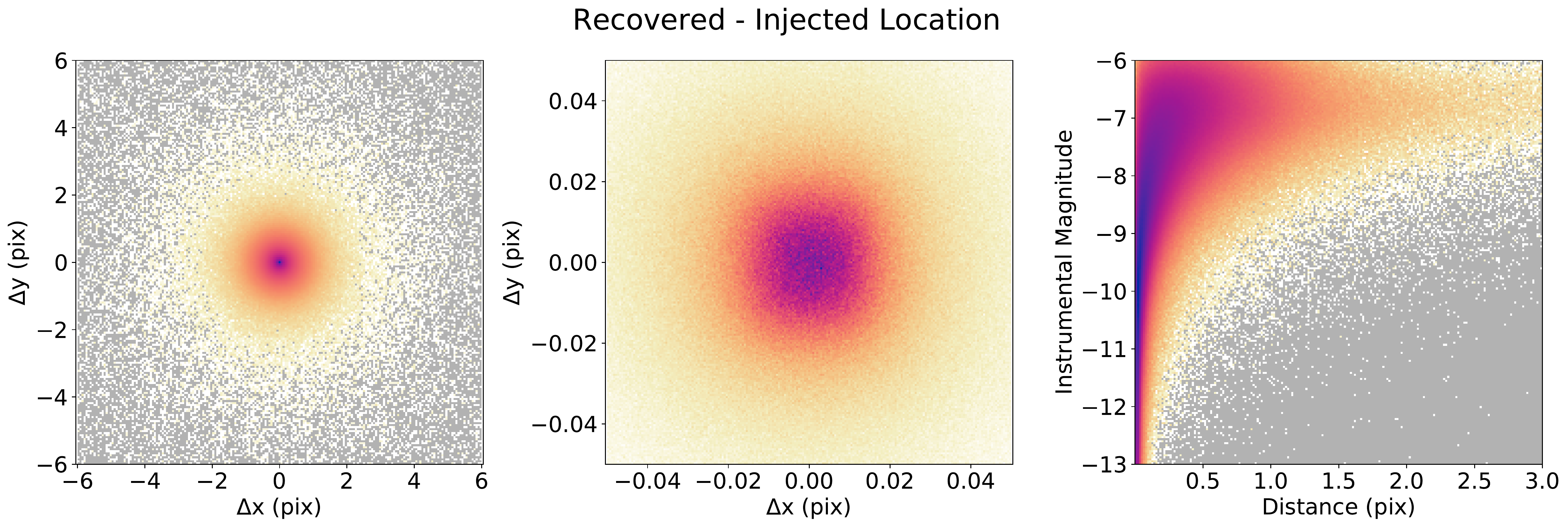}
\caption{(Left): Histogram (2D) of the position errors for recovering injected sources. Color scale shows log density (light, white, low; dark, blue, high). (Center): Same as (Left), but with linear density and focused on the central peak. (Right): Histogram (2D) of the distance error between the recovered and injected source locations as a function of the brightness of the injected source. Color scale shows log density (light, white, low; dark, blue, high).}
\label{fig:loc_prec}

\includegraphics[width=\linewidth]{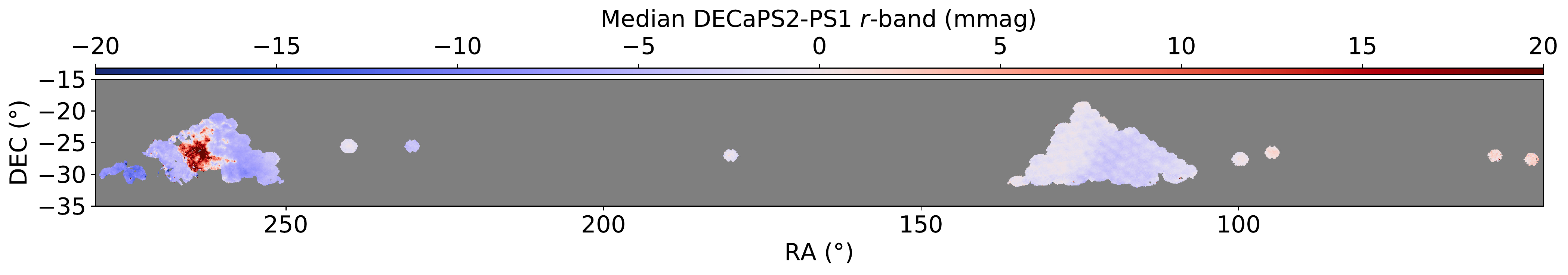}
\caption{Median offset between DECaPS2-PS1 in $r$-band for stars between between $15-17$th mag, as defined using the color transform between the PS1 and DECaPS filter systems which was derived using the calibration field at $({\rm RA,DEC})=(100\degree,-27\degree)$.
}
\label{fig:ps1color}
\end{figure*}

For the \crowdsource runs, a first pass limiting memory to 3.8 G/exposure succeeded on $\sim80\%$ of the survey footprint. The remaining exposures, which represent fields with high stellar densities, were rerun with 9.8 GB of memory. The total \crowdsource runtime was 230k core-hours. The \cloudcoverr runs used $3.6-4.8$ GB per exposure (median 3.7 GB) and completed in 203k core-hours (see \cite{Saydjari:2022:arXiv:} for more details). The runtimes demonstrate high linearity with respect to the number of stars per exposure (Figure \ref{fig:runtime}). Building images for the Legacy Survey Viewer using \texttt{legacypipe} took $\sim 2.4$k core-h each for the images and model images, with $95\%$ of the run completing with 7.7 GB of RAM (the rest completing with a 35 GB/core allocation). Processing the raw images by the DECam Community Pipeline was largely I/O bound.

A bug in the \crowdsource injection module was found after the run, which impacts the ``true'' source locations saved in the \texttt{MCK} extension of the \crowdsource injection catalogs. The pixelization of the PSF and the integer location for its injection into the image was rounded in one place and truncated in another. Thus, the ``true'' injected source coordinate is -1 pixel from the reported coordinate in the injection catalog for sources with positions mod 1 $> 0.5$. An example of how to convert the injection catalog to the true positions is provided in Section \ref{sec:dataavil}.

\section{Location Precision} \label{sec:locprec}

We use the injection tests described in Section \ref{sec:inject} to evaluate the precision with which \crowdsource can recover source locations, focusing on the unblended limit. Figure \ref{fig:loc_prec} shows histograms of the difference between the recovered and injected locations of the synthetic sources added to the image. We restrict to unblended sources by requiring \texttt{fracflux} $> 0.75$, that no source in the original image within 2 FWHM of the injected location, that the recovered source location is not pulled toward the nearest neighboring source by more than 1 pixel, and that the source/background flux be positive.

The left panel shows the logarithmic density over a larger range and exhibits rotational symmetry. The median offset in both x and y are $<1 \times 10^{-3}$ pixels, illustrating the accuracy of the recovered source locations from \crowdsourcens. The middle panel focuses on the central peak with a linear density scale, indicating slight distortions from full rotational symmetry. The right panel shows a logarithmic histogram of the recovered-injected distance as a function of the injected source magnitude. The astrometric precision is clearly a function of the source flux, decreasing for fainter sources. Characterizing the faint astrometry would require a more careful consideration of the cuts used to exclude blended or misidentified sources.

To compare to Figure \ref{fig:gaia_gcdist}, we report the median recovered-injected distance for bright injected sources (brighter than $\sim19$th $g$-band mag, $-11$ instrumental mag) which is 0.016 pixels, or 4 mas given the DECam pixel scale. This would be the astrometric performance if sources had exactly the \crowdsource PSF model and the WCS solution across the CCD were perfect. In DECaPS2, the WCS solution used is that from the CP, without any modification. Even over the portion of the DECaPS2 footprint with WCS solutions from Gaia eDR3, the typical median astrometric errors per NSide=512 HEALPix pixel were 17.5 mas. This suggests that \crowdsource is not the limiting factor on astrometry and improvements on astrometry beyond 17.5 mas will require careful treatment of higher-order terms in the WCS, including effects such as those considered by \cite{Bernstein:2017:PASP:} and \cite{Fortino:2021:AJ:}.

\section{PS1 Color Transform Variation} \label{sec:colorTvary}

The DECaPS2 absolute calibration is tied to PS1 via a color transform described in Section \ref{sec:PS1}. The color transformation was derived on a low-reddening field so that it could capture the effects of varying stellar temperature without confusion caused by the effect of varying reddening. 

While the color transformation used explicitly does not account for how dust affects colors, we can nevertheless apply it even in highly extinguished areas to measure how PS1 and DECam colors compare across the survey footprint. Figure \ref{fig:ps1color} shows this comparison for $r$-band, restricting to stars measured in PS1 $g,r,i$-band, DECaPS $r$-band, and between $15-17$th mag in $r$-band. The crossmatch requires sources in PS1 to be within 0.5" of the source location in DECaPS2 and that there be no more than one match for a given source.

In the outer Galaxy, small offsets in DECaPS2-PS1 of $<5$ mmag are observed. In the inner Galaxy, outside of highly extinguished regions in the Pipe Nebula, typical offsets are negative and $<15$ mmag. Over the entire overlap, the standard deviation of the offsets, excluding regions of high extinction around the Pipe Nebula, is 3.7 mmag. We consider this to be excellent agreement between two different surveys performed by different systems and tied together only by a single field. A few regions with offsets of $\sim 50$ mmag are are observed in regions of high extinction, reflecting our color transformation's neglect for the effect of extinction.

Another DECam program, the Blanco DECam Bulge Survey (BDBS), suggested that DECaPS was inconsistent with PS1 by 200 mmag on several bulge fields and in several filters, including r-band \citep{Johnson:2020:MNRAS:}.  However, this large offset was due to the combination of not applying a color transformation and performing the comparison in regions of large reddening.  Accounting for these two effects resolves the apparent discrepancy \citetext{private communication, Johnson, 2022}.

\onecolumngrid
\section{Detection Threshold Bias} \label{sec:threshbias}

We derive the photometric bias due to the presence of a detection threshold as a function of the true SNR of the source. Then, we apply an approximation to obtain a functional form which can be fit to the faint-limit bias in synthetic injection tests. The single free parameter in this fit is exactly the detection limit in SNR of the photometric pipeline.

Suppose we are considering an isolated source (star) with true flux $g f_0$ on a constant background with flux $g b_0$, where $g$ is the detector gain so that we convert to and work entirely in photoelectrons. The point spread function (PSF) of the source through the imaging system is $p (x,y)$, which has an effective area $A_{\rm{eff}} = \frac{\int p}{\int p^2} = \frac{1}{\int p^2}$.

We work in the faint limit where the measurement noise is dominated by the Poisson statistics of the background counts. We will further work in the large count limit of all Poisson distributions so that we can take the Gaussian approximation. In this limit, we can define the SNR of the (true) source $s_0$ and the flux of a source just at the detection threshold ($f_d$) imposed by a detection limit $s_d$ on (observed) SNR.

\begin{ceqn}
\begin{align} \label{eq:s0}
s_0 = \frac{g f_0}{\sqrt{A_{\rm{eff}} g b_0}} = \frac{g f_0}{\sigma} \quad \text{and} \quad  s_d = \frac{g f_d}{\sqrt{A_{\rm{eff}} g b_0}} = \frac{g f_d}{\sigma}
\end{align}
\end{ceqn}

In addition to $s_0$ and $s_d$ considered here, both the apparent SNR given the recovered flux $s_r$ and the SNR of the injected source accounting for the realization of the Poisson noise $s_i$ are distinct quantities important for future analyses.

To measure the bias, we compute the expectation value of $y = g f - g f_0 + \epsilon$, where $f$ is a draw of the source flux and $\epsilon$ is a draw of the background-induced noise on the PSF photometry. In the background-dominated noise limit, $y \sim \mathcal{N}(\mu = 0,\,\sigma^{2} = A_{\rm{eff}} g b_0)$. However, we only detect a source when the estimated flux has sufficient SNR, or as a flux condition, when $g f + \epsilon > g f_d$. Thus, we take the expectation value only over values of $y$ above threshold.

\begin{ceqn}
\begin{align} \label{eq:expect}
    \left< g f - g f_0 \right> = \frac{\int_{f_d - f_0}^\infty dy\, h(y) y}{\int_{f_d - f_0}^\infty dy\, h(y)} \quad \text{where} \quad h(y) = \frac{1}{\sqrt{2\pi}\sigma} e^{-\frac{1}{2}\left(\frac{y-\mu}{\sigma}\right)^2}
\end{align}
\end{ceqn}
is the usual Gaussian PDF given $y \sim \mathcal{N}(\mu = 0,\,\sigma^{2} = A_{\rm{eff}} g b_0)$. These are just improper Gaussian integrals and, normalizing by $g f_0$ so we can talk about fractional errors, we obtain

\begin{ceqn}
\begin{align} \label{eq:expectRes}
    \frac{\left< g f - g f_0 \right>}{g f_0} = \sqrt{\frac{2}{\pi}}\frac{\sigma}{g f_0}\frac{e^{-x^2}}{\erfc(x)} \quad \text{where} \quad x = \frac{g f_d - g f_0}{\sqrt{2}\sigma}
\end{align}
\end{ceqn}
and $\erfc$ is the complementary error function.

Using Equation \ref{eq:s0}, we can transform this result to be only in terms of the true SNR of the source $s_0$ and the detection threshold SNR $s_d$. Then it is clear that all factors of the gain, effective area of the PSF, and the background level drop out.

\begin{ceqn}
\begin{align} \label{eq:expectResSNR}
    \frac{\left< f - f_0 \right>}{f_0} = \sqrt{\frac{2}{\pi}}\frac{1}{s_0}\frac{e^{-x^2}}{\erfc(x)} \quad \text{where} \quad x = \frac{s_d - s_0}{\sqrt{2}}
\end{align}
\end{ceqn}

Under the limits taken, this result is exact, but difficult to work with given that $\erfc$ is not an elementary function. Since we are interested in the faint limit where the sources are near threshold ($x = f_d - f_0 \approx 0$), we use the approximation for $\erfc$ due to \cite{karagiannidis2007improved}, often called the KL approximation.

\begin{ceqn}
\begin{align} \label{eq:KLapprox}
    \erfc(x) \approx \frac{\left(1-e^{-Ax}\right)e^{-x^2}}{B\sqrt{\pi}x} \quad \text{where} \quad (A,B) = (1.98, 1.135)
\end{align}
\end{ceqn}

Substituting into Equation \ref{eq:expectResSNR}, we obtain a functional form for the threshold-induced bias with a single free parameter that can be easily fit to data from synthetic injection tests. That single free parameter $s_d$ is the effective detection threshold of the photometric pipeline which can differ from the targeted threshold and is not known a priori. Given that the KL approximation is best for $x>0$ and $\erfc(-x) = 2 - \erfc(x)$, one could define Equation \ref{eq:KLapprox} piecewise for higher accuracy. Further, one could simply evaluate $\erfc$ at every point and use the exact form (Equation \ref{eq:expectResSNR}) for the fit. However, the KL approximation, which overestimates the bias at $x<0$ ($s_0 > s_d$) is a slightly better fit (by approximately a factor of 2 in the residuals) to the data than the true functional form using Equation \ref{eq:expectResSNR}). We expect this is the result of unmodeled biases, such as in the selection we apply to the injection tests, which will be the focus of future work.

\begin{ceqn}
\begin{align} \label{eq:simpleExpect}
    \frac{\left< f - f_0 \right>}{f_0} = \frac{B\left(\frac{s_d}{s_0} - 1\right)}{1 - e^{-\frac{A}{\sqrt{2}}\left(s_d-s_0\right)}} \quad \text{where} \quad (A,B) = (1.98, 1.135)
\end{align}
\end{ceqn}

At exactly $s = s_0$, Equation \ref{eq:simpleExpect} has a removable singularity and we find

\begin{ceqn}
\begin{align} \label{eq:rmSing}
    \frac{\left< f - f_0 \right>}{f_0}\Bigr|_{s_0 = s_d} = \frac{B \sqrt{2}}{A s_d} \quad \text{where} \quad (A,B) = (1.98, 1.135)
\end{align}
\end{ceqn}

The case where $s_0 = s$ is exactly when $f_0 = f_d$, when the threshold falls in the middle of the Gaussian PDF for the observed flux distribution. That is when a source will be detected exactly 50\% of the time as a result of the detection threshold. Since the source flux where that occurs is usually taken to be the definition of the depth, we can interpret $f_d$ as the photometric depth. In fact, many photometric surveys estimate imaging depth by assuming a photometric pipeline can retrieve $s_d$-sigma (generally 5-6) sources and then using Equation \ref{eq:s0} to solve for the flux limit given the (average) PSF and (average) background counts of the image. However, this measure of depth is limited by that averaging as well as accurately knowing $s_d$.

Instead, using injection tests, we can empirically measure the real $s_d$ achieved by a given pipeline by fitting the faint-limit bias and then use the predicted fractional error at $s_d$ to determine the depth by the flux at which such a bias is observed. In the case of an ideal isolated source with perfect PSF modeling and a constant background, the definition of photometric depth via bias in synthetic injections is identical to the usual definition. 

In the crowded limit, it is not exactly clear how to extend the usual definition of depth. How do we measure when 50\% of injected sources at a given flux are recovered? In the crowded limit, there will often be a detection within a small radius around the injected source location. That detection could be a source that is a ``good'' model of the injected source, a model that combines the flux of the injected source with a neighboring (pre-existing) source, or a detection of a completely different neighboring source and the injected source was simply not detected. One must then make a quality cut on the recovered detections, such as a maximum fractional flux error, to restrict to detections that are ``good'' models of the injected source. However, in the highly crowded limit, the percentage of injected sources that have a high-quality detection might be small and the choice of quality cuts effectively arbitrary, though those choices have a large impact on the reported depth. 

The faint-limit bias approach to measuring photometric depth proposed here does not suffer from these problems. In the crowded limit, where all injected sources may end up absorbing flux from pre-existing sources, this definition very cleanly extends to give the magnitude such that the flux estimates of sources are as biased as they would be by the detection threshold alone in an isolated field. This retains the key notion of photometric depth as an indicator of the faintest reliable star in the catalog.

In the $s_0 << s_d$ limit (source fainter than threshold) of Equation \ref{eq:simpleExpect}, the bias diverges positively. In the limit $s_0 >> s_d$ (source brighter than threshold), the bias becomes exponentially small as a function of $s_0$. This rapid falloff means the detection threshold bias is only important for sources near the threshold. However, the other faint-limit bias discussed in the main text, which results from using the maximum likelihood position, approaches zero fractional bias much more slowly as a function of $s_0$. As such, when fitting the faint-limit bias, we find it necessary to model both sources of bias. We will make the naive assumption that the two biases are independent and simply sum them. Taking the highest-order form in \cite{Portillo:2020:AJ:}, we obtain the combined functional form that we actually fit 

\begin{ceqn}
\begin{align} \label{eq:ExpectAndML}
\boxed{
    \frac{\left< f - f_0 \right>}{f_0} = \frac{B\left(\frac{s_d}{s_0} - 1\right)}{1 - e^{-\frac{A}{\sqrt{2}}\left(s_d-s_0\right)}} + \frac{1}{s_0^2} + \frac{1}{s_0^4} \quad \text{where} \quad (A,B) = (1.98, 1.135)
}
\end{align}
\end{ceqn}

and define the photometric depth as the flux such that the fractional bias is 

\begin{ceqn}
\begin{align} \label{eq:rmSingML}
\boxed{
    \frac{\left< f - f_0 \right>}{f_0}\Bigr|_{s_0 = s_d} = \frac{B \sqrt{2}}{A s_d} + \frac{1}{s_d^2} + \frac{1}{s_d^4} \quad \text{where} \quad (A,B) = (1.98, 1.135).
}
\end{align}
\end{ceqn}

As mentioned in the text, we do not apply a ``correction'' for this bias to the DECaPS2 catalog. The main requirement for such a correction is the need to use a prior on the distribution of true source fluxes (even several $\sigma$ below the detection threshold) so that one can convert the bias on true SNR to a bias on observed SNR. In addition, there is a bias-variance trade-off in applying such a correction (as in \citealt{Portillo:2020:AJ:}) which would need to be computed. However, near the detection limit, the selection of only sources above the threshold subsamples the usual Poisson distribution and can impact the scatter of the source distribution. Finally, especially in the crowded limit, not detecting a source can bias neighboring sources and a low SNR bias correction should depend on the source density as well.

\end{appendices}

\bibliography{DECaPS2.bib}{}
\bibliographystyle{aasjournal}
\end{document}